\documentclass[twocolumn,showpacs,aps,prd,nobibnotes,floatfix]{revtex4-1}

\newcommand{\beq}{\begin{equation}}
\newcommand{\eeq}{\end{equation}}
\newcommand{\beqn}{\begin{eqnarray}}
\newcommand{\eeqn}{\end{eqnarray}}

\newcommand \tb {\tilde{\beta}}
\newcommand \ta {\tilde{\alpha}}

\usepackage{graphicx}
\usepackage{epsf}
\usepackage{longtable}
\usepackage{graphics,epsfig,placeins,subfigure,wrapfig}
\usepackage[usenames]{color}
\usepackage{amsmath}
\usepackage{soul}
\DeclareMathOperator\erf{erf}
\DeclareMathOperator\Erf{Erf}

\begin{document}
\title{Improved Moving Puncture Gauge Conditions for Compact Binary Evolutions}

\author{Zachariah B.~Etienne$^{1,2,3,4}$, John G.~Baker$^{1,3}$, Vasileios  Paschalidis$^5$, Bernard J.~Kelly$^{1,6,7}$, Stuart~L.~Shapiro$^{5,8}$}
\affiliation{$^1$ Joint Space-Science Institute, University of Maryland, College Park, MD 20742}
\affiliation{$^2$ Department of Physics, University of Maryland, College Park, MD 20742}
\email{zetienne@umd.edu}
\affiliation{$^3$ Gravitational Astrophysics Laboratory, NASA Goddard Space Flight Center, Greenbelt, MD 20771}
\affiliation{$^4$ Department of Mathematics, West Virginia University, Morgantown, WV 26506}
\affiliation{$^5$ Department of Physics, University of Illinois at Urbana-Champaign, Urbana, IL 61801}
\affiliation{$^6$ Department of Physics, University of Maryland, Baltimore County, Baltimore, MD 21250}
\affiliation{$^7$ CRESST and Gravitational Astrophysics Laboratory, NASA Goddard Space Flight Center, Greenbelt, MD 20771}
\affiliation{$^8$ Department of Astronomy and NCSA, University of Illinois at Urbana-Champaign, Urbana, IL 61801}

\begin{abstract}
Robust gauge conditions are critically important to the stability and
accuracy of numerical relativity (NR) simulations involving compact
objects. Most of the NR community use the highly robust---though
decade-old---moving-puncture (MP) gauge conditions for such
simulations. It has been argued that in binary black hole (BBH) evolutions
adopting this gauge, noise generated near adaptive-mesh-refinement
(AMR) boundaries does not converge away cleanly with increasing resolution,
severely limiting gravitational waveform accuracy at computationally
feasible resolutions. We link this noise to a sharp (short-wavelength), initial outgoing
gauge wave crossing into progressively lower resolution AMR grids, and
present improvements to the standard MP gauge conditions that
focus on stretching, smoothing, and more rapidly settling this
outgoing wave. Our best gauge choice greatly reduces gravitational
waveform noise during inspiral, yielding less fluctuation in
convergence order and $\sim 40\%$ lower waveform phase and
amplitude errors at typical resolutions. Noise in
other physical quantities of interest is also reduced, and constraint
violations drop by more than an order of magnitude. We expect these
improvements will 
carry over to simulations of all types of compact binary systems, as
well as other $N$+1 formulations of gravity for which
MP-like gauge conditions can be chosen.
\end{abstract}

\pacs{04.25.D-,04.25.dg,04.30.-w,04.30.Db,04.70.Bw}

\maketitle

\section{Introduction}

With the first direct observations of incident gravitational waves
(GWs) expected in only a few years, an
exciting new window on the Universe is about to be opened. But our
interpretation of these observations will be limited by our
understanding of how information about the GW sources is encoded within the
waves themselves. The parameter space of likely sources is large, and
compact binary systems consisting of two black holes (BHs), two neutron stars,
and one of each are likely to be the most promising
sources. However, filling this parameter space with the corresponding
theoretical gravitational waveforms
through merger---when GWs are strongest and many binary parameters
are most distinguishable---will require a large number of
computationally expensive numerical relativity (NR) simulations.

At the heart of the computational challenge lies an arguably even
greater theoretical one, which seeks to find an optimal approach for
solving Einstein's equations on the computer. These approaches generally decompose
the intrinsically four-dimensional set of Einstein's field equations
into time-evolution and constraint equations---similar to
Maxwell's equations \cite{Baumgarte_2010}. Once the data on the initial 3D spatial
hypersurface are specified, the time-evolution equations are
repeatedly evaluated, gradually building the four-dimensional
spacetime one 3D hypersurface at a time. 
Einstein's equations guarantee the freedom to choose coordinates in
the 4D spacetime, and these are specified via a set of gauge
evolution equations. Devising robust gauge conditions,
particularly when black holes inhabit the spacetime, has been a
problem at the forefront of NR for decades. In fact, discovery of gauge
conditions leading to stable evolutions in the presence of moving black holes played a large
role in the 2005 numerical relativity revolution, culminating in the
first successful binary black hole (BBH) inspiral and merger
calculations~\cite{Pretorius:2005gq,Campanelli:2005dd,Baker:2005vv}. 

The most widely adopted formulation for compact binary evolutions
is the highly robust ``BSSN/moving puncture'' (hereafter BSSN/MP)
formulation \cite{Campanelli:2005dd,Baker:2005vv}, which combines the
Baumgarte-Shapiro-Shibata-Nakamura (BSSN) 3+1 decomposition of
Einstein's equations
\cite{Nakamura:1987zz,Shibata:1995we,Baumgarte:1998te} with the
``moving puncture'' (MP) gauge conditions that implement the {\tt 1+log}/$\Gamma$-freezing gauge
evolution equations \cite{Alcubierre:2002kk,vanMeter:2006vi}. It 
has been about a decade since BSSN/MP was first developed, and it
remains in use by much of the NR community, largely unmodified. This
is a testament to the robustness of the MP gauge choice, as well as the
difficulties associated with devising stable techniques for evolving
spacetimes containing BHs.

The BSSN/MP equations are typically solved on the computer with
high-order finite-difference approaches on adaptive-mesh refinement
(AMR) numerical grids
\cite{Zlochower:2005bj,Baker:2006yw,Bruegmann:2006at,Campanelli:2007ew,Herrmann:2007ac,Sperhake:2006cy,Etienne:2007jg,Pollney:2007,Pollney:2009yz}. Such
grids are essential for computational efficiency, as the simulation
domain must be hundreds to thousands of times larger than the initial binary
separation to causally disconnect the outer boundary from the physical
domain of interest for the duration of the simulation. This effectively prevents errors linked to the
enforcement of approximate outer boundary conditions from propagating
inward and lowering the convergence order of our gravitational waveforms. In
addition, very high resolution must be used in the strongly curved 
spacetime near the BHs, while resolution requirements are much lower far from
the binary, where GWs with wavelengths of order $10M$
must be accurately propagated ($M$ is the total ADM mass of the
system). Thus, AMR grids in MP calculations
typically resolve the strong-field region best, with progressively
lower resolutions away from this region to the outer boundary. Typical
gridspacings at the outer boundary can be $\sim 1000$ times larger
than in the strong-field region.

The BSSN/MP+AMR paradigm has been most thoroughly tested in the
context of BBH evolutions, and a large number of theoretical BBH
GWs have been produced with this paradigm that widely sample parameter
space (e.g., \cite{Lousto:2013wta,Pekowsky:2013ska,Hinder:2013oqa}).
However, when the BSSN/MP+AMR paradigm is pushed to very high
accuracies or to difficult regions of BBH parameter space, it has been
found (e.g., \cite{Zlochower:2012fk}) to yield inconsistent
gravitational waveform convergence with increasing resolution,
making error estimates difficult, and effectively limiting waveform
accuracy. In \cite{Zlochower:2012fk} it was hypothesized that these
convergence issues might be linked to noise from
high-frequency waves generated at grid refinement boundaries.

\begin{figure}
\includegraphics[angle=270,width=0.45\textwidth]{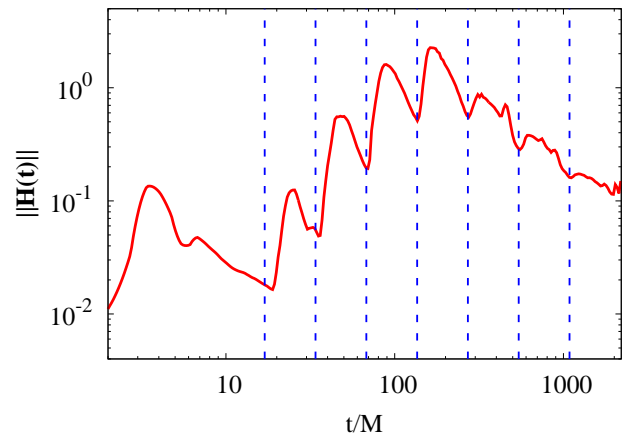}
\caption{L2 norm of Hamiltonian constraint violation
  $||\mathcal{H}(t)||$ outside black hole horizons (Eq.~\ref{L2_Ham}),
  versus time (solid red curve), for ``standard'' BSSN/MP+AMR BBH
  calculation. Dashed blue vertical lines denote the times at which an
  outgoing wave traveling at $\sqrt{2}c$---starting from the origin at
  the onset of the calculation---encounters a grid refinement
  boundary.}
\label{Intro_Ham_constraint_bumps}
\end{figure}

This is a compelling hypothesis, especially in light of
Fig.~\ref{Intro_Ham_constraint_bumps}, which shows
regularly-spaced spikes in L2 norm Hamiltonian constraint violations
on a logarithmic time scale. Interestingly, the onset of each spike
in this BBH evolution is timed almost perfectly to grid refinement
crossings of a wave that starts from the strong-field region at the
onset of the calculation and propagates outward {\it superluminally at
  speed $\sqrt{2}c$} (vertical lines). This propagation speed is a
smoking gun, as linearized analyses \cite{Alcubierre:2002iq} indicate only one propagation mode
with that speed in the standard BSSN/MP formulation. This mode
primarily involves the lapse and is governed by the lapse evolution
gauge condition. The connection between spikes in Hamiltonian
constraint violations and the initial outgoing lapse wave has been
noted previously~\cite{Baker:2006yw}.

Expanding on this idea, we note that all BSSN evolution equations are
coupled to the lapse and its derivatives, so noise
from reflections in this {\it gauge} quantity can be easily converted
into noise in {\it physical} modes (e.g., noise in GWs) and even constraint
violations. Further,
differencing errors from the part of the wave that is transmitted into
the coarser grid are converted into constraint violations and errors in
non-gauge waves as well. 
We expect improved resolution would mitigate this
problem, but even a factor of two increase---the
typical range in most BSSN/MP convergence studies---would enable
evolution of a sharp (short-wavelength) feature through only one
additional coarser level before it succumbs to dissipation through
under-resolution. Additionally, it is likely that zero- or low-speed
modes in Hamiltonian constraint violations (see, e.g.,
\cite{Gundlach:2004jp}) may exacerbate any errors generated as the
initial sharp lapse wave propagates into a coarse grid.

Since sharp features in the initial outgoing lapse wave are likely
responsible for spikes in Hamiltonian constraint violations, as well
as noise and other errors in physical quantities, we can conceive of
at least two strategies for mitigating this effect: (1) adjusting the
initial conditions to minimize gauge dynamics, in a similar vein
to~\cite{Hannam:PRD80:2009}, or (2) modifying the standard MP lapse
evolution equation so that the initial sharp outgoing lapse wave is
stretched and smoothed as it propagates outward from the binary. We
chose the latter strategy. Strikingly, we find our stretching 
and smoothing modifications to the lapse evolution equation are
sufficient to reduce constraint violations by {\it more than an order
  of magnitude}, confirming our hypothesis, {\it except near the
  beginning of our calculations}, where early-time spikes in
constraint violations remain.

We remove these stubborn spikes by initially reducing the timestep on
the BSSN field equations by a factor of $\sim 10^3$, while the
gauge evolution equations are evolved according to the original time
coordinate. This time reparameterization technique increases the gauge
wave speeds to about 30 times the coordinate speed of light, allowing
the gauge evolution to very quickly relax near $t=0$ in response to
the physical fields. As the gauge fields relax, evolution of the BSSN
field variables is slowly accelerated so that their timestep matches
that of the MP gauge evolution, after about $10M$ of evolution in the
un-reparameterized time coordinate. We show this time
reparameterization is, on an analytic level, equivalent to a
time-dependent rescaling of the lapse and shift, leaving the BSSN
field equations unmodified in the reparameterized time coordinate.

When combined, these gauge improvements significantly reduce
constraint violations {\it from the onset of the
calculation through BBH merger}. For example, the L2 norm
Hamiltonian constraint violation outside the BHs is reduced on average
by a factor of 20 at $t>0$ at lowest resolution. Hamiltonian
constraint violations converge at a much higher order in the presence
of our improved gauge conditions, so this factor increases to $\sim
50$ at the highest resolutions attempted.

Turning to gravitational waveforms, we find that in the ``standard''
BSSN/MP+AMR paradigm, noise appears to dominate GW power at wave periods
$\sim 10M$, and this spike in power does not diminish with increasing
resolution. Meanwhile, our ``best'' gauge choice appears to
drop this noise-generated GW power spike by about an order of
magnitude, and its power spectrum at short wave periods {\it does}
drop with increasing resolution. In conjunction with this
removal of waveform noise---and perhaps most significantly---we find that
gravitational waveform convergence is far cleaner when using the
new techniques, enabling more reliable Richardson extrapolations and
possibly paving the way to higher accuracies than the standard
BSSN/MP+AMR paradigm might allow. In fact, our ``best''
gauge choice reduces Richardson-extrapolation-based error estimates of
GW phase and amplitude error during inspiral by about
40--50\%. However, we observe in these waveforms that only the
phase error is reduced significantly at merger.

Our gauge improvements are presented exclusively in the 
challenging context of an 11-orbit, equal-mass BBH calculation in
which one BH is nonspinning and the other possesses a spin of
dimensionless magnitude 0.3, aligned with the orbital angular
momentum. Despite our focus on a single physical
scenario, we expect these new techniques will improve the accuracy and
reliability of simulations couple the MP+AMR
paradigm to BSSN and other $N$+1 NR formulations (e.g.,
\cite{Bona:2003fj,Gundlach:2005eh,Bernuzzi:2009ex,Zilhao:2011}),
not only for BBH and single-BH evolutions, but also for binary systems
with matter, such as neutron star--neutron star and black
hole--neutron star binaries. Finally, we anticipate that careful
reparameterization of the time coordinate when the gauge and physical
quantities change most significantly---both at the beginning of
evolutions {\it and during merger}---may help stabilize compact
binary simulations, making it easier to cover difficult regions of
parameter space.

The rest of the paper is organized as
follows. Section~\ref{Basic_Equations} summarizes our basic equations
and gauge improvements, Sec.~\ref{Num_Algs_Techs} outlines numerical
algorithms and techniques, Sec.~\ref{Results} details results from
these improved gauge conditions, and Sec.~\ref{Conclusions} summarizes
both our conclusions and plans for future improvements.

\section{Basic Equations}
\label{Basic_Equations}

In the following sections, all quantities are given in geometrized
units: $G=c=1$. Also, with the exception of Sec.~\ref{Num_Algs_Techs},
we designate quantities in terms of $M$, the total initial ADM mass of
our binary.

\subsection{Initial Data}

The Bowen-York initial data \cite{Bowen:1980yu,Brandt:1997tf}
prescription is used to generate an equal-mass, low-eccentricity BBH
system approximately 11 orbits prior to merger. One BH is nonspinning,
and the other is spinning with initial spin parameter $0.3$ aligned
with the orbital angular momentum. We choose this case because it (1)
avoids many of the symmetries of an equal-mass, aligned-spin system
(though is still symmetric about the orbital plane), (2) requires a
significantly long inspiral calculation to merger, and (3) is one of
the cases featured by the Numerical Relativity/Analytical
Relativity (NRAR) \cite{NRARwebsite,Hinder:2013oqa} collaboration
(NRAR case label {\tt U1+30+00} in \cite{Hinder:2013oqa}).

We adopt standard initial values for the lapse and shift, choosing
$\beta^i(t=0)=0$ and $\alpha(t=0)=\Psi^{-1}_\text{BL}$, where
$\Psi_\text{BL}$ is the Brill-Lindquist conformal factor~(Eq.~5
in~\cite{Bruegmann:2006at}). We devised other choices for initial
lapse and shift that more closely approximate their final, relaxed
values. These choices either had no effect or resulted in amplified
constraint violations. We attribute this to the fact that in the
process of settling, the gauge responds to early gravitational-field
dynamics, and \textit{vice versa}. Thus if one wishes to minimize early
gauge dynamics, a different strategy for specifying the initial gauge that
accounts for the early $t>0$ evolution of the spacetime might be
more effective.

\subsection{Evolution Equations and Diagnostics}

\subsubsection{Evolution Equations for Gravitational Fields}
We employ the standard set of BSSN equations for gravitational field
evolution \cite{Baumgarte:1998te}, except $W=e^{-2 \phi}$
\cite{vanMeter:2006g2n,Marronetti:2007wz} is chosen as the
evolution variable instead of the original BSSN conformal variable
$\phi$.
Such a choice or variant thereof (e.g., $\chi=W^2$; see \cite{Hinder:2013oqa} for
other possibilities) is motivated by the the fact that
finite-difference techniques generally approximate functions by
overlapping polynomials when taking derivatives, and $W$ (or $\chi$)
is better approximated by polynomials than $\phi$, particularly in the
strong-field region ($\phi\to \infty$ at black-hole punctures). Thus
choosing $W$ or $\chi$ as an evolution variable results in less
truncation error at a given resolution than using $\phi$.

\subsubsection{Reparameterization of the Time Coordinate}
\label{sec:TR}

We present a new modification of the BSSN/MP equations, originally
aimed at reducing an early spike of constraint violations in our BBH
evolutions. Understanding one source of these violations requires some
insights into how our AMR driver Carpet \cite{Schnetter:2003rb} works.

AMR drivers are generally faced with the problem of interpolating data
between grids at different resolutions. Carpet solves this problem by
interpolation ({\it prolongation}) in both space and time. Suppose we
wish to evolve data on a fine (high-resolution) AMR grid with spatial
resolution $\Delta x$ from time $t$ to 
$t+\Delta t$. Also suppose data already exist on the surrounding coarser grid with
spatial resolution $2 \Delta x$ at $t$ and $t \pm2 \Delta t$. After the fine grid
has been updated, Carpet needs to fill the buffer cells between the
fine and coarser levels. However, data do not exist on the coarser
level at $t+\Delta t$, so coarse-grid data at $t$ and $t \pm2 \Delta t$ are
interpolated in space and time to provide data at $t+\Delta t$.

At the start of the calculation, the initial data solver provides data
at only a single time, $t=0$, but we have chosen second-order time
interpolation to fill buffer zones at grid refinement boundaries. Thus
we need data at three timesteps to proceed. Carpet offers two
algorithms for filling in these two missing timesteps. The first is
to simply copy the data from $t=0$, and the second evolves data at
$t=0$ backward and forward in time to fill them in. The latter
technique reduces truncation errors at a higher order of convergence
than the former, but constraint violations in either case just after
the evolution begins will generally be much larger than at $t=0$, as
the initial data are extremely accurate and are usually generated on
high-resolution spectral grids.

To mitigate the errors at the start of the calculation, we choose to
copy the data from $t=0$ {\it and} reparameterize the time coordinate
so that the timesteps near $t=0$ are incredibly small, compared to
our original time coordinate. The coordinate timestep then very
gradually and smoothly grows to the usual value. In particular, we
find the following prescription works well.

All the BSSN/MP evolution equations can be written in the following
form:
\beq
\partial_t \mathcal{E} = {\rm RHS},
\label{RHS}
\eeq
where $\mathcal{E}$ is any of the evolved variables,
$\partial_t$ denotes the usual partial time-derivative and RHS the
right-hand side of the evolution equation. Our first time
reparameterization scheme, which we call ``TR1'', simply multiplies
\textit{every} ``RHS'' by 
\beq
\label{f_of_t}
f(t) = \Erf(t;10,5).
\eeq
Here, we define $\Erf(x;x_0,w)$ as follows
\beq
\label{Erf_of_x}
\Erf(x;x_0,w) = \frac{1}{2} \left[\erf\left(\frac{x-x_0}{w}\right)+1\right],
\eeq
where $\erf(x)$ is the standard, $C^{\infty}$ error
function, $x_0$ marks the center of the error function distribution and
$w$ the width.

Note that although $f(t)$ yields extremely small timesteps early in
the evolution, the timesteps return to the fiducial value
($f(t)\to 1$) smoothly and rapidly. The net effect of TR1 is an
increase in total wallclock time of only $0.5\%$ for our 11-orbit BBH
calculations, equivalent in cost to evolving without time
reparameterization an additional $11M$. Though our time
reparameterization techniques do appear to modify the RHS of the BSSN
field equations, we demonstrate in Appendix~\ref{TR_Appendix} that
Einstein's equations are still satisfied, as both of our time
reparameterization techniques are in fact equivalent to gauge
choices.

Using our ``TR1'' strategy, the initial timestep is $\sim 10^{-3}$
times the  initial timestep without time reparameterization, and
we observe a significant reduction in evolution errors when compared
to copying the initial data alone without time reparameterization,
particularly in the momentum constraint. This finding is demonstrated in
Sec.~\ref{Results}.

Though our main objective in reparameterizing the time coordinate was
to reduce errors induced by copying data from $t=0$ to earlier times,
we find we can reduce errors at early times {\it significantly more} 
by choosing {\it not} to reparameterize time in the gauge evolution
equations (i.e., the lapse and shift). We call this our ``second''
time reparameterization choice, or ``TR2''. TR2
increases the characteristic speeds of the lapse evolution equation,
of order 30 times the coordinate speed of light initially \footnote{This can be
seen by combining Eq.~\ref{trK_Evol_Eq} after time reparameterization
with Eq.~\ref{oldlapse}, then taking the principal part, as in
Sec.~\ref{basiceqs:gauge_evol_description}.}. The net 
effect is accelerated evolution of the gauge relative to the
gravitational fields, causing the coordinates to relax much faster.

We apply two other major modifications to the gauge evolution equations,
including one aimed at stretching outgoing gauge waves and another
focused on adding dissipation terms that significantly reduce
short-wavelength noise. These modifications are described in the next
section.

\subsubsection{Gauge Evolution Equations}
\label{basiceqs:gauge_evol_description}

The usual MP gauge-evolution equations consist of the
``Gamma-driver'' shift condition \cite{Alcubierre:2002kk}, of which we choose the
first-order ``nonshifting-shift'' variant \cite{vanMeter:2006vi}, and the {\tt 1+log}
slicing condition for the lapse~\cite{Alcubierre:2002kk}. 
\beqn
\label{gamma_driver_first_order}
  \partial_t \beta^i &=& \frac{3}{4} \tilde{\Gamma}^i - \eta \beta^i,\ \ \ \eta=1.375 \\
\label{oldlapse}
\partial_t \alpha &=& \beta^i \partial_i \alpha - 2 \alpha K.
\eeqn
%
We refer to these gauge conditions as the ``OldGauge''.
The motivation behind this {\tt 1+log} slicing condition becomes
apparent when combined with the evolution equation for $K$, the trace
of the extrinsic curvature,
\beq
\label{trK_Evol_Eq}
\partial_t K = \beta^i \partial_i K - D^i D_i \alpha + \alpha \left(\tilde A_{ij}\tilde A^{ij} +\frac{1}{3}K^2\right).
\eeq
Taking a time-derivative of Eq.~\eqref{oldlapse}, combining with 
Eq.~\eqref{trK_Evol_Eq}, and keeping only the principal parts yields
\beq
\label{opl_lapse_wave_eq}
\partial_t^2 \alpha \simeq 2 \alpha \gamma^{ij}\partial_i\partial_j \alpha,
\eeq
a wave equation with wave speed $\sqrt{2 \alpha}$ (see also
\cite{Alcubierre:2002iq}).

As mentioned previously, our strategy is based on the observation that
the sharp initial outgoing lapse wave appears to generate spikes in the
time evolution of the L2 norm of the Hamiltonian constraint violations when it crosses coarser and
coarser AMR grid-refinement boundaries. Thus we focus on modifying the
{\tt 1+log} slicing condition for the lapse to stretch this initial
wave packet, as well as damp short-wavelength noise in $\alpha$
generated by grid refinement boundary crossings. Gauge freedom
guarantees our ability to do this, and the job is made easier by its
interpretation as a wave equation (Eq.~\ref{opl_lapse_wave_eq}). We
review our improvements to the {\tt 1+log} slicing condition below.

Our improved {\tt 1+log} slicing condition is as follows
\beq
\label{lapse_master_eq}
\partial_t \alpha = 
        {\underbrace {\textstyle 
            f(t) \beta^i \partial_i \alpha
          }_{\text{Term (1)}}}
        -
        {\underbrace {\textstyle 
            g(t,r) \alpha K 
          }_{\text{Term (2)}}}
        +
        {\underbrace {\textstyle 
            h(r) \nabla_{\rm flat}^2 \alpha
          }_{\text{Term (3)}}}
        +
        \text{KO}.
\eeq
Here $r=\sqrt{x^2+y^2+z^2}$.
We now specify the unknown functions and explain each term.

\textbf{Term (1)}: $f(t)=f_0(t)\equiv\Erf(t;10,5)$ (consistent with
Eq.~(\ref{f_of_t})). {\it Note that this is the only term in the
  lapse condition that is multiplied by $f(t)$ in the} TR2 {\it time
  reparameterization, while all other RHS terms in
  Eq.~\ref{lapse_master_eq} are multiplied by $f(t)$ as well in TR1.}

\begin{figure}
\includegraphics[angle=270,width=0.45\textwidth]{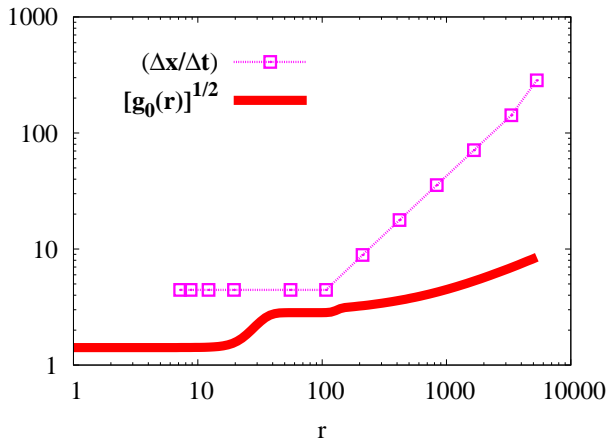}
\caption{Local lapse speed parameter $\sqrt{g_0(r)}$ (solid red line),
  as compared to the maximum allowed speed by the CFL condition at
  each AMR level's most distant point from the origin, at $t=0$ (magenta
  squares). Note that this curve is indistinguishable from
  $g_1(t\gtrsim50,r)$, as $g_1(t,r) \to g_0(r)$ rapidly at
  $t>40$. Note also that in cases for which time reparameterization
  techniques are applied, $t$ refers to the reparameterized time
  coordinate.}
\label{Fig:g0_of_r}
\end{figure}

\textbf{Term (2)}: As in Eq.~(\ref{opl_lapse_wave_eq}),  $\sqrt{g(t,r)
  \alpha}$ provides the wave speed. We stretch outgoing lapse waves by
making $g(t,r)$ a monotonically increasing function of
coordinate radius $r$ (i.e., by accelerating the wave as it propagates
outward). In so doing, one must be careful to avoid exceeding the
maximum local speed allowed by the local Courant-Friedrichs-Lewy
(CFL) condition on each AMR grid. 
In practice, however, this is not a highly restrictive constraint, as
the shift evolution equation's $\eta$ term
(\ref{gamma_driver_first_order}) also sets a stringent upper bound on
the timestep \cite{Schnetter:2010cz}, forcing us to
evolve the outermost AMR levels with a very small timestep
anyway. So without significant modification of current techniques,
we may already stably evolve waves in the outer regions of our grid
that propagate much faster than the local coordinate speed of
light. This enables us to choose the lapse acceleration function
$g(t,r)$ to be a steeply increasing
function  of $r$, and we are careful to select a function that is
easily extensible to arbitrary binary separations,
choosing the following functional form:
\beqn
\label{g0_of_r}
g_0(r) &\equiv& v_i + (v_o-v_i) \Erf(r;x_1,w_1) \\
       &&+ r e^{-4.42} \Erf(r;x_2,w_2), \nonumber
\eeqn
where we set inner wave speed $v_i=2$, outer wave speed
$v_o=8$, $x_1=30$, $w_1=10$, $x_2=130$, and
$w_2=15$. The variable $x_1$ may be adjusted proportionally up (down) for binaries
with larger (smaller) initial separations. Figure~\ref{Fig:g0_of_r}
shows how this function satisfies
the CFL condition $\sqrt{g(t,r)}<\Delta x/\Delta t$ on all
grids. Notice that this choice leads to a safety factor $\Delta
x/\Delta t/\sqrt{g(t,r)}$ of $\sim 2-50$ that reduces errors in time
and ensures that a moving AMR grid (tracking the trajectory of a BH)
does not cross into a CFL-unstable 
region. Notice also that when the lapse wave hits the outer boundary,
its coordinate speed is nearly $10$. In principle, outer boundary
conditions should be adopted to accommodate this radially dependent
superluminal speed, though our outer boundary conditions for the lapse
were not adjusted for the runs presented here, with no
apparent ill effects.

\begin{figure*}
\includegraphics[angle=270,width=0.45\textwidth]{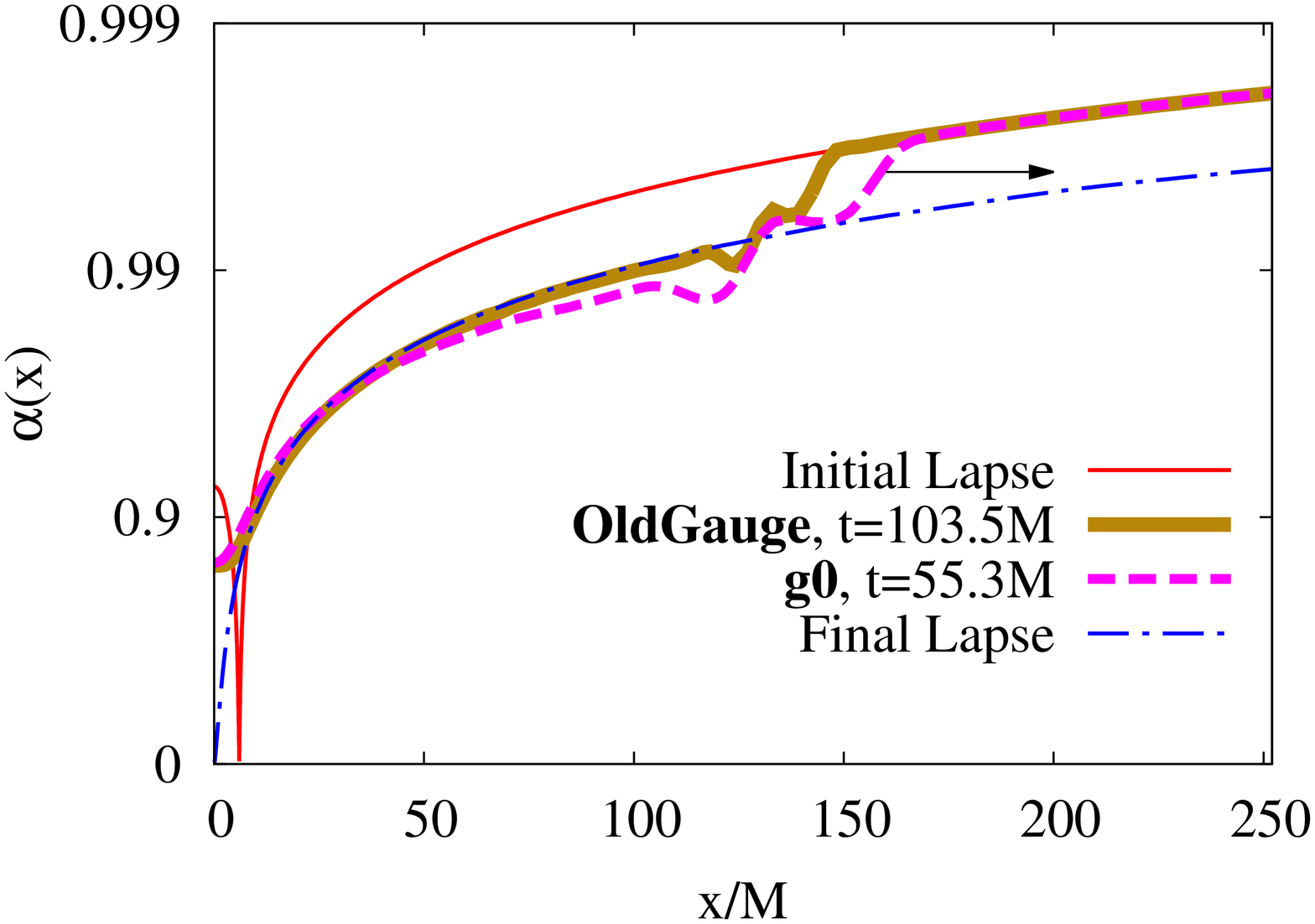}
\includegraphics[angle=270,width=0.45\textwidth]{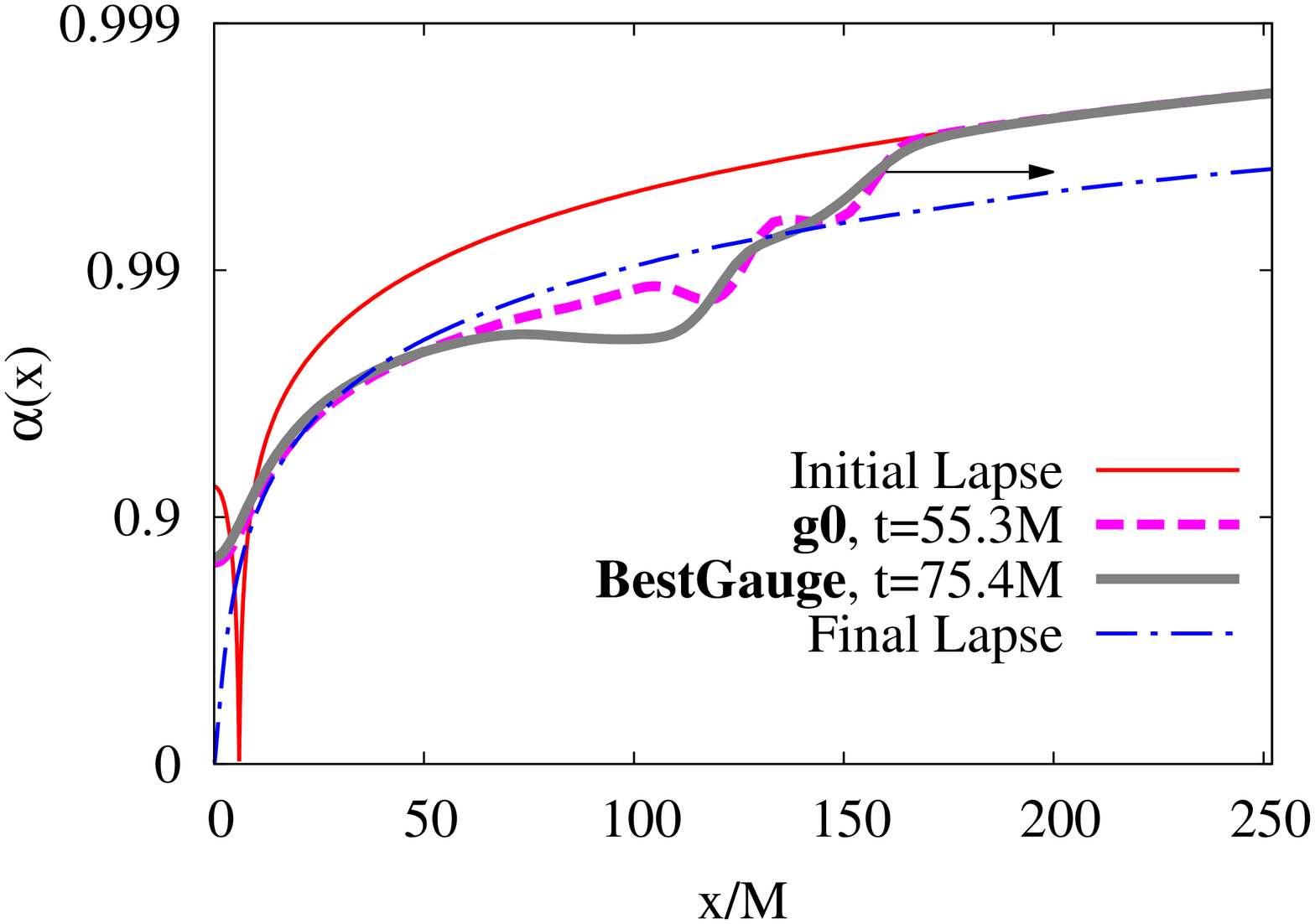}
\caption{Effect of replacing the lapse speed factor of $2$ in the
  ``standard'' BSSN/MP {\tt 1+log} slicing condition (OldGauge,
  Eq.~\ref{oldlapse}) with other functions. {\bf Left panel:}
  Evolved OldGauge lapse along $x$-axis after $\approx 100M$ of evolution
  (thick solid brown) is compared with the g0 lapse
  (thick dashed magenta) as it crosses the same position, at $t\approx 55M$. For
  comparison, initial lapse (thin solid red) is plotted alongside the
  settled lapse of the merged BH remnant (thin dot-dashed blue). The black
  arrow shows the direction of lapse wave propagation.
  {\bf Right panel:} Same as left panel, except the g0
  case (thick dashed magenta) is compared to our ``best'' gauge choice
  BestGauge (g1+TR2+h+d5d7); thick solid
  gray; see Table~\ref{Table_of_Runs} for full details).
}
\label{Fig:lapse_stretch_demonstration}
\end{figure*}

The left panel of Fig.~\ref{Fig:lapse_stretch_demonstration} plots lapse along
the $x$-axis at different times, demonstrating that choosing
$g(r,t)=g_0(r)$ instead of the standard choice $g(r,t)=2$ stretches
the initial outgoing lapse wave. The red-solid curve shows the initial
lapse (the puncture is located at $x/M\approx6$), and the cyan dash-dotted
curve the final, settled lapse. The other curves demonstrate how the
initial lapse wave settles as it propagates away from the strong-field
region (direction of propagation shown by black arrow). In particular, we
find the outgoing lapse wave indeed possesses sharp features. Also, in
the OldGauge case, the wave front crosses $r=140M$ at $t\approx100$,
corresponding to a propagation speed of very nearly $\sqrt{2}$,
consistent with the linear analysis of the lapse evolution
equation. Notice that the choice $g(r,t)=g_0(r)$ does in fact
accelerate and stretch the initial outgoing lapse pulse, leading the
wavefront to the same location as the OldGauge case in about half the
time.

We explore another realization of $g(t,r)$, which, independent of
other terms in Eq.~(\ref{lapse_master_eq}),
introduces a time dependence:
\beqn
\label{g1_of_t_r}
g_1(t,r) &\equiv& [v_i(t) + (v_o-v_i(t)) \Erf(r;x_1,w_1)] \\
  &&+ r e^{-4.42} \Erf(r;x_2,w_2). \nonumber
\eeqn
Here, all parameters are as in $g_0(r)$, except the inner wave speed
$v_i$ is chosen to be a function of time,
\beq 
v_i(t)=0.5 + 1.5 \Erf(t;40,10). 
\eeq
Notice that at early times, $v_i(t)\approx 0.5$, yielding sub-luminal
lapse speeds in the inner region, providing
a greater stretch, and at times $t \gtrsim 40$, $v_i(t)$ rapidly
increases to $2.0$, which brings $g_1(t,r)\to g_0(r)$. We have
attempted evolutions that maintain sub-luminal lapse speeds in the
inner region, and Hamiltonian constraint violations are significantly
reduced. However, maintaining sub-luminal lapse speeds also resulted
in a large increase in the BHs' irreducible masses as well as a very
strong spike in constraint violations near merger, perhaps due to a
gauge shock (see, e.g., \cite{Alcubierre:2002iq}).

We also attempted a much steeper function for $g(t,r)$ in the outer
regions, keeping a more constant safety factor, leading to wave
speeds $\sim 100$ at the outer boundary. This greatly reduced
constraint violations (Hamiltonian constraint in particular) through
much of the inspiral, but the constraints spiked after the
lapse wave returned from the outer boundary and crossed into the
strong-field region. No such reflection was observed with $g_0(r)$ or
$g_1(t,r)$, as we anticipate that the much slower lapse wave speeds
near the outer boundary enabled dissipation terms to act longer
and more effectively, resulting in significantly less reflection.

\begin{figure}
\includegraphics[angle=270,width=0.45\textwidth]{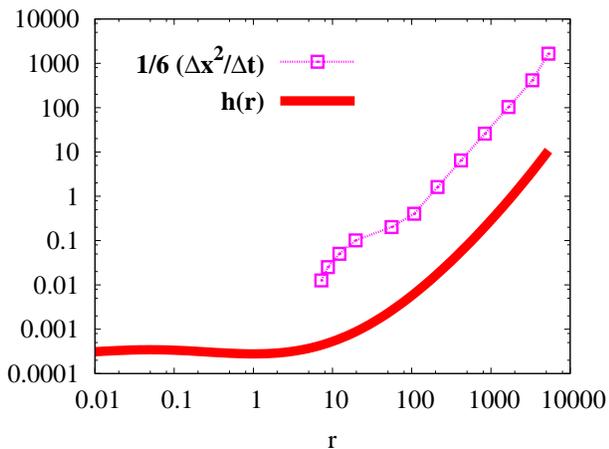}
\caption{Parabolic damping factor $h(r)$ (solid red line), as compared
  to the maximum factor allowed by the CFL condition
  (Eq.~\ref{h_of_r}) at each AMR level's most distant point from the
  origin, at $t=0$ (magenta squares). $\Delta x$ corresponding to the
  ``hhr'' resolution (see Table~\ref{Table_of_Runs}) for which the
  smallest grid spacing is $1/58.\bar{6}$, as the highest resolution
  provides the most stringent constraint on the parabolic damping
  strength.}
\label{Fig:h_of_r}
\end{figure}

\textbf{Term (3)}: The flat-space Laplacian acts as a dissipative filter
  that damps high-frequency oscillations. These oscillations include 
  the sharp features in the problematic wave pulse, and may include noise
  resulting from reflections or interpolation errors at refinement
  boundaries.
  In constructing an $h(r)$ for our
  AMR grids, care must be taken so that the following numerical
  stability criterion for parabolic terms will be satisfied:
\beq
\label{parabolic_CFL_inequality}
h(r) \lesssim \frac{1}{6} \frac{(\Delta x)^2}{\Delta t},
\eeq
where $\Delta x=\Delta y=\Delta z$ is assumed, consistent with our
numerical grids. This is not a strict inequality, as it was
derived from a CFL analysis of the standard heat equation, using
centered second-order finite-differencing and Euler time
integration. Our numerical scheme is more sophisticated, so we reduce
$h(r)$ so that the inequality is comfortably satisfied. In addition,
$\Delta t/\Delta x$ is held fixed as we vary $\Delta x$ in our
resolution studies, so with sufficiently high resolution and
$h(r)>0$, the above inequality will always be violated for explicit 
time integration. To protect against this, $h(r)$ is reduced
even more when higher resolution is deemed necessary. In general, we
choose $h(r)$ such that it is roughly $1/100$ the right-hand side of
the inequality (\ref{parabolic_CFL_inequality}) at all
radii. Specifically, we choose the following functional form for
$h(r)$:
\beq
\label{h_of_r}
h(r) = \exp\left(\sum_{n=0}^8 a_n \log(r)^n\right),
\eeq
where, in standard floating point notation:
\begin{eqnarray*}
a_i &=& \{ \text{\tt
-8.18859,
-2.43507e-3,
8.19733e-2,}\\ && \text{\tt
2.05629e-2,
6.78098e-4,
-2.97857e-4,}\\ && \text{\tt
1.69112e-5,
1.74269e-6,
-1.23471e-7}\},
\end{eqnarray*}
and $r=0$ is mapped to $r=0.01$. This function guarantees the above
inequality is comfortably satisfied at all resolutions in our
simulations. Figure~\ref{Fig:h_of_r} demonstrates how this function
has been adapted to our chosen AMR grid structure to ensure
inequality~(\ref{parabolic_CFL_inequality}) is satisfied. The gap between the
curves indicates the chosen safety factor of $\sim100$. Note that this
function can be adapted to other binary separations, though not as
easily as $g_0(r)$ and $g_1(t,r)$.

\textbf{KO}: This refers to Kreiss-Oliger (KO) dissipation
  terms on the lapse, which are high-order derivatives that behave as
  an artificial viscosity, except unlike the parabolic term, the
  strength of these terms drops to zero as
  $\Delta x \to 0$. We choose the following form for the KO terms:
\beq
\text{KO terms} = d_5(t,r) \text{KO}_5 + d_7(t,r) \text{KO}_7.
\eeq
$\text{KO}_5$ and $\text{KO}_7$ are the standard fifth- and
seventh-order KO derivative operators, respectively \cite{Kreiss73}. The strength
factors $d_5(t,r)$ and $d_7(t,r)$ are normally set to values
$<1$. In fact, when using $i^{\rm th}$-order finite differencing
stencils for spatial derivatives, we typically choose $d_{i+1}=0.3$
and zero for all other $d_j$. We find that stronger, mixed-order KO
dissipation on the lapse reduces errors in our BBH calculations beyond
the typical choice. However, we find empirically that care must be
taken to avoid very strong, mixed-order KO dissipation at the puncture
or in the binary region. A significant reduction in constraint
violations is observed when $d_5(t,r)$ and $d_7(t,r)$ are set as
follows: 
\beqn
\label{d5}
d_5(t,r)&=& 0.98 \Erf(t;25,5) \Erf(r;15,5) \\
\label{d7}
d_7(t,r)&=& 0.3 A(\phi) + 0.68 \Erf(t;25,5) \Erf(r;15,5),
\eeqn
where $\phi$ is the BSSN conformal variable, and $A(\phi)=e^{-10 (\phi/0.3)}$
for $\phi>0.3$ (the region just surrounding the apparent
horizons) and 1 otherwise.

In the right panel of
Fig.~\ref{Fig:lapse_stretch_demonstration} we plot the initial
outgoing wave pulse, comparing a case with all the
new gauge features just described (BestGauge) to a case that only
includes one new feature (g0); see Table~\ref{Table_of_Runs} for
details. The difference is dramatic, especially
when compared to the original gauge choice (left plot).

We conclude this section with two comments. First, a note on
well-posedness and hyperbolicity of the new terms. The introduction of
terms (1), (2) and (3) modify the principal part of the BSSN system
and in general a hyperbolicity analysis with terms (1) and (2) would
be needed to demonstrate that the underlying system admits a
well-posed initial value problem. Once strong hyperbolicity is
established with terms (1) and (2), well-posedness of the Cauchy
problem including term (3) can be studied using standard methods that
apply to mixed hyperbolic--parabolic PDEs (see,
e.g.,~\cite{Gustafsson:1995,Gundlach:2006}), because performing a
low-order reduction of Eq.~\ref{lapse_master_eq} yields a second-order
operator which has parabolic-like properties. However, a
well-posedness analysis of this system goes beyond the scope of our
paper. Nevertheless, we have indirect evidence to believe that
mathematical well-posedness can be established because our simulations
are stable, and convergent, and because the solutions obtained with
our new gauge are close to the solutions obtained with the old gauge,
for which the BSSN system becomes strongly hyperbolic and hence admits
a well-posed Cauchy problem.

Second, we note that $g_0(r)$, $g_1(t,r)$,
and $h(r)$ are $C^{\infty}$ for $r>0$, but all three functions possess
a small kink at $r=0$. In addition, $A(\phi)$ has a significant kink
just outside the apparent horizons. This lack of differentiability in
the evolution equations could potentially translate to non-convergent
errors in simulations. To determine if non-smoothness in $g_1(t,r)$
has any significant effect, we devised a ``smoothed'' version of
$g_1(t,r)$,
\begin{eqnarray}
g_{1s}(t,r)&=&g_0(t,r)-r\frac{\partial g_1}{\partial r}(t,0)\nonumber\\
         &=&g_0(r)-r\frac{v_0-v_i(t)}{w_1\sqrt{\pi}}\exp\left(-\frac{x_1^2}{w_1^2}\right)
\end{eqnarray}
nearly identical to the
original function at $r/M \gg 1$ but differentiable at $r=0$, and
performed a run at ``lr'' (low) resolution with this new
function. Though our results were unchanged, we would generally
recommend choosing functions that are at least once differentiable
everywhere, including $r=0$.

\section{Numerical Algorithms and Techniques}
\label{Num_Algs_Techs}

To maintain round numbers in the following section, we specify all
quantities in code units. For conversion to physically
meaningful numbers, note that in code units the summed masses of the
BH apparent horizons is 1.0 and the total ADM mass $M$ of the
initial system is 0.99095, though we stress that these BBH
calculations are scale-free in terms of mass.

\subsection{Initial Data Parameters}

To generate BBH initial data, we use the {\tt TwoPunctures} code
\cite{Ansorg:2004ds}, with spectral grids of 
resolution $(n_A,n_B,n_{\phi})=(64,64,44)$. These spectral data are
then mapped onto our finite-difference grids via the new spectral
interpolation scheme of \cite{Paschalidis:2013oya}. The full set of
{\tt TwoPunctures}/Bowen-York \cite{Bowen:1980yu}
parameters used to generate these BBH initial data are as specified
for case {\tt U1+30+00} in Table~A2 of \cite{Hinder:2013oqa}. Note that these
parameters were first generated by the AEI group (cf. case {\tt
  A1+30+00} in the same table). With these parameters, the bare mass
of the \{spinning,nonspinning\} BH is found to be {\tt
  \{0.4698442439908046,0.48811120218500131\}}, and these bare masses
yield ADM masses of \{0.5,0.5\} as measured at spatial infinity on the
individual punctures' interior worldsheets (see,
e.g.,~\cite{Baker:2002gm,Tichy:2003qi} for how this is computed).

\subsection{AMR Grid Parameters}
\label{nummethods:AMRgridparams}

To minimize computational expense, all grids impose symmetry across
the orbital plane ($z=0$), and only the upper-half plane ($z>0$) is
evolved. Evolution equations are evaluated on two nested sets of
Carpet-generated AMR grids, with one set tracking the centroid of each
apparent horizon with half-side-lengths $0.75 \times 2^n$, where
$n=\{0,\dots,3\},\{5,\dots,12\}$ ($n=4$ is skipped, so that the
region around the binary is better resolved). The lowest-resolution
grid (containing the outer boundary) has a half-side-length of 3072,
which is out of causal contact from all gravitational wave extraction
radii, for waves propagating at or below the speed of light, for the
duration of the simulation.

To reduce errors due to our low-order time-evolution (fourth-order)
and AMR time prolongation (second-order) below the level of
higher-order spatial finite differencing and prolongation errors, the
Courant-Friedrichs-Lewy (CFL) factor is set at most to 45\% of its
maximum value for BSSN evolutions (i.e., 0.225 instead of 0.5). 
Specifically, the Carpet parameter {\tt
  time\_refinement\_factors} is set to {\tt
  "[1,1,1,1,1,1,1,2,4,8,16,32]"}, which corresponds to local CFL
factors $\Delta t/\Delta x$ of 0.225 for the highest six levels of
refinement, and $0.225/2^n$ for the $n$th level beyond that, where
$n=\{1,...,6\}$. 

For each gauge choice, we perform simulations with four different
maximum grid resolutions, labeled \{lr,mr,hr,hhr\} in order of increasing resolution.
For \{lr,mr,hr,hhr\} runs, the most refined AMR grid (at each
BH) has resolution
$\Delta x=\{1/42.\bar{6},1/48,1/53.\bar{3},1/58.\bar{6}\}$,
corresponding to an average of
$\{37.5,42.1,46.8,51.5\}$ gridpoints across the average diameter of
the apparent horizons (at $t\approx800$), respectively. (Note that the
apparent horizon diameters of the two black holes differ by approximately 4\%). These grids
and resolutions have been carefully chosen so that at all 
resolutions, the physical extent of each grid-refinement level is the
same. Unless grids and resolutions are carefully chosen, Carpet will
not respect the desired physical extent of grid-refinement levels,
instead rounding the physical size to the nearest gridpoint, which can
potentially render a convergence study inconsistent.

Spatial and temporal prolongation (i.e., interpolation between AMR
boundaries to fill buffer zones) are set to fifth- and second-order,
respectively. Also, the standard technique for reducing AMR buffer
zones as described in, e.g., \cite{Bruegmann:2006at}, is not applied
here, as there are indications that reducing AMR buffer zones may
result in inaccuracies \cite{Zlochower:2012fk}. 
Thus, we require the use of 16 spatial buffer zones at AMR boundaries
(i.e., four ghostzones due to sixth-order upwinded finite-difference
stencils and seventh-order Kreiss-Oliger dissipation, multiplied by
four Runge-Kutta time evolution substeps).
 
\subsection{Evolution Parameters}

We use the Illinois group's finite-difference (FD) BSSN sector of
their AMR GRMHD code \cite{Etienne:2007jg} for spacetime evolutions,
with sixth-order-accurate upwinded FD stencils for all shift advection
terms, and sixth-order centered FD stencils on all other spatial
derivatives. Explicit fourth-order Runge-Kutta (RK4) timestepping is
used for time-evolution, and the CFL factors on each refinement level
are specified in Sec.~\ref{nummethods:AMRgridparams}.

Seventh-order Kreiss-Oliger dissipation with strength
parameter $\epsilon=0.3$ is applied to all BSSN gravitational field
and shift evolution equations. Kreiss-Oliger dissipation on the lapse
is applied on a case-by-case basis, as specified in
Table~\ref{Table_of_Runs}.

\subsection{Diagnostics}

Apparent horizon tracking and diagnostics (including the monitoring of
irreducible masses) are handled via the {\tt AHFinderDirect} thorn
\cite{Thornburg:2003sf} for all BHs in runs presented here. 
We also employ an isolated horizon formalism
\cite{Ashtekar:1998sp,Dreyer:2002mx} diagnostic to monitor black hole
spins and masses throughout the evolutions.

Hamiltonian and momentum constraint violations are monitored via the
following L2-norms:
\beqn
\label{L2_Ham}
||\mathcal{H}||   &=& \sqrt{\int_\mathcal{V} \mathcal{H}^2\ d^3x},\\
\label{L2_Mom}
||\mathcal{M}^i|| &=& \sqrt{\int_\mathcal{V} \left(\mathcal{M}^i\right) ^2\ d^3x},
\eeqn
where $\mathcal{H}$ ($\mathcal{M}^i$) is the locally computed
Hamiltonian (momentum) constraint 
violation (as defined in, e.g., \cite{Etienne:2007jg}) and $\mathcal{V}$
includes the entire simulation volume, 
excluding spheres of radius $2.2M$ about each apparent horizon's
centroid. As a point of comparison, the maximum apparent horizon
radius of all BHs in these calculations varies between $\approx 0.23M$
and $0.77M$. Thus in all cases this integral excludes a region
both in and around each BH.

Gravitational waves are extracted at ten radii, equally spaced in $1/r$,
from $r=45.19M$ to $r=192.7M$. In particular, we compute the spin-weight
$-2$ spherical harmonic decomposition of the Newman-Penrose Weyl scalar
$\psi^4_{l,m}$ for all $(l,m)$ modes up to $l=4$. This scalar is
computed using our own heavily-modified version of the {\tt
  PsiKadelia} thorn.
In this paper, we focus primarily on results for waveforms measured
at extraction radius $r=68.6M$, for the dominant ($l=m=2$) mode. We
find that for this dominant mode, our qualitative results are
unchanged at other extraction radii. The effect on other modes is
discussed in Sec.~\ref{Psi4noisereduction}.

The energy and angular-momentum ($z$-component) content of the non-radiative part of the
spacetime is estimated through ADM surface integrals at finite radius
\beqn
\label{M_ADM_def}
M(t) &=& \frac{1}{16\pi} \oint_r \left( \gamma_{ij,i} - \gamma_{ii,j} \right) dA^j,\\
\label{J_ADM_def}
J(t) &=& \frac{1}{16\pi} \varepsilon_{ijk} \oint_r \left( x^j K^k_m - x^k K^j_m \right) dA^m, 
\eeqn
where $\gamma_{ij}$ and $K_{ij}$ are the three-metric and extrinsic
curvature, respectively.

\section{Results}
\label{Results}

Table~\ref{Table_of_Runs} lists all runs presented in this paper, and
this section is ordered as follows. First, Sec.~\ref{sec:stepbystep}
demonstrates how constraint violations are reduced as we add our
gauge/evolution modifications, one step at a time, at the lowest
(``lr'') resolution.
Next, Sec.~\ref{sec:noise_reduction} shows how our 
improved gauge conditions reduce short-wavelength, high-frequency
noise both in GWs extracted from these calculations, and in the
ADM mass and angular momentum surface integral diagnostics. 
Finally, we perform convergence studies at four resolutions, primarily
comparing results using the ``standard'' moving-puncture gauge
conditions (OldGauge) to our most-improved gauge choice
BestGauge (g1+TR2+h+d5d7). These
studies, presented in Sec.~\ref{sec:convergence}, focus on convergence
of BH irreducible masses, constraint violations, and
waveform amplitude, phase, and noise.

\begin{table*}[t]
\caption{Summary of runs performed. See Eq.~(\ref{lapse_master_eq})
  and ensuing discussion for explanations of how functions $g(t,r)$,
  $h(r)$, $d_5(t,r)$, and $d_7(t,r)$ are generated in our extensions
  to the standard {\tt 1+log} slicing condition. ``TR Type'' 
  refers to the chosen time reparameterization, where TR Type 0
  denotes no reparameterization, TR Type 1 implies that the RHS of
  {\it all} evolved variables is multiplied by $f(t)$
  (Eq.~(\ref{f_of_t})), and TR Type 2 indicates that all RHS are
  multiplied by $f(t)$ except for the shift RHS and most of the lapse
  RHS (as specified in Eq.~(\ref{lapse_master_eq})). Resolutions
  \{lr,mr,hr,hhr\} are as defined in
  Sec.~(\ref{nummethods:AMRgridparams}).}
\begin{tabular}{|c|c|c|c|c|c|c|}
\hline
Case Name & $g(t,r)$ & TR Type & $h(r)$ & $d_5(t,r)$ & $d_7(t,r)$ & Resolutions \\
\hline\hline
OldGauge 
& 2         & 0 & 0          & 0          & 0.3 & lr,mr,hr,hhr\\
\cline{1-2}\cline{7-7}
g0 
& $g_0(r)$; Eq.~(\ref{g0_of_r}) & &       &            &     & lr \\
\cline{1-1}\cline{3-3}\cline{7-7}
g0+TR1 
&           & 1 &            &            &     & lr \\
\cline{1-1}\cline{3-3}\cline{7-7}
g0+TR2 
&           & 2 &            &            &     & lr \\
\cline{1-1}\cline{4-4}\cline{7-7}
g0+TR2+h 
&           &   & Eq.~(\ref{h_of_r})&     &     & lr,mr,hr,hhr \\
\cline{1-1}\cline{5-7}
g0+TR2+h+d5d7
&           &   &            &  Eq.~(\ref{d5}) & Eq.~(\ref{d7}) & lr,mr,hr\\
\cline{1-2}\cline{7-7}
g1+TR2+h+d5d7
& $g_1(t,r)$; Eq.~(\ref{g1_of_t_r})&   &            &                 &                & lr,mr,hr,hhr\\
{\it alias:} {\bf BestGauge}        
&                                  &   &            &                 &                & \\
\hline
\end{tabular}
\label{Table_of_Runs}
\end{table*}

\subsection{Step-by-Step Addition of New Features}
\label{sec:stepbystep}
This section adds our evolution and gauge changes step-by-step
at lowest (``lr'') resolution, unfolding the benefits of each
modification. Table~\ref{Table_of_Runs} contains a complete reference
of case names.

\subsubsection{Reduction of Constraint Violations}
\label{sec:stepbystep:constraints}
\begin{figure*}
\includegraphics[angle=270,width=0.45\textwidth]{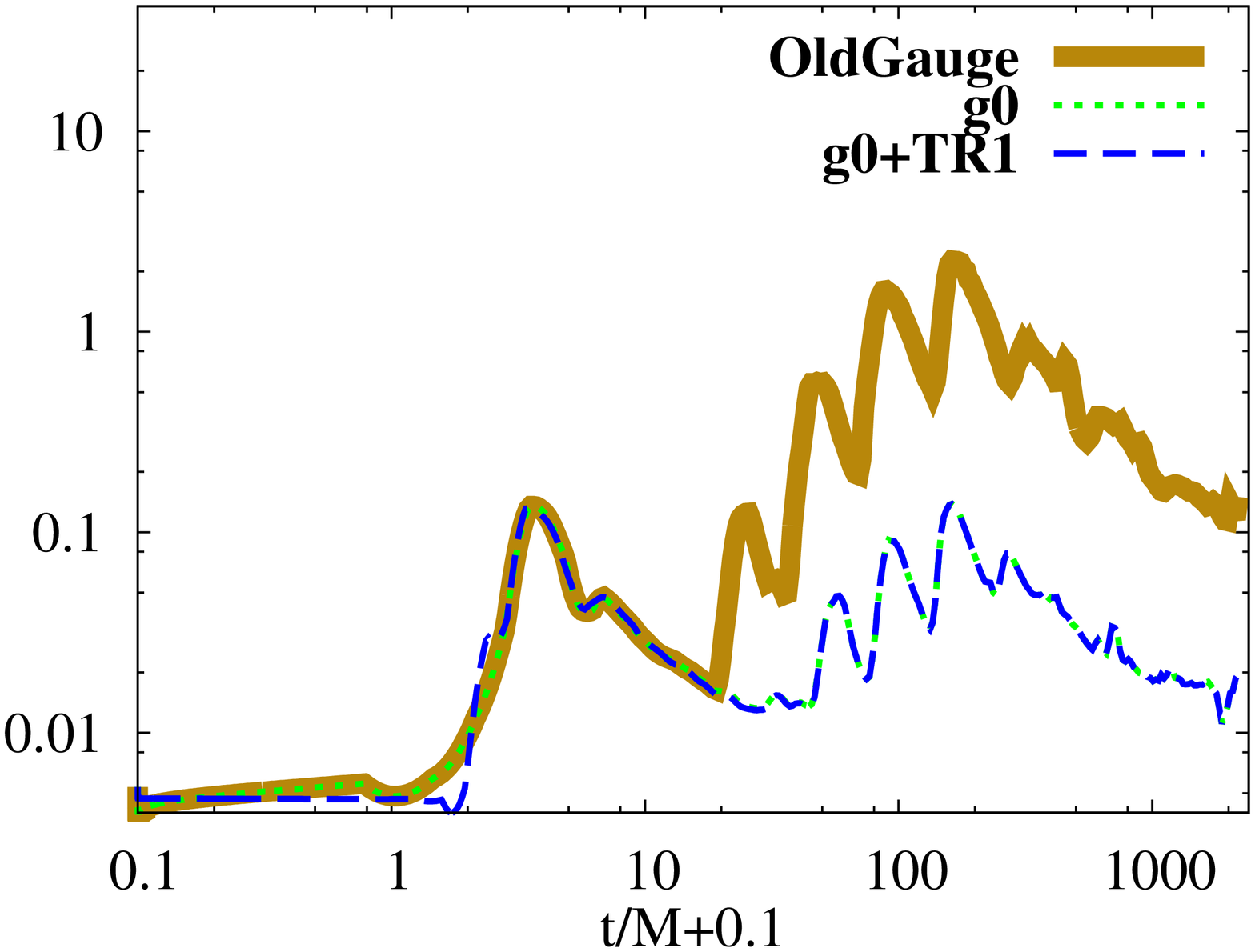}
\includegraphics[angle=270,width=0.45\textwidth]{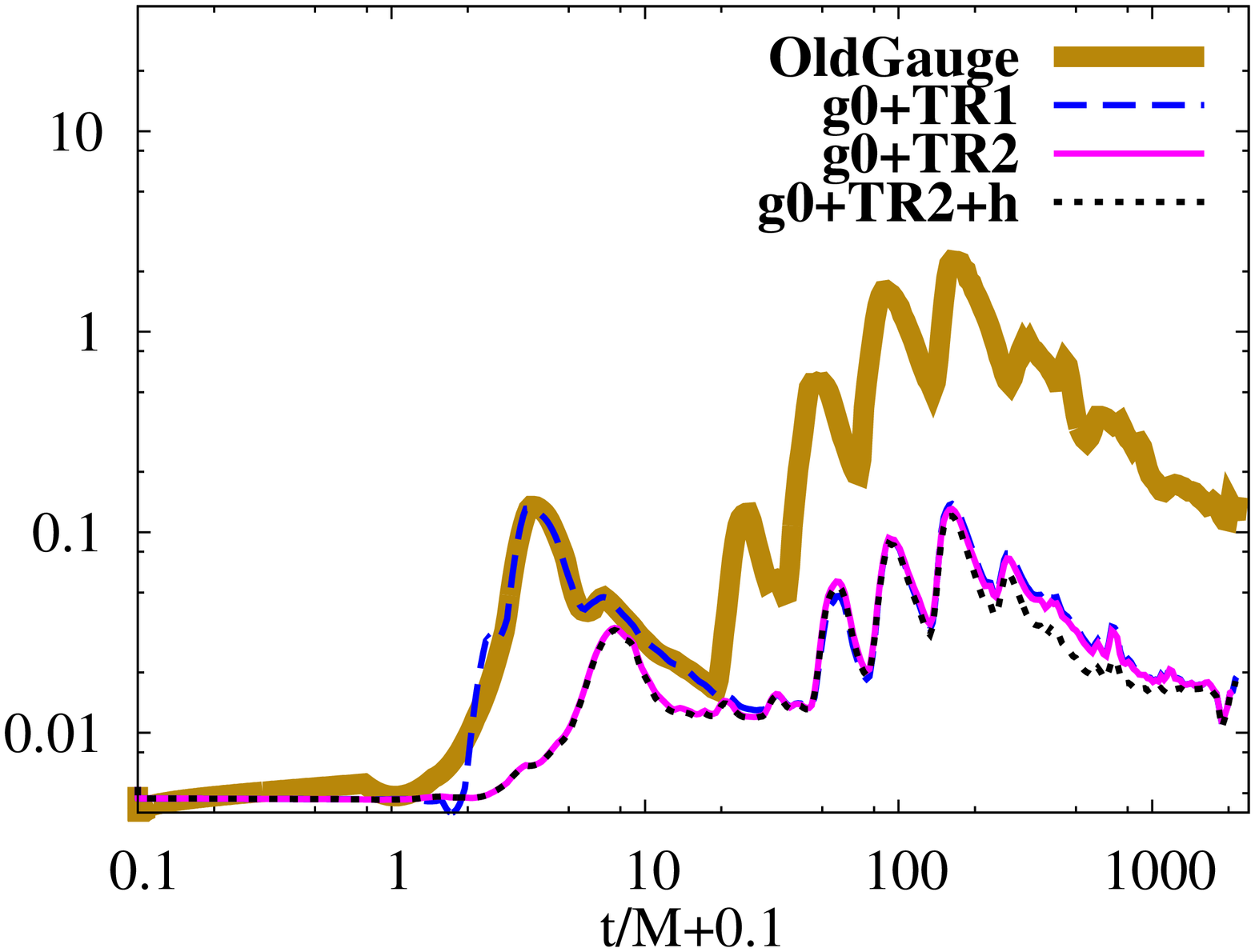} \\
\vspace{0.3cm}
\includegraphics[angle=270,width=0.45\textwidth]{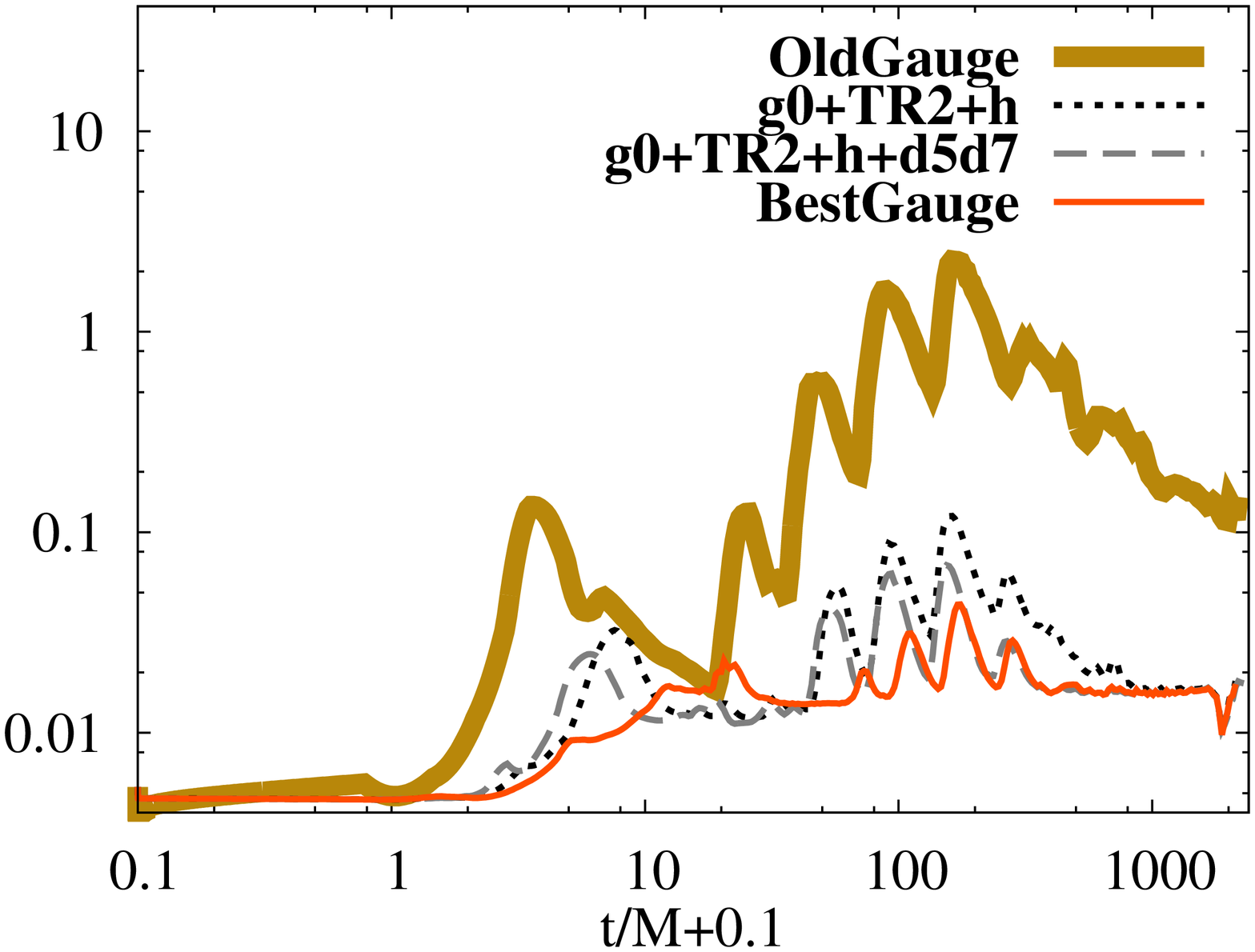}
\includegraphics[angle=270,width=0.45\textwidth]{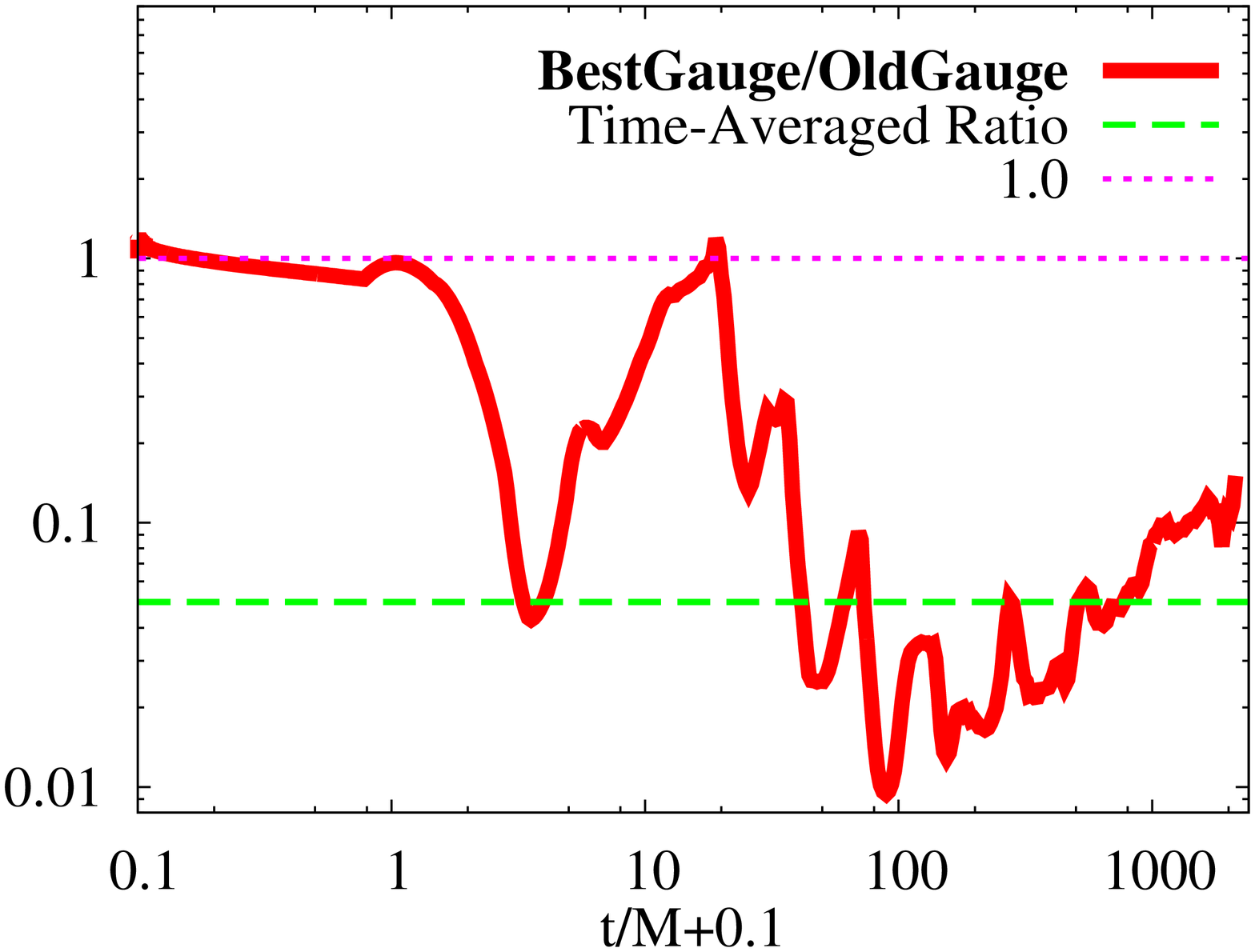}
\caption{Excised L2 norm of Hamiltonian constraint
  violation vs time, $||\mathcal{H}||(t)$ (Eq.~\ref{L2_Ham}),
  as new techniques are added one-by-one at fixed ``lr''
  resolution. Cases are as defined in Table~\ref{Table_of_Runs}. 
  {\bf Upper-left panel}: $||\mathcal{H}||(t)$ for cases OldGauge (thick brown),
  g0 (green dotted), and g0+TR1 (blue dashed). 
  {\bf Upper-right panel}: $||\mathcal{H}||(t)$ for cases OldGauge (thick
  brown), g0+TR1 (blue
  dashed), g0+TR2 (magenta solid),
  g0+TR2+h (black dotted).
  {\bf Lower-left panel}: $||\mathcal{H}||(t)$ for cases OldGauge (thick
  brown), g0+TR2+h   (black dotted),
  g0+TR2+h+d5d7 (gray dashed), and 
  BestGauge (g1+TR2+h+d5d7; orange solid).
  {\bf Lower-right panel}: $||\mathcal{H}||(t)$ for the
  BestGauge case, divided by
  $||\mathcal{H}||(t)$ for the OldGauge case
  (solid red). This ratio is compared to its time-averaged value
  ($\approx 0.05$, green dashed) and unity (magenta dotted). Values
  below unity indicate an improvement (reduction in
  $||\mathcal{H}||(t)$).}
\label{Fig:Ham_step_by_step}
\end{figure*}

\begin{figure*}
\includegraphics[angle=270,width=0.45\textwidth]{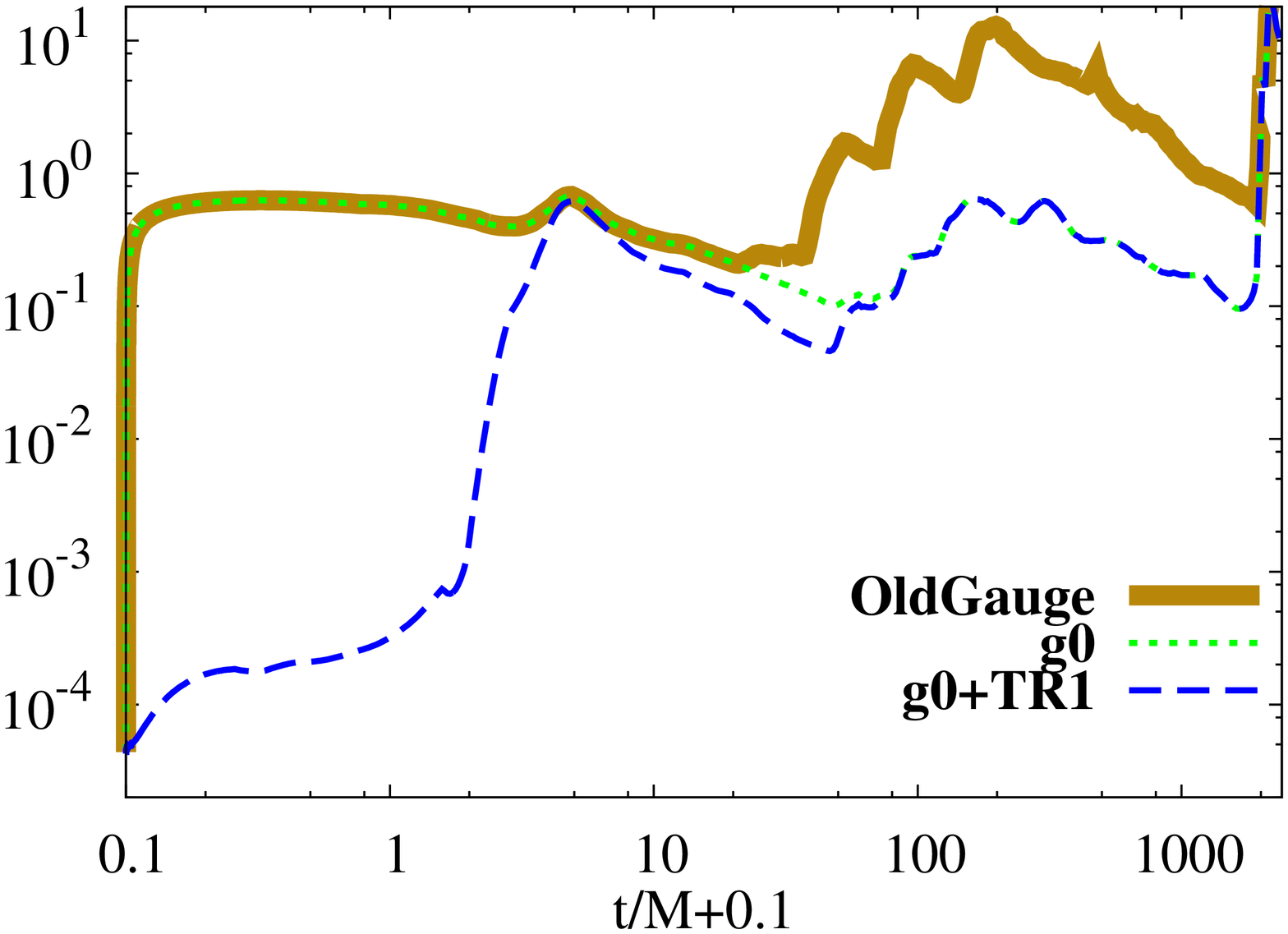}
\includegraphics[angle=270,width=0.45\textwidth]{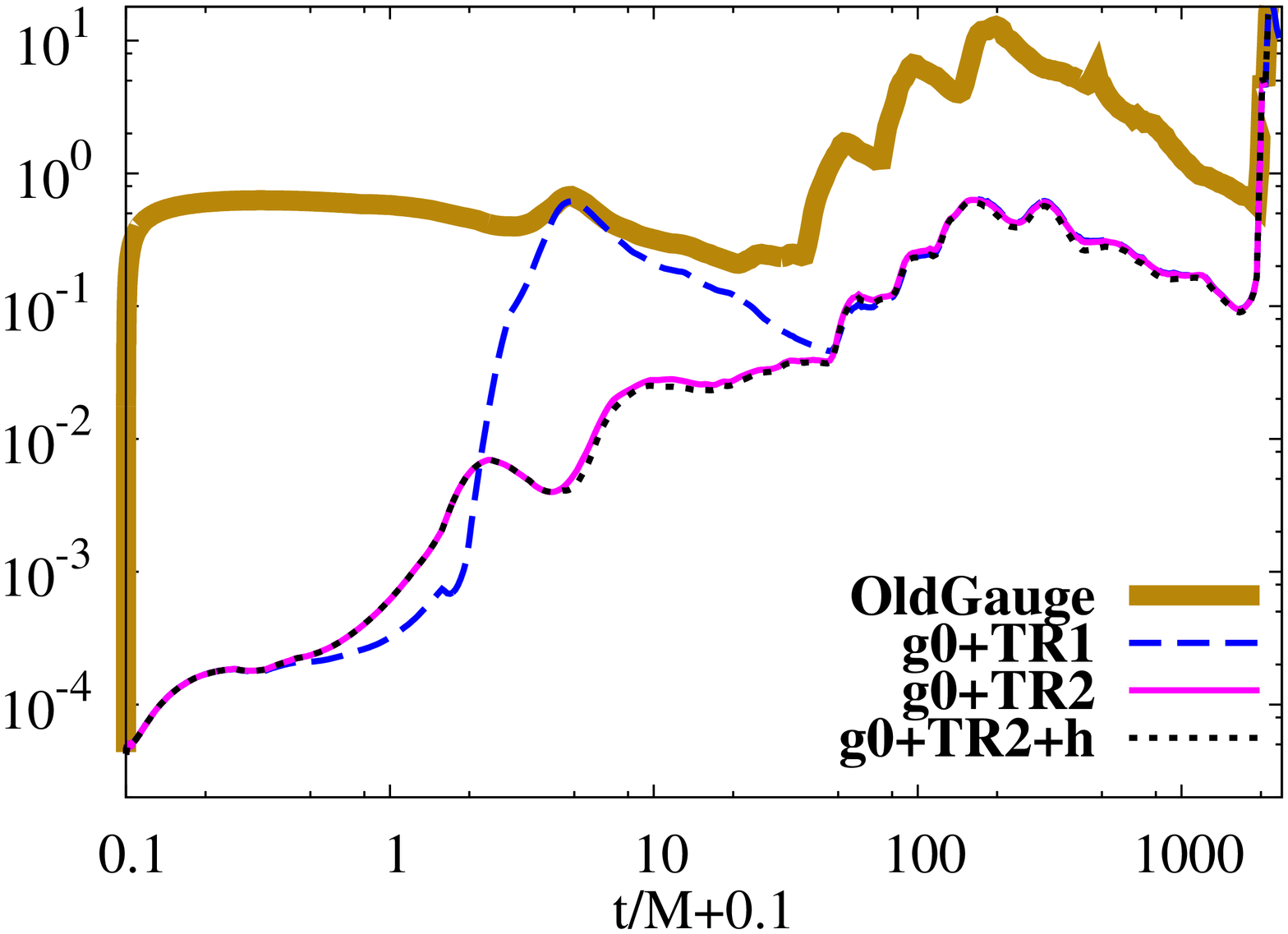} \\
\vspace{0.3cm}
\includegraphics[angle=270,width=0.45\textwidth]{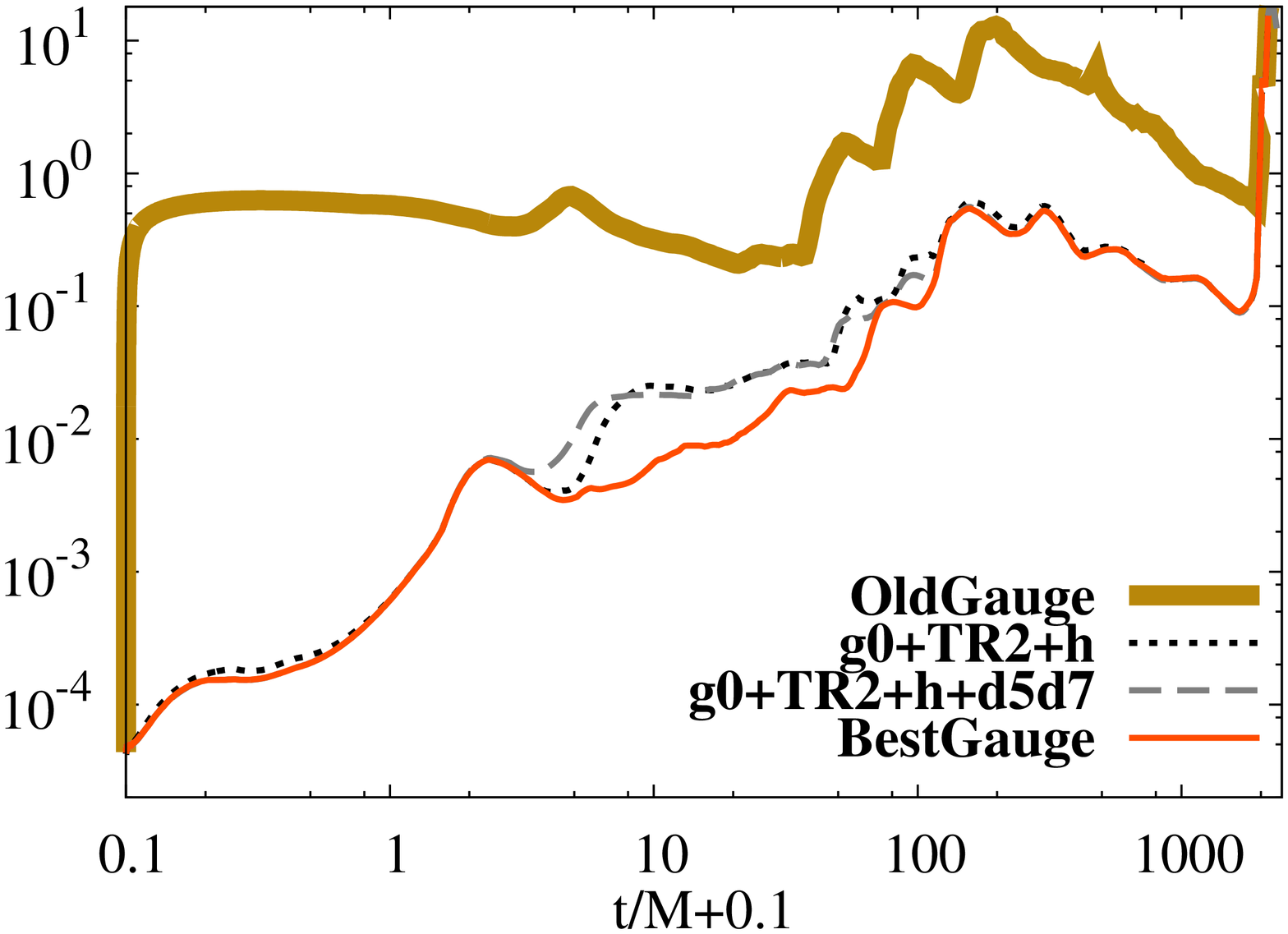}
\includegraphics[angle=270,width=0.45\textwidth]{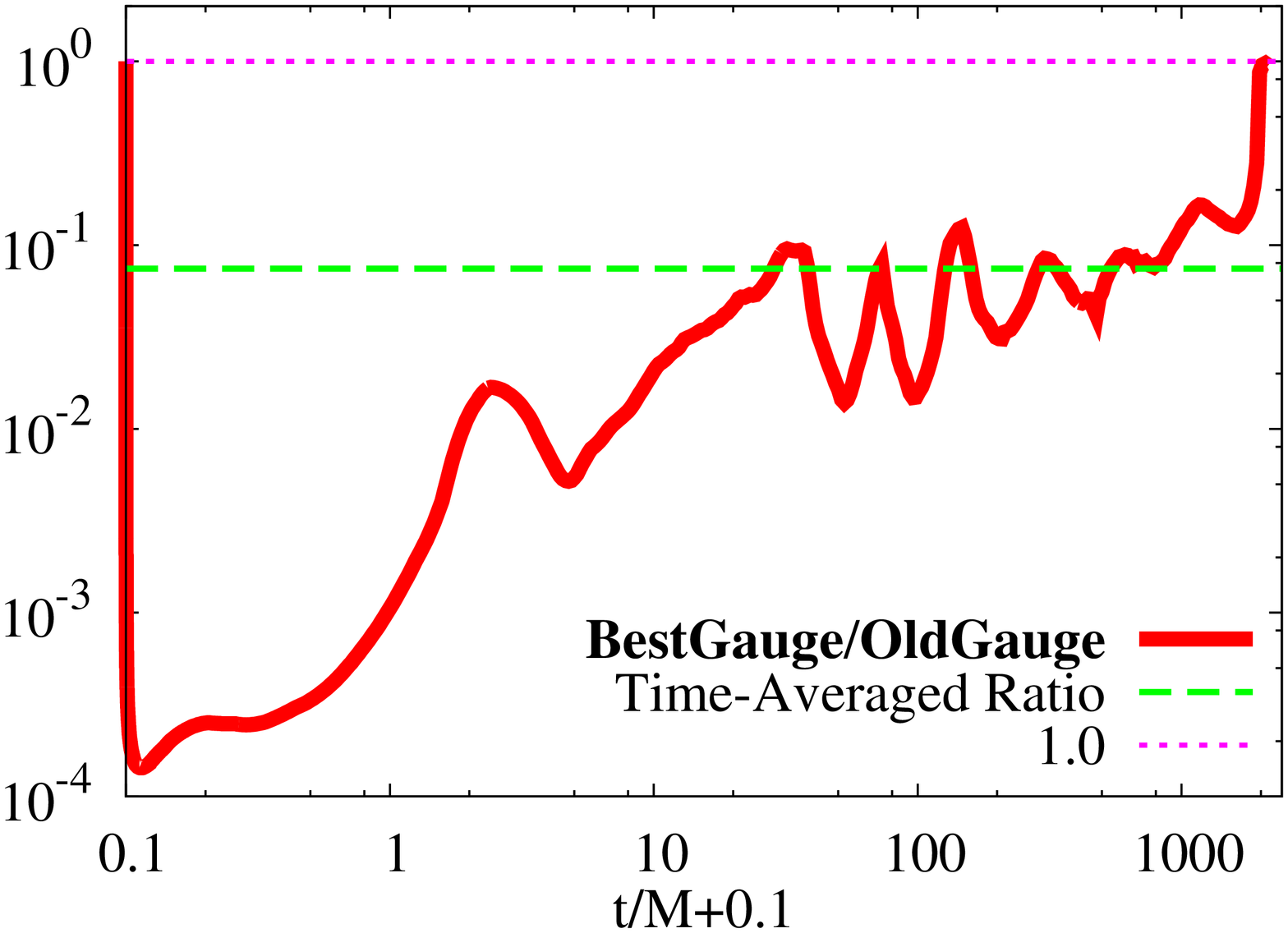}
\caption{Excised L2 norm of $x$-component of Momentum constraint
  violation vs time, $||\mathcal{M}^x||(t)$ (Eq.~\ref{L2_Mom}),
  as new techniques are added one-by-one at fixed ``lr''
  resolution. Cases are as defined in Table~\ref{Table_of_Runs}. 
  {\bf Upper-left panel}: $||\mathcal{M}^x||(t)$ for cases OldGauge (thick brown),
  g0 (green dotted), and g0+TR1 (blue dashed). 
  {\bf Upper-right panel}: $||\mathcal{M}^x||(t)$ for cases OldGauge (thick
  brown), g0+TR1 (blue dashed), g0+TR2
  (magenta solid), g0+TR2+h (black dotted).
  {\bf Lower-left panel}: $||\mathcal{M}^x||(t)$ for cases OldGauge (thick
  brown), g0+TR2+h (black dotted),
  g0+TR2+h+d5d7 (gray dashed), and
  BestGauge (g1+TR2+h+d5d7; orange solid).
  {\bf Lower-right panel}: $||\mathcal{M}^x||(t)$ for the
  BestGauge case, divided by
  $||\mathcal{M}^x||(t)$ for the OldGauge case
  (solid red). This ratio is compared to its time-averaged value
  ($\approx 0.075$, green dashed) and unity (magenta dotted). Values
  below unity indicate an improvement (reduction in
  $||\mathcal{M}^x||(t)$).}
\label{Fig:Momx_step_by_step}
\end{figure*}

Figures~\ref{Fig:Ham_step_by_step} and~\ref{Fig:Momx_step_by_step}
show how Hamiltonian and momentum ($x$-component; $y$- and
$z$-components are similar) constraint violations are diminished as new
gauge and evolution techniques are added one at a time, at resolution
``lr''. Notice the most dramatic decrease in constraint violations
occurs between cases OldGauge and g0, except at early
times, where a very early spike in momentum constraint violations
remains (note that $\mathcal{M}^x(t=0)\sim 10^{-5}$). The
upper-left panel in Fig.~\ref{Fig:Momx_step_by_step} demonstrates that
this spike in early-time momentum constraint violation is dramatically
reduced when reparameterizing the time 
coordinate such that {\it all} BSSN and gauge variables are evolved
very slowly at early times (i.e., {\it all} evolution equation RHSs
are multiplied by the $f(t)$ of Eq.~\ref{f_of_t}). Such a reduction
demonstrates that this early spike in momentum constraint violation is
likely due to copying data from $t=0$ as described in
Sec.~\ref{sec:TR}. However, there still exists a significant spike
slightly later, at
$t/M\approx 4$, in both Hamiltonian and momentum constraint violations
that was not reduced by multiplying {\it all} evolution equation RHSs
by $f(t)$.

This remaining spike might be related to the initial
settling of coordinates in the strong-field region, so we choose a time
reparameterization such that the lapse {\it and shift} evolution is
greatly accelerated initially (upper-right panels) with respect to the
evolution of all other variables, again as described in
Sec.~\ref{sec:TR}. In this way, the coordinates are able to settle
significantly faster than otherwise. Notice that although this
strategy significantly modifies the timestep only until about $t/M=10$
in these plots, the reduction in constraint violations is
longer-lasting, until $t/M\approx40$
(Fig.~\ref{Fig:Momx_step_by_step}), when apparently other
errors start to dominate and constraint violations match those of
the g0 technique alone. With just these improvements, the
momentum constraint violation maximum during inspiral is on par with
the {\it initial} momentum constraint peak in OldGauge.

About $150M$ after merger, at $t/M\gtrsim 2000$, a large spike in
momentum constraint violation appears, regardless of gauge choice. The
cause of this spike may be similar to that of the initial spikes, which
were significantly reduced by application of a time reparameterization
such that the lapse and shift evolution was significantly
accelerated relative to the BSSN field variables. So it is possible
this late-time spike might be reduced
via another time reparameterization at merger. Alternatively, this
spike may be caused by under-resolution of outgoing, short-wavelength
{\it physical} (i.e., non-gauge) waves generated in the strong-field
region during merger, common horizon formation, and ringdown.

The upper-right panel of Figs.~\ref{Fig:Ham_step_by_step}
and~\ref{Fig:Momx_step_by_step} demonstrate that addition of a
parabolic damping term ($h(r)$ as defined in Eq.~(\ref{h_of_r})
reduces Hamiltonian constraint violations from mid-inspiral to
merger by about 30\%, and momentum constraint violations midway
through evolutions up to 10\%. We anticipate that stronger damping
(corresponding to smaller safety factors for $h(r)$) may be possible
in the outer, weak-field regions of the grid, resulting in further
reductions of constraint violations.

The lower-left panel of Figs.~\ref{Fig:Ham_step_by_step}
and~\ref{Fig:Momx_step_by_step} show that addition of stronger
Kreiss-Oliger damping (as defined in Eqs.~\ref{d5} \& \ref{d7})
reduces momentum constraint violations over time by another 10\% at
most, but Hamiltonian constraint violations are reduced by up to a
factor of 2 early in the inspiral, and roughly 3\% for late inspiral
through merger ($t\gtrsim1000M$). Finally, we attempt an alternative
lapse acceleration function $g_1(t,r)$, which decreases the
lapse propagation speed significantly in the strong-field region
at early times. This change further reduces average constraint violations
during the early inspiral ($t\lesssim 300M$). It is conceivable that
residual constraint violations may be reduced beyond the levels we
were able to attain with gauge improvements alone,
either by parabolically smoothing the constraint-violating degrees of
freedom as in~\cite{Paschalidis:2007ey,Paschalidis:2007cp}, or by
adopting a constraint-damping formalism such as
Z4c~\cite{Bernuzzi:2009ex,Hilditch:2013} or CCZ4~\cite{Alic:2012}.

The overall reduction in constraint violations is shown in the
lower-right panel of Figs.~\ref{Fig:Ham_step_by_step}
and~\ref{Fig:Momx_step_by_step}. Averaged over time, Hamiltonian constraint
violations are reduced by about a factor of 20 and momentum constraint
violations by about a factor of 13. 

We conclude this section by stressing that despite nearly two orders
of magnitude reduction in constraint violation, the addition of these
new features 
has virtually no impact on computational expense. The next section
describes how these gauge improvements affect noise in important quantities,
such as gravitational waveforms and ADM (mass and angular momentum) surface
integrals. 

\subsection{Noise Reduction Features of New Gauge Conditions}
\label{sec:noise_reduction}

In addition to constraint-violation reductions, our new gauge
conditions also act to reduce short-wavelength noise in the GWs
($\psi^4(t)$) as well as the surface-integral representations of
ADM mass and angular momentum. We present these improvements here.

\subsubsection{Reduction of Noise in $\psi^4_{2,2}$}
\label{Psi4noisereduction}

\begin{figure*}
\includegraphics[angle=270,width=0.45\textwidth]{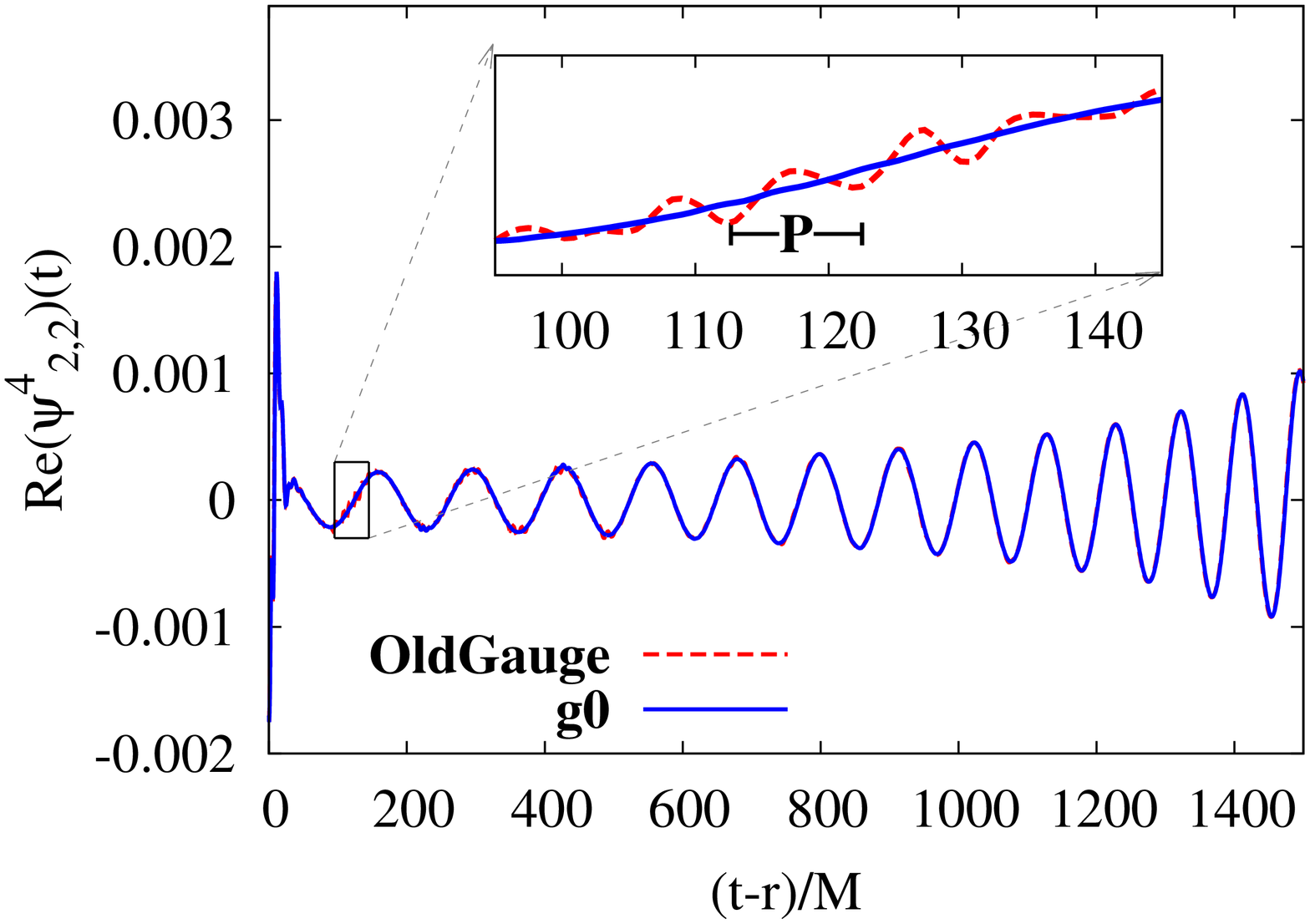}
\includegraphics[angle=270,width=0.45\textwidth]{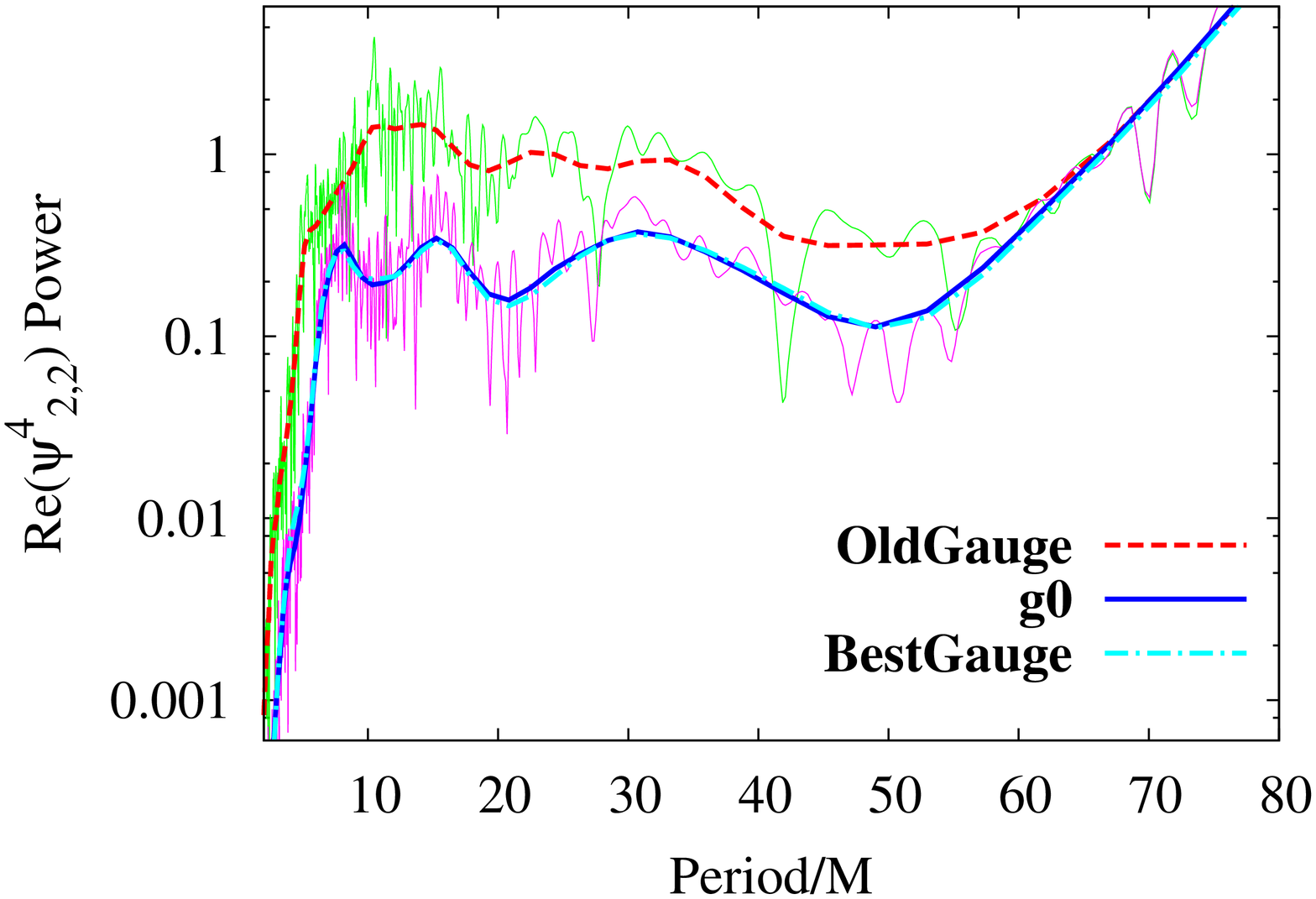}
\caption{Inspiral waveform noise analysis. {\bf Left panel}:
  Re($\psi^4_{2,2}$) versus retarded time, for cases OldGauge
  (dashed red) and g0 (solid blue). Inset
  magnifies noisiest part of waveform, demonstrating that the dominant
  noise frequency corresponds to a wave period of $P\approx 10M$ in
  the OldGauge case.  
  {\bf Right panel}: Power spectrum of Re($\psi^4_{2,2}$) versus wave period
  during inspiral, including data from $(t-r)/M\approx58$ to $1160$,
  for cases OldGauge (thin green line: raw data, thick dashed red
  line: B\'{e}zier-smoothed data; i.e., fit to a degree-$n$ B\'{e}zier
  curve for $n$ data points), g0 (thin magenta line: raw data,
  thick solid blue line: B\'{e}zier-smoothed data), and
  BestGauge (g1+TR2+h+d5d7; thick dot-dashed cyan line:
  B\'{e}zier-smoothed data). All data plotted in this figure are from
  ``lr'' resolution runs, and Re($\psi^4_{2,2}$) data are from extraction
  radius $r=68.6M$. Re($\psi^4_{2,2}$) time series data were
  multiplied by the tapering function $\Erf(t;200,50) \times
  \Erf(-t;-1300,150)$ prior to the Fourier transform, to smoothly
  suppress junk radiation before $\sim 100M$ and the high-frequency
  late-inspiral/merger signals after $\sim 1500M$.}
\label{Fig:Psi4_noise_reduction}
\end{figure*}

The left panel of Fig.~\ref{Fig:Psi4_noise_reduction} shows how
short-period noise in $\psi^4_{2,2}(t-r)$ is significantly reduced
when we choose a gauge in which the lapse speed is accelerated
radially outward (g0), keeping all else identical to
a standard moving-puncture gauge choice (OldGauge). Notice that
the characteristic period for this noise is $P\sim10M$, which is
made clear in the right panel, which plots the power spectra of
Re($\psi^4_{2,2}(t)$) for cases OldGauge, g0, and
BestGauge (g1+TR2+h+d5d7) during early inspiral 
($50\lesssim(t-r)/M\lesssim1160$), when the signal-to-noise ratio of
our numerical waveforms after junk radiation is lowest. Notice that
accelerating the outgoing lapse wave speed has the largest impact on
reducing high-frequency noise in Re($\psi^4_{2,2}(t)$), diminishing
the noise by nearly an order of magnitude at the peak noise frequency
(period of $\approx 10M$). Also, although additional tricks have some
noticeable impact on constraint violations, they appear to contribute
only a small amount to waveform noise at the shortest wavelengths.

\begin{figure}
\includegraphics[angle=270,width=0.45\textwidth]{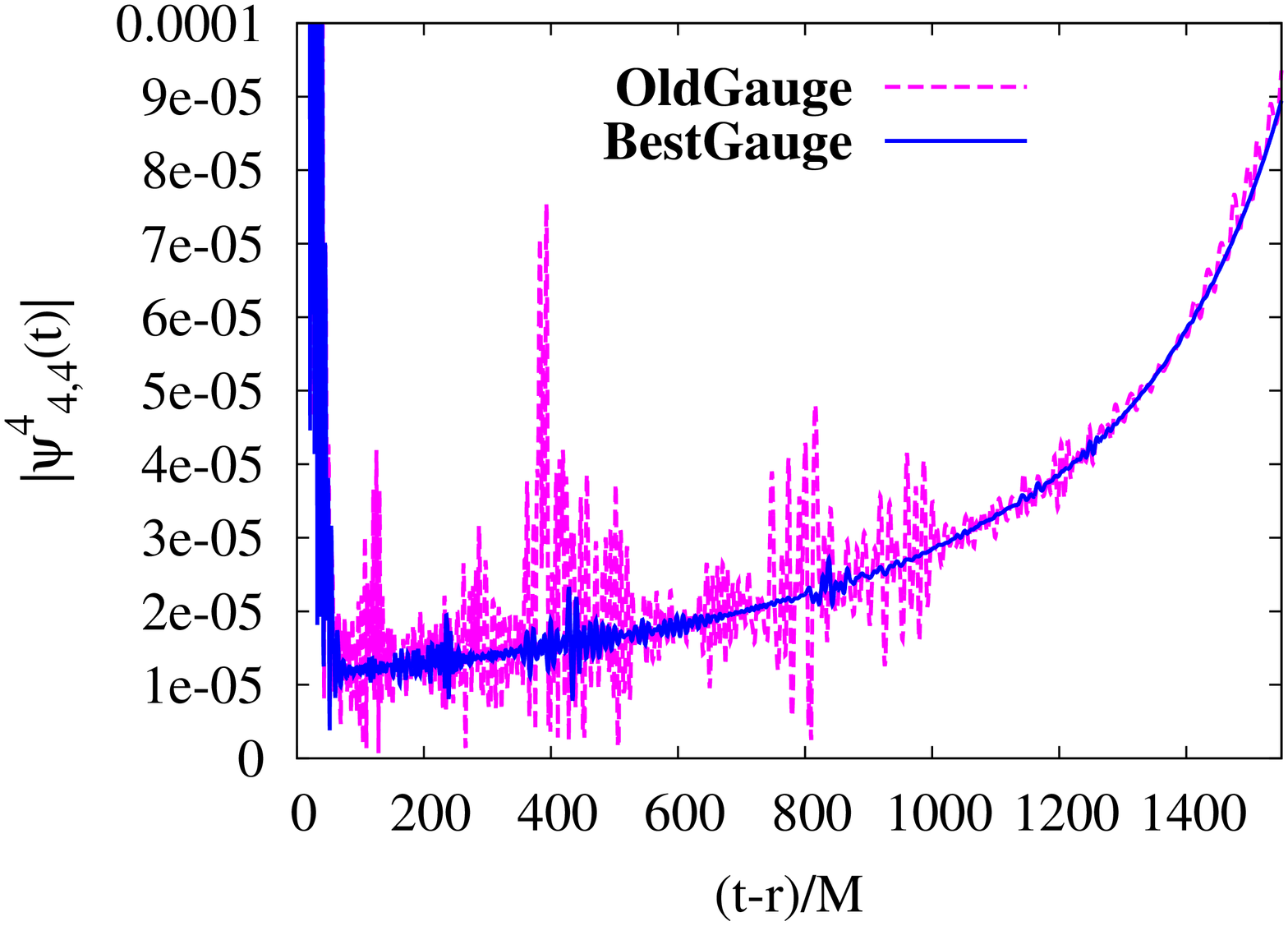}
\caption{Amplitude of $\psi^4_{4,4}$ versus retarded time at ``hhr''
  resolution, for cases OldGauge (dashed magenta) and BestGauge (solid blue).}
\label{Fig:Psi4_44_noise_reduction}
\end{figure}

Although we concentrate here on the dominant, ($l,m$)=(2,2), mode, we
find significant noise reduction in sub-dominant modes as well. As an
example, Fig.~\ref{Fig:Psi4_44_noise_reduction} demonstrates that in
OldGauge, the amplitude of the ($l,m$)=(4,4) mode of $\psi^4$ is
dominated by noise throughout a significant fraction of the
inspiral. However, in BestGauge the noise is mostly non-dominant. Based on
this and observations of other non-dominant modes, we conclude that
these gauge improvements stand to improve the usefulness of
sub-dominant modes from BSSN/MP evolutions.

\subsubsection{Reduction of Noise in ADM Surface Integrals}
\label{sec:ADMnoise}

\begin{figure*}
\includegraphics[angle=270,width=0.45\textwidth]{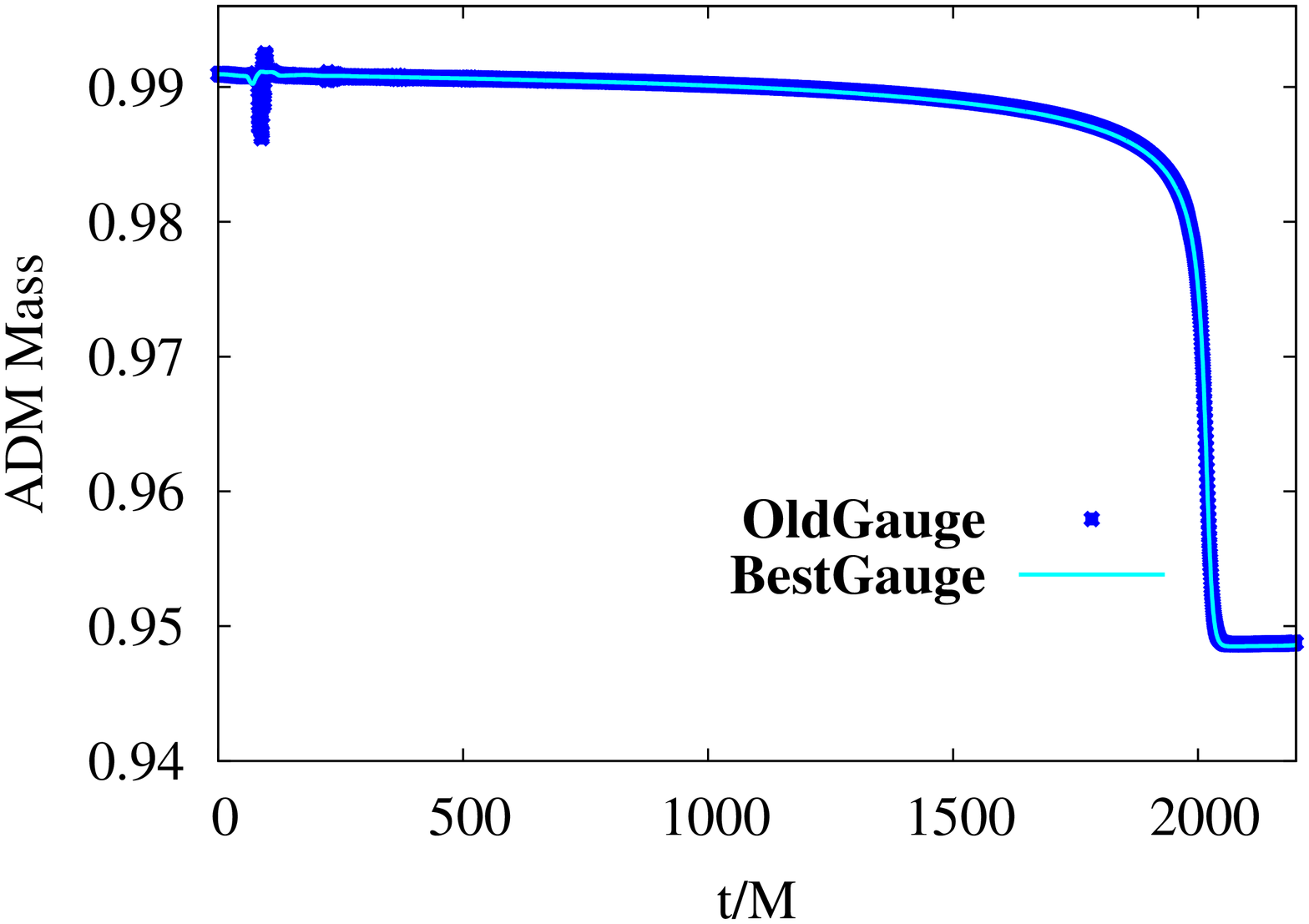}
\includegraphics[angle=270,width=0.45\textwidth]{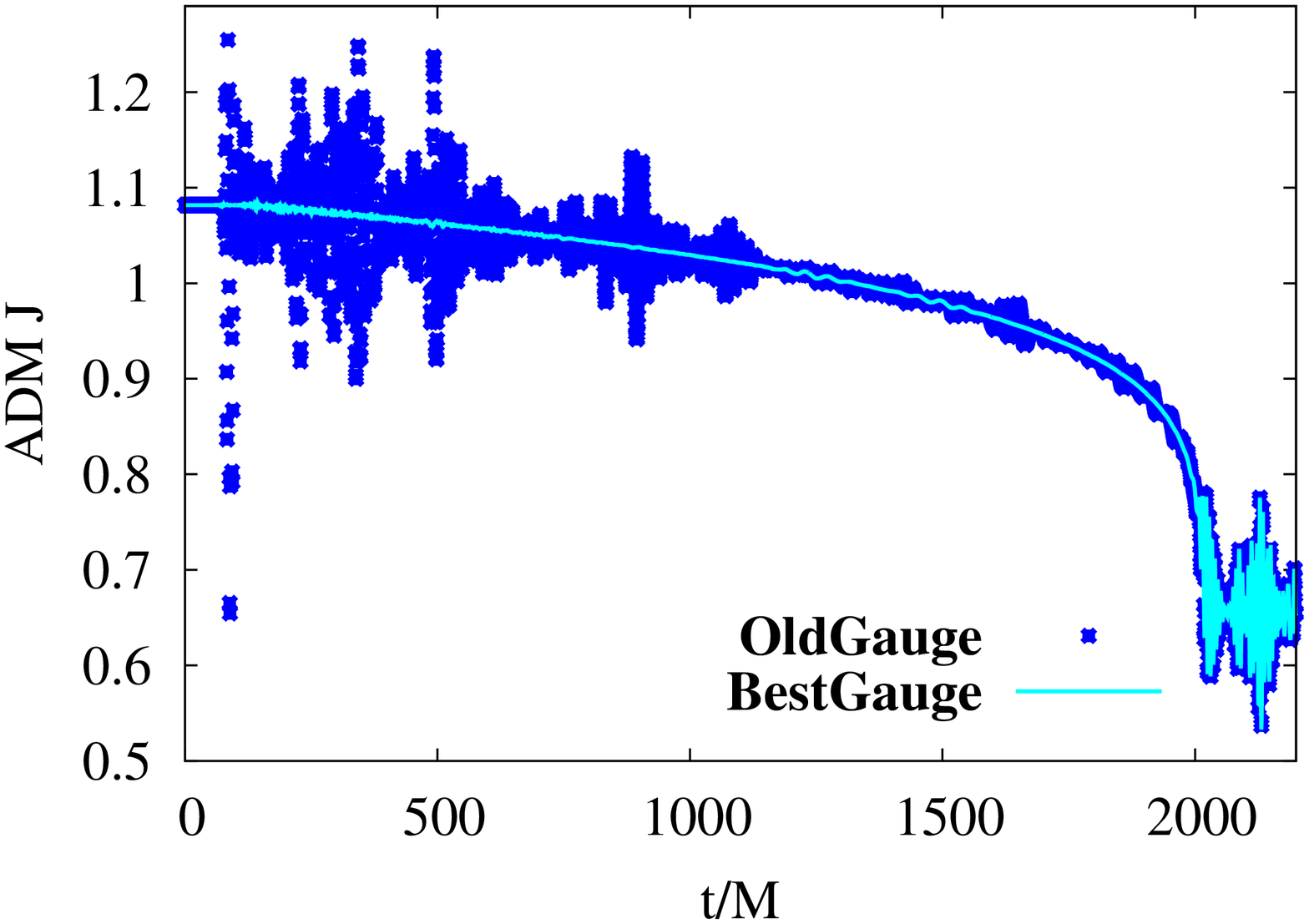}
\caption{ADM surface integral noise analysis. {\bf Left panel}:
  ADM mass $M(t)$ (Eq.~\ref{M_ADM_def}), as measured by a surface integral at
  $r=114.2M$. Large blue points denote OldGauge case and thin cyan
  line BestGauge (g1+TR2+h+d5d7). Note that all data are
  taken from the ``hhr'' resolution runs. {\bf Right panel}: Same as
  left plot, except ADM angular momentum $J(t)$ (Eq.~\ref{J_ADM_def}) is plotted.}
\label{Fig:diag_integral_noise_reduction}
\end{figure*}

It has been found that surface integral representations of ADM mass
$M(t)$ and angular momentum $J(t)$ (Eqs.~\ref{M_ADM_def} and \ref{J_ADM_def})
are noisy and suffer from large drifts in AMR moving-puncture BBH
evolutions~\cite{Marronetti:2007ya}. Though getting around the problem
is usually a matter of recasting the distant surface integral into a
sum of surface plus volume integrals, we find that this may be
unnecessary with suitable gauge
improvements. Figure~\ref{Fig:diag_integral_noise_reduction} plots
$M(t)$ and $J(t)$ surface integrals as measured at $r=114.2M$,
comparing our original moving-puncture gauge choice (OldGauge) with our most
sophisticated modification, g1+TR2+h+d5d7 (BestGauge),
all at ``hhr'' resolution. The overall pattern is a
drop in $M(t)$ and $J(t)$ as the GWs pass through the surface integral
radius. However, our new gauge choices reduce noise in these
surface integral diagnostics by a huge factor during inspiral,
particularly $J(t)$. This demonstrates that noise in the region far
from the BHs is not restricted to GWs, at least throughout the
inspiral. However, after merger a large amount of noise in $J(t)$
remains, despite our gauge improvements.

We believe the lack of noise reduction in $J(t)$ at late times may be
related to the spike in momentum constraint violation observed at
roughly the same (retarded) time (see
Sec.~\ref{sec:stepbystep:constraints}), which we hypothesized
may have something to do with either the rapid settling of the gauge
or the outgoing short-wavelength physical waves at and after merger.

\subsection{Convergence Study}
\label{sec:convergence}

In a mixed-order finite-difference code such as ours, in which some
components converge at different orders, we must be particularly
careful about error estimates, as lower-order errors can sometimes
be non-dominant, leading to higher-order convergence than otherwise
expected. In such a scheme, numerical results for
a quantity $Q_\text{numerical}$ at a given gridpoint, time, and grid
spacing $\Delta x$, are related to the exact quantity
$Q_\text{exact}$, via the following equation
\beq
\label{convergence_exact}
Q_\text{numerical}(\Delta x) = Q_\text{exact} + C \Delta x^n +
\mathcal{O}(\Delta x^{m})
\eeq
where $C$ is a constant, $n$ is the dominant convergence order,
$m \neq n$ is the next dominant convergence order, and $m<n$
is possible. Generally we would aim to have $C\Delta
x^n\gg\mathcal{O}(\Delta x^{m})$, so the $\mathcal{O}(\Delta x^{m})$
term can be neglected. When non-dominant convergence order terms are
neglected in this way, we call $Q_\text{exact}$ the {\it
  Richardson-extrapolated} $Q$, $Q_\text{RE}$, as it
provides an estimate for $Q$ that removes the dominant error term,
similar to \cite{Richardson1911}:
\beq
\label{convergence}
Q_\text{RE} = Q_\text{numerical}(\Delta x) - C \Delta x^n.
\eeq

AMR prolongations in time and space are performed with second- and
fifth-order-accurate stencils, respectively, and temporal and spatial
derivatives via fourth- and sixth-order-accurate finite-difference
stencils, respectively. Thus the {\it dominant} convergence order $n$
is {\it a priori} unknown in our BBH calculations, as they are of
mixed order. As described in Sec.~\ref{nummethods:AMRgridparams}, we
have reduced the timestep on all grids well below the CFL limit to
prevent the relatively low-order temporal differencing and
interpolations from dominating our errors.

To determine the dominant convergence order $n$, we perform runs at
three resolutions, obtaining three values for
$Q_\text{numerical}(\Delta x)$. With these three knowns, we then
employ a bisection technique to solve Eq.~(\ref{convergence}) for
unknowns $n$, $Q_\text{RE}$ and $C$. 

For quantities $Q_\text{numerical}(\Delta x)$ that are also
functions of time (e.g., $\psi^4(t)$), we obtain estimates for $n$ each
time $Q_\text{numerical}(\Delta x)$ is output by the code. We then
define the dominant convergence order as the average of all
bisection-calculated estimates for $n$, rounded to the nearest
integer. This averaged, rounded value for $n$, combined with
$Q_\text{numerical}(\Delta x)$ at the two lowest resolutions of the
three used to obtain $n$, are chosen as inputs to solve the linear
$2\times 2$ system for $C$ and $Q_\text{RE}$. We then find another
estimate for $Q_\text{RE}$ combining the same value for $n$ with the
two highest resolutions of the three $Q_\text{numerical}(\Delta x)$
used to estimate $n$. The difference between these two values for
$Q_\text{RE}$ is used as an error estimate, as deviations from zero
will be caused by variations in $n$ as well as higher-order terms
(cf. Eqs.~(\ref{convergence}) and~(\ref{convergence_exact})).

In the special case where $Q_\text{RE}\to0$, as is
expected for quantities that should converge to zero (e.g., constraint
violations), we may solve Eq.~(\ref{convergence}) for $n$ with only
two realizations of $Q_\text{numerical}(\Delta x)$. In this special
case, $n$ can be found analytically.


\subsubsection{Irreducible Mass Convergence}
\label{sec:Mirr_conv}

\begin{figure*}
\includegraphics[angle=270,width=0.45\textwidth]{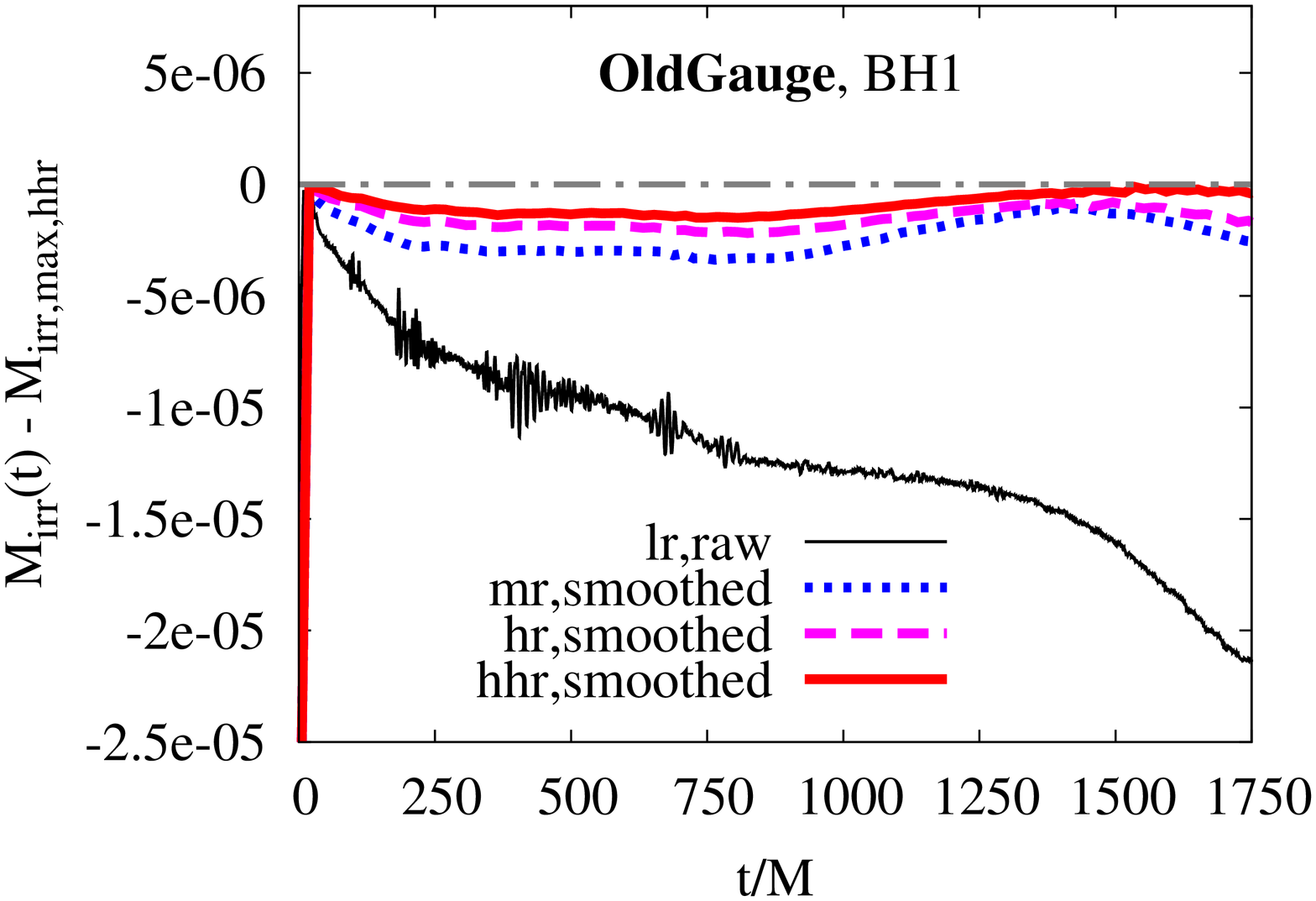}
\includegraphics[angle=270,width=0.45\textwidth]{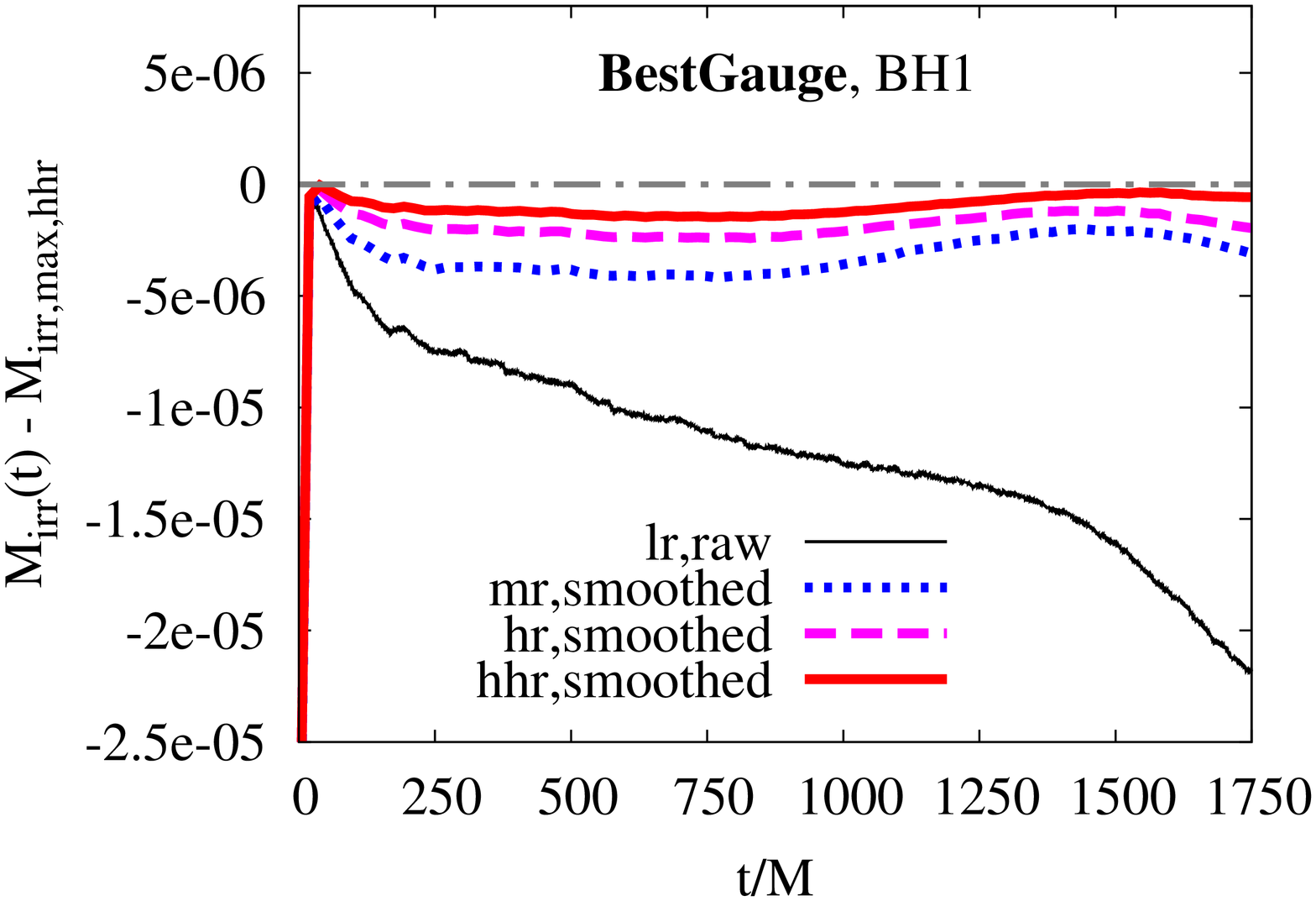}
\caption{Convergence study: $M_\text{irr}(t)$.
  {\bf Left panel}: $M_\text{irr}(t)-M_\text{irr,hhr,max}$ over time for BH1 (i.e., the
  BH with initial spin parameter$=0.3$), where $M_\text{irr,hhr,max}$
  is the maximum value of $M_\text{irr}$ in the ``hhr'' run. This
  panel plots data for the OldGauge case at
  resolutions ``lr'' (solid red), ``mr'' (medium-dashed blue), ``hr''
  (long-dashed magenta), and ``hhr'' (dotted green). 
  {\bf Right panel}: Same as left panel, but for case
  BestGauge (g1+TR2+h+d5d7). 
  ``smoothed'' denotes that the $n$ data points for a given
  dataset plotted in this figure have been fit to degree-$n$
  B\'{e}zier curve, and the resulting B\'{e}zier curve is shown.}
\label{Fig:M_irr_conv_study}
\end{figure*}

Figure~\ref{Fig:M_irr_conv_study} presents a convergence study of
 $M_\text{irr}(t)$ for the spinning BH (i.e., the BH with initial spin
parameter$=0.3$) at various resolutions. For ease of comparison (it is
difficult to distinguish a dotted line from a dashed one if data are
exceedingly noisy), the ``mr'', ``hr'', and
``hhr'' data are B\'{e}zier-smoothed (i.e., fit to a degree-$n$
B\'{e}zier curve for $n$ data points). However, to show how the
high-frequency noise in $M_\text{irr}(t)$ is affected by gauge choice,
we do not smooth the ``lr'' run data. Notice how $M_\text{irr}(t)$
data are {\it significantly} less noisy in the improved gauge case
BestGauge (right panel) than in the OldGauge case
(left panel), demonstrating that gauge-generated noise affects the
strong-field region as well. We observe the same relative noise in the
higher-resolution runs prior to smoothing. We also note that our 
gauge improvements have little impact on the secular drift in
$M_\text{irr}(t)$. As for the initially {\it nonspinning} BH, similar
results are observed, except the ``lr'' run is not as much an outlier.

\begin{figure}
\includegraphics[angle=270,width=0.45\textwidth]{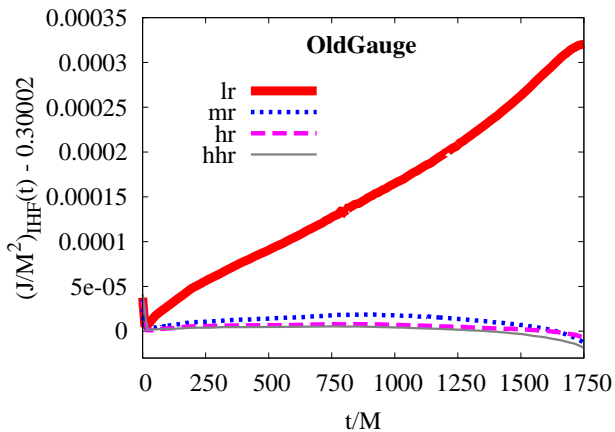}
\caption{Spin parameter versus time for BH1 (i.e., ``the spinning
  BH'', with initial spin parameter $\approx0.3$), as given by isolated horizon
  formalism. Data are plotted at resolutions ``lr'' (thick solid red),
  ``mr'' (dotted blue), ``hr'' (dashed magenta), and
  ``hhr'' (thin solid gray). Note that the offset 0.30002 is chosen
  for $(J/M^2)_\text{IHF}(t)$, which is the value all runs converge to
  after a rapid initial settling. We attribute this settling to the
  fact that our initial data are conformally flat. Note also that only
  case OldGauge is shown, as all other cases share the same
  qualitative behavior.}
\label{Fig:IHF_spin_bh1_oldgauge}
\end{figure}

To understand why the ``lr'' resolution run appears to be such a
significant outlier, we take a look at the isolated horizon formalism
spin \cite{Ashtekar:2004cn} of the initially spinning BH, shown in
Fig.~\ref{Fig:IHF_spin_bh1_oldgauge}. Notice the
secular increase in dimensionless spin parameter $(J/M^2)_\text{IHF}$ at ``lr'' resolution, which
completely dwarfs spin parameter drifts in the highest three
resolution runs. Figure~\ref{Fig:IHF_spin_bh1_oldgauge} shows data
from the OldGauge case only. For comparison, we report that at
``lr'' resolution, g0+TR2+h shares the same drift
as OldGauge to within a percent, and
g1+TR2+h+d5d7 (BestGauge) exhibits about
6\% less drift. Finally we note that for the initially nonspinning BH,
we observe no significant drift in spin parameter over time. Based on
these observations, we conclude that the ``lr'' run is at insufficient
resolution to properly resolve the spinning BH. 

\begin{figure}
\includegraphics[angle=270,width=0.43\textwidth]{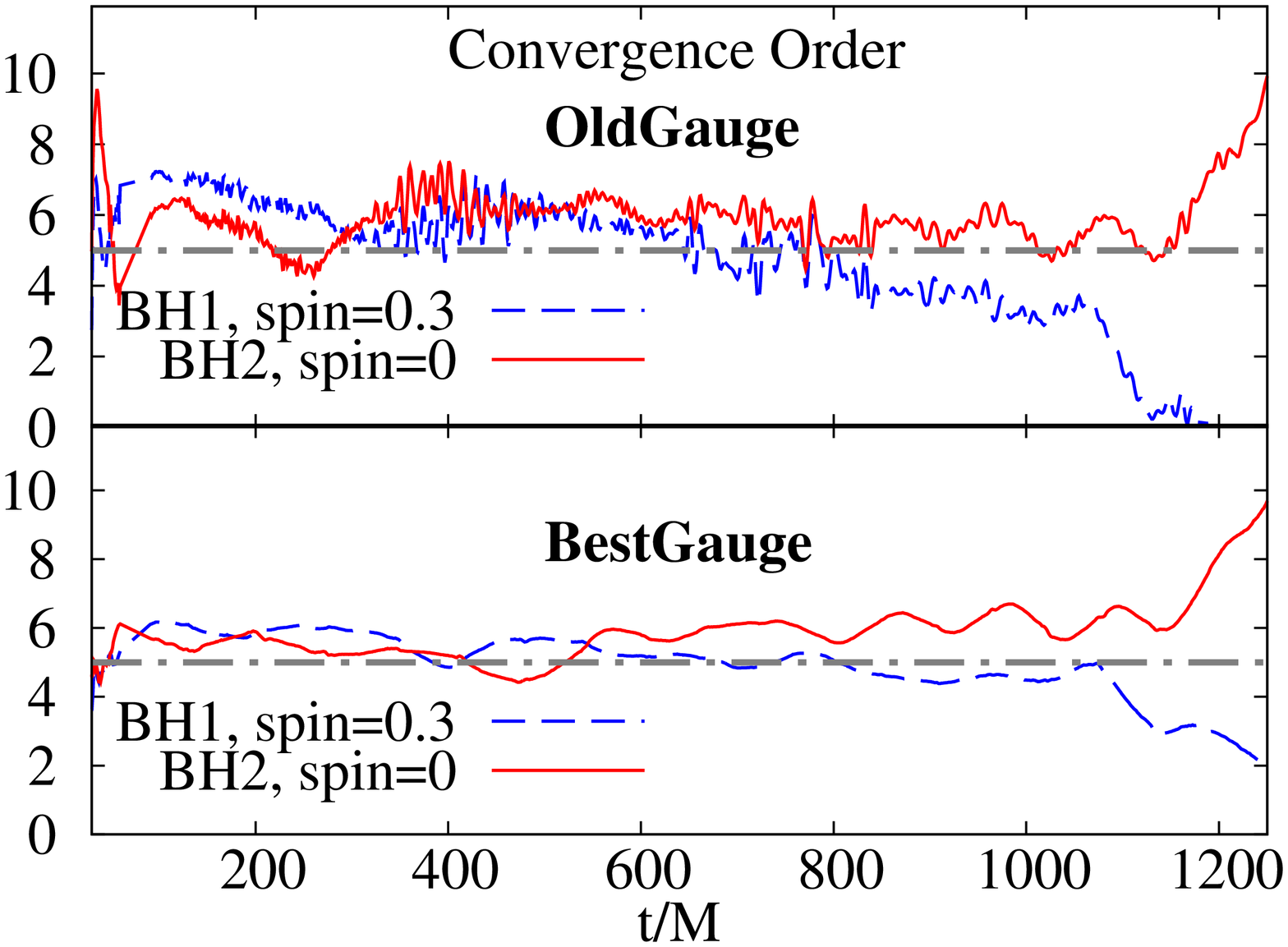}
\caption{Convergence study: Implied convergence order $n$
  (Eq.~\ref{convergence}) for $M_\text{irr}(t)$, using
  $M_\text{irr}(t)$ data at three highest resolutions \{mr,hr,hhr\}
  for both BHs during inspiral. Convergence order for BH1 (initial
  spin parameter$=0.3$; solid red) and BH2 (initially nonspinning;
  dashed blue) are compared to $n=5$ (dot-dashed gray). The top plot
  shows data from the OldGauge case the middle plot
  g0+TR2+h, and the bottom BestGauge
  (g1+TR2+h+d5d7). For ease of analysis, all raw $M_\text{irr}(t)$
  data in this plot were first smoothed with a moving-window linear
  least-squares algorithm prior to computing convergence order. Also,
  data in the range $t/M<30$ and $t/M>1250$ have been removed, as
  $M_\text{irr}(t)$ at all three resolutions overlaps at $t/M=0$ and
  $t/M\approx1300$ in the nonspinning BH (BH2) data, so near these
  points convergence order estimates for BH2 are unreliable.}
\label{Fig:M_irr_conv_study_conv_order}
\end{figure}

Throwing out data at the ``lr'' resolution, and using data at
resolutions \{mr,hr,hhr\}, Fig.~\ref{Fig:M_irr_conv_study_conv_order}
shows convergence order $n$ for $M_\text{irr}$ (Eq.~\ref{convergence}) versus time. This
panel compares observed convergence order $n$ for both BHs in all
three cases throughout much of the inspiral. Convergence is poor for
$t\gtrsim1100M$, as BH2 $M_\text{irr}$ data at all three resolutions
cross near $t\approx1300M$, regardless of gauge choice. Apparently
this clear non-convergence in BH2 negatively impacts BH1 convergence as
well. Outside of this non-convergent episode, we find average
convergence orders slightly higher than fifth order (our chosen
spatial prolongation order), independent of the gauge choice,
although the results using the BestGauge choice are clearly less
noisy.

In the next section, we show that this nonconvergence in irreducible
mass drift at low resolution does not translate to nonconvergence at
low resolution in constraint violations.

\subsubsection{Convergence to Zero of Constraint Violations}

\begin{figure*}[t]
\includegraphics[angle=270,width=0.45\textwidth]{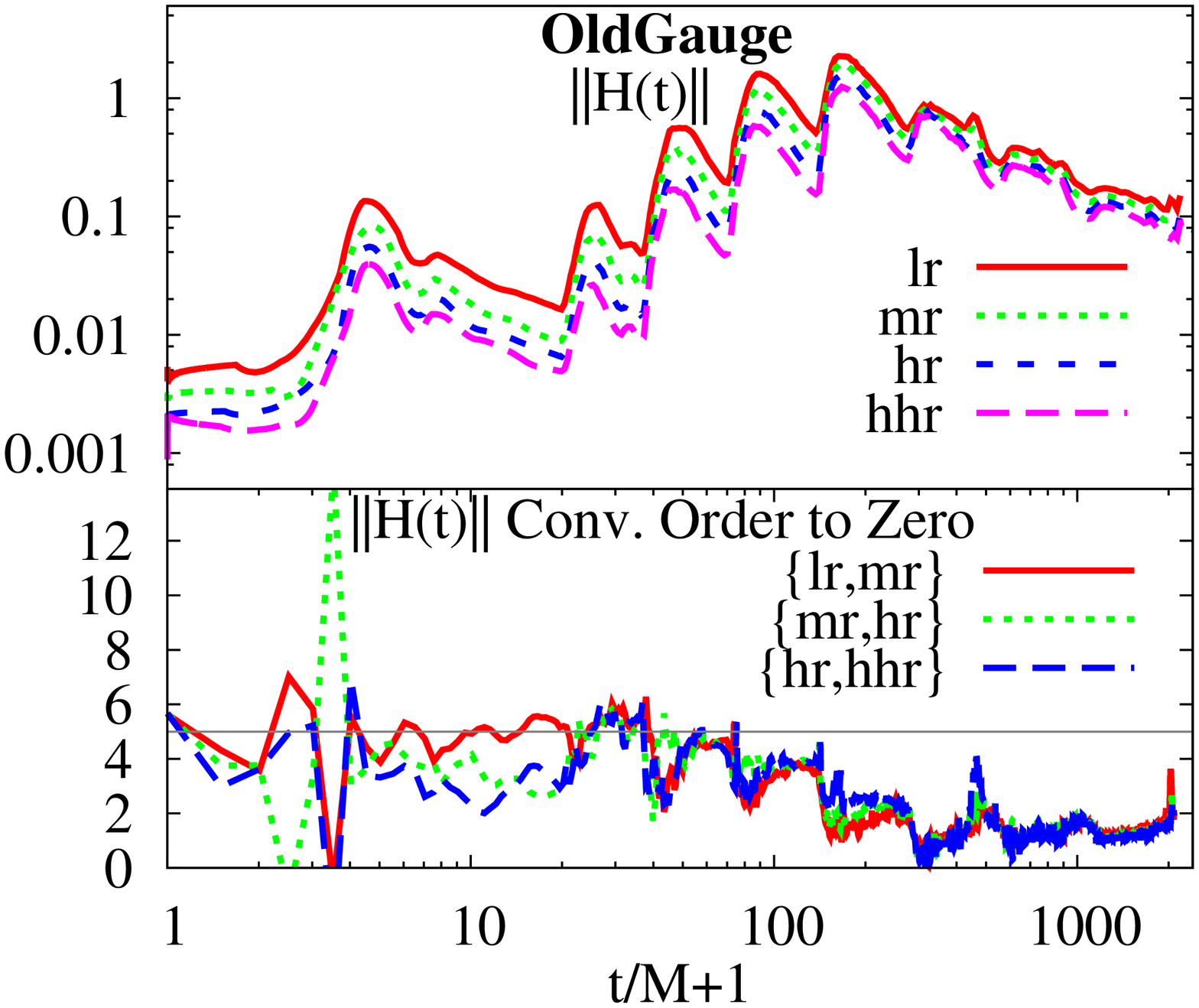}
\includegraphics[angle=270,width=0.45\textwidth]{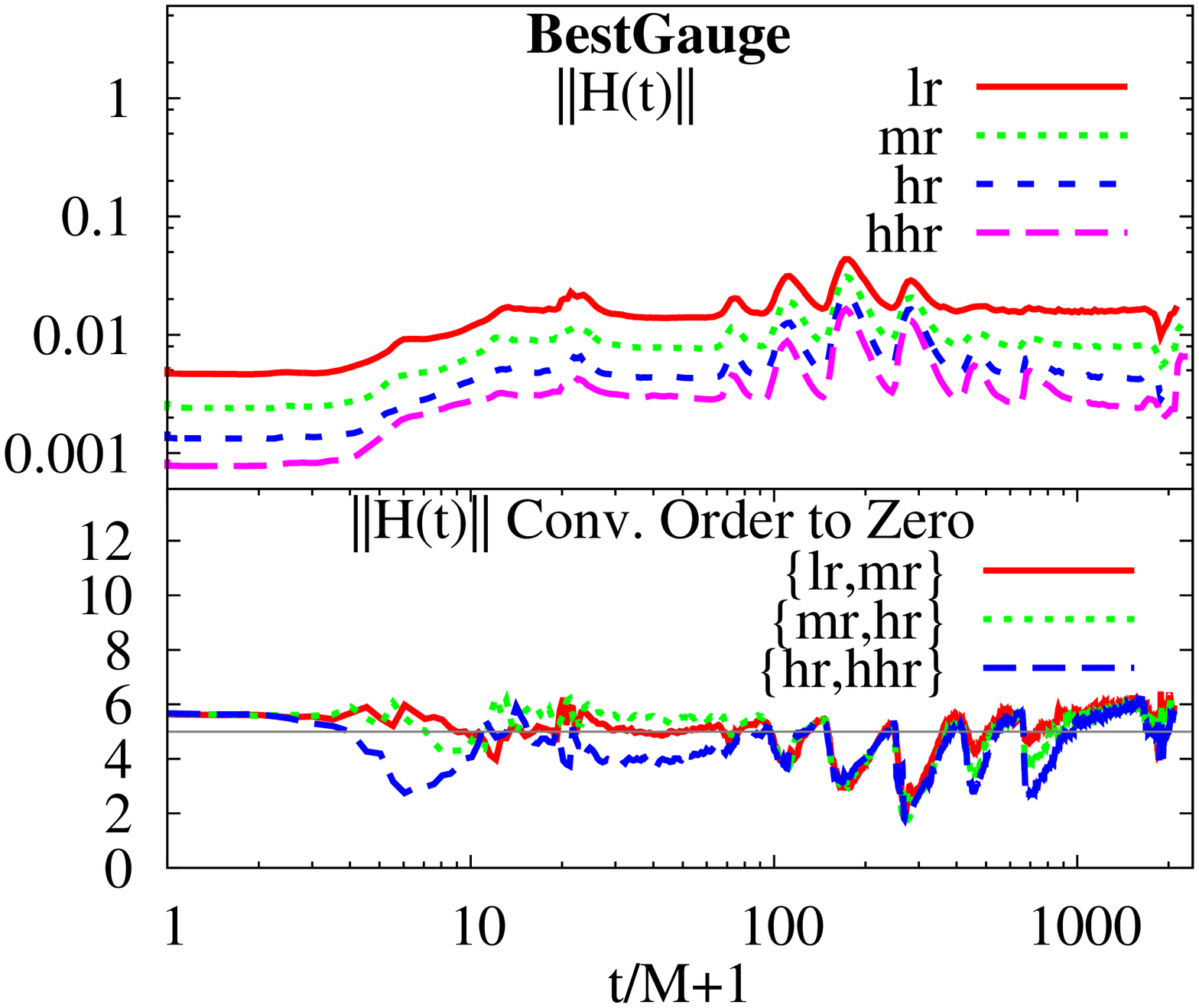}\\
\vspace{0.3cm}
\includegraphics[angle=270,width=0.45\textwidth]{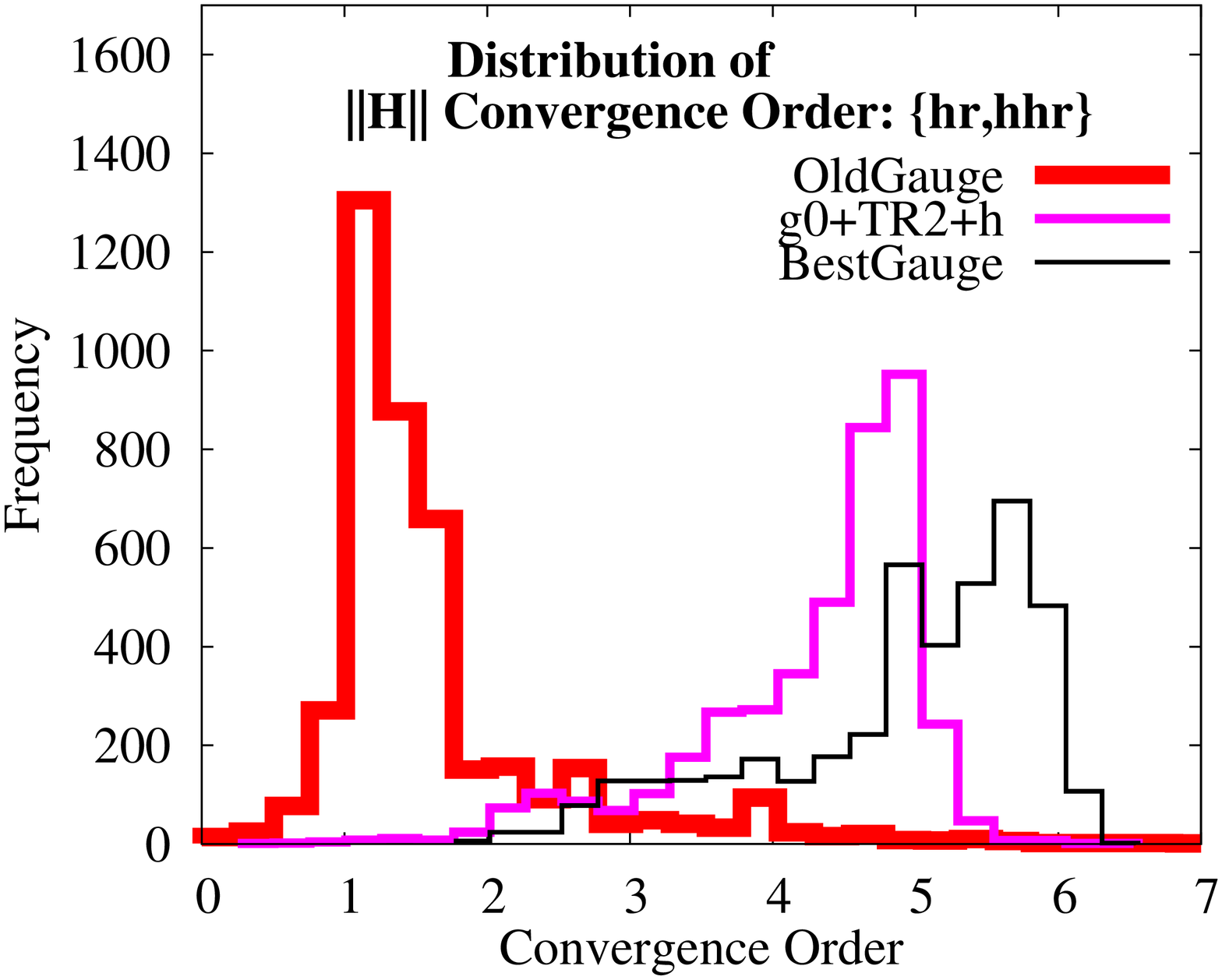}
\includegraphics[angle=270,width=0.45\textwidth]{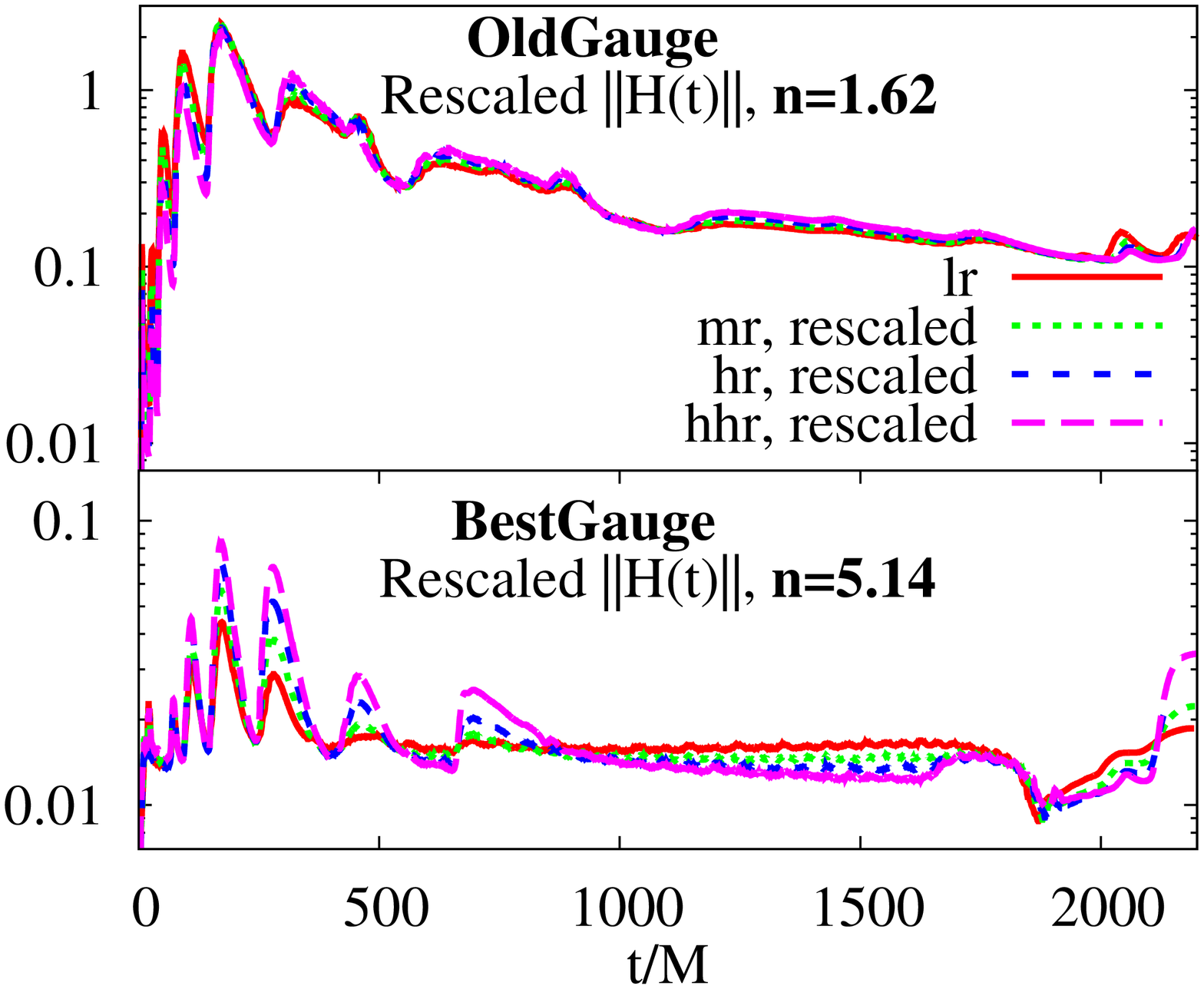}
\caption{Convergence to zero of Hamiltonian constraint
  violations. 
  {\bf Upper-left panel}: Top plot shows raw $||\mathcal{H}(t)||$ at
  resolutions ``lr'' (solid red), ``mr'' (dotted green), ``hr''
  (short-dashed blue), and ``hhr'' (long-dashed magenta) for OldGauge
  case. Bottom plot shows implied convergence order $n$
  (Eq.~\ref{convergence}) to zero for resolution pairs \{lr,mr\}
  (solid red); \{mr,hr\} (dotted green); and \{hr,hhr\} (dashed
  blue).
  {\bf Upper-right panel:} Same as upper-left panel, except for
  BestGauge case (g1+TR2+h+d5d7). 
  {\bf Lower-left panel:} Distribution of $||\mathcal{H}(t)||$
  convergence order, using \{hr,hhr\} data for
  OldGauge case (thick red), 
  g0+TR2+h (medium thickness magenta), and
  BestGauge (g1+TR2+h+d5d7; thin black). Note
  that the thick red line corresponds to the distribution of convergence
  order, as shown by the red line in the bottom plot of the upper-left
  panel.
  {\bf Lower-right panel:} Top plot: $||\mathcal{H}(t)||$ for
  OldGauge case, rescaled according to the observed average
  convergence order to zero over all resolutions, $n=1.62$. Bottom
  plot: same as top, except for BestGauge (g1+TR2+h+d5d7) case, where
  the average convergence order to zero over all resolutions is found
  to be $n=5.14$. All cases are rescaled to ``lr'' resolution.}
\label{Fig:Ham_constraint_order_description}
\end{figure*}

\begin{table*}
\caption{Convergence to zero of Hamiltonian and $x$-component of
  Momentum constraint violations. ``Median'' (``Mean'') numbers denote
  median (average) convergence order using an evenly-spaced sample of
  integrated L2-norm of constraint violations in time. This
  integral excludes regions around the BHs (Eq.~\ref{L2_Ham}), and for
  $||\mathcal{H}_\text{inner}||$, the region $r>32.3M$ as well. Note
  that for consistency, all data are truncated at time $t=2086.4M$,
  which is the earliest time at which a run was stopped ($\approx
  210M$ after merger).}
\begin{tabular}{|c|c|c|c|}
\hline
Case Name & $||\mathcal{H}||_\text{lr,mr}$ Median (Mean) &
$||\mathcal{H}||_\text{mr,hr}$ Median (Mean) &
$||\mathcal{H}||_\text{hr,hhr}$ Median (Mean) \\
\hline
OldGauge 
& 1.43 (1.64) & 1.30 (1.62) & 1.46 (1.60) \\
\hline
g0+TR2+h 
& 5.17 (4.70) & 5.97 (4.31) & 5.27 (4.24) \\
\hline
g0+TR2+h+d5d7 
& 5.57 (5.27) & 5.86 (4.97) & No hhr \\
\hline
BestGauge (g1+TR2+h+d5d7)
& 5.63 (5.38) & 5.92 (5.16) & 5.41 (4.87) \\
\hline
\hline
 & $||\mathcal{H}_\text{inner}||_\text{lr,mr}$ Median (Mean) &
$||\mathcal{H}_\text{inner}||_\text{mr,hr}$ Median (Mean) &
$||\mathcal{H}_\text{inner}||_\text{hr,hhr}$ Median (Mean) \\
\hline
OldGauge 
&5.77 (5.75) &5.72 (5.46) &5.98 (5.87) \\
\hline
g0+TR2+h 
&5.82 (5.80) &5.86 (5.67) &5.96 (5.74)\\
\hline
g0+TR2+h+d5d7 
&5.78 (5.75) &5.61 (5.69) & No hhr \\
\hline
BestGauge (g1+TR2+h+d5d7)
&5.81 (5.79) &5.94 (5.77) &5.76 (5.71) \\
\hline
\hline
& $||\mathcal{M}^x||_\text{lr,mr}$ Median (Mean) &
$||\mathcal{M}^x||_\text{mr,hr}$ Median (Mean) &
$||\mathcal{M}^x||_\text{hr,hhr}$ Median (Mean) \\
\hline
OldGauge 
& 1.37 (1.81) & 1.37 (1.62) & 1.29 (1.96) \\
\hline
g0+TR2+h 
& 1.09 (1.71) & 1.00 (1.98) & 1.37 (2.00) \\
\hline
g0+TR2+h+d5d7 
& 1.27 (1.86) & 1.31 (1.86) & No hhr \\
\hline
BestGauge (g1+TR2+h+d5d7)
& 1.24 (1.94) & 1.26 (1.84) & 1.78 (1.91) \\
\hline
\end{tabular}
\label{Convergence_order_table}
\end{table*}

Table~\ref{Convergence_order_table} presents the average and median order
at which Hamiltonian and ($x$-component) momentum constraints converge to zero for
various cases. To better understand how these numbers are generated,
Fig.~\ref{Fig:Ham_constraint_order_description} outlines the
procedure. The top left (right) plots in the upper panels show the raw
Hamiltonian constraint data $||\mathcal{H}(t)||$ (Eq.~\ref{L2_Ham})
for the OldGauge (BestGauge) case, and the bottom plots of the
upper panels use data at pairs of resolutions to solve for the implied
convergence order $n$ (Eq.~\ref{convergence}, where $Q_\text{RE}\to0$,
leaving only $C$ and $n$ as unknowns). Using many data points for $n$,
uniformly spaced in time, we can then construct histograms of
convergence order from this plot. The lower-left panel shows the
histogram of convergence order for ``hr'' and ``hhr'' resolutions,
where the thick red line corresponds to the solid red curve in the
bottom plot of the upper-left panel. This histogram is compared to the
same histogram for cases g0+TR2+h and BestGauge (g1+TR2+h+d5d7). We
point out two clear patterns: (1) the g0+TR2+h and BestGauge cases
possess significantly higher average convergence order than OldGauge,
(2) BestGauge appears to converge at a higher average convergence
order than g0+TR2+h. Notice the distribution widens with increasing
average convergence order. This is a reflection of the fact that after
t~300M, the OldGauge ||H|| converges below second-order and remains
there until merger. In BestGauge, convergence order drops
significantly as the sharp lapse wave crosses progressively
lower-resolution refinement boundaries. However, this subconvergent
spike in noise does not translate to sub-convergence later on in the
evolution. We attribute this to the lapse stretching and smoothing
features of BestGauge. Instead of plotting histograms for all cases at
all resolutions, we simply report the median and mean convergence
orders for all cases for which multiple resolution data are available
(Table~\ref{Convergence_order_table}).

According to Table~\ref{Convergence_order_table}, we expect
$||\mathcal{H}(t)||$ to converge to zero on average at order $n=1.62$
in the OldGauge case, and $n=5.14$ in the BestGauge
(g1+TR2+h+d5d7) case. 
The lower-right panel of
Fig.~\ref{Fig:Ham_constraint_order_description} demonstrates how well
data at different resolutions overlap when rescaled according to these
observed mean convergence orders. Notice that although
convergence order drops to as low as second order at each
spike in the BestGauge $||\mathcal{H}(t)||$ (cf. upper-right and
lower-right panels), BestGauge convergence order equilibrates at
late times to $n\approx6$. In the OldGauge case,
$||\mathcal{H}(t)||$  convergence order steadily drops from the outset
of the calculation (upper-left panel) as the initial sharp lapse wave
crosses into progressively lower resolution AMR grid boundaries.

Recall that in Fig.~\ref{Fig:Ham_step_by_step}, we found that
$||\mathcal{H}(t)||$ is reduced by a factor of 20 on average in the
BestGauge case, compared to a standard moving-puncture gauge
choice (OldGauge). Table~\ref{Convergence_order_table}
demonstrates that the {\it convergence order} in $||\mathcal{H}(t)||$
is also significantly improved with our gauge improvements, increasing
from below second-order to nearly sixth-order convergence. Thus at
higher resolutions, we expect our gauge improvements to reduce
$||\mathcal{H}(t)||$ by an even higher factor than that observed
at ``lr'' resolution. In fact ``BestGauge, hhr'' drops
$||\mathcal{H}(t)||$ by a factor of approximately {\it 58}, as
compared to the ``OldGauge, hhr'' case. 

When the L2 norm of $\mathcal{H}(t)$ (Eq.~\ref{L2_Ham}) excises
the region outside $r=32.3M$ (i.e., $||\mathcal{H}_\text{inner}||$ in
Table~\ref{Convergence_order_table}), we observe nearly sixth-order
convergence in {\it all cases}. Thus a bulk of sub-convergent
Hamiltonian constraint violations comes from the outer regions, where
gravitational waves are measured, and the under-resolution of the
spinning BH at ``lr'' resolution (as discussed in
Sec.~\ref{sec:Mirr_conv}) does not appear to translate to inconsistent
convergence at low resolutions in constraint violations.

We did not have such an ``inner region'' diagnostic for momentum
constraint violations (i.e., an $||\mathcal{M}^x_\text{inner}||$
diagnostic). Recall that at ``lr'' resolution, $||\mathcal{M}^x||$ has
been reduced on average by a factor of $\approx 13.4$ with our new
gauge choices (Fig.~\ref{Fig:Momx_step_by_step}). However, its {\it
  convergence to zero} does not improve significantly and remains
between first and second-order regardless of gauge choice. 
This may be due to the fact that even in our most advanced
gauge choice (BestGauge), some high-frequency noise remains and
further efforts may be required to tamp down this remaining noise.

So do drastic constraint violation reductions in the
outer regions of our grid translate to improvements in gravitational
waveform (i.e., $\psi^4$) convergence, or does the relatively large
drift in the spinning BH's spin parameter over time imply poor waveform
convergence at ``lr'' resolution? The next section examines this
question in detail.


\subsubsection{Waveform Convergence}

\begin{figure*}
\includegraphics[angle=270,width=0.45\textwidth]{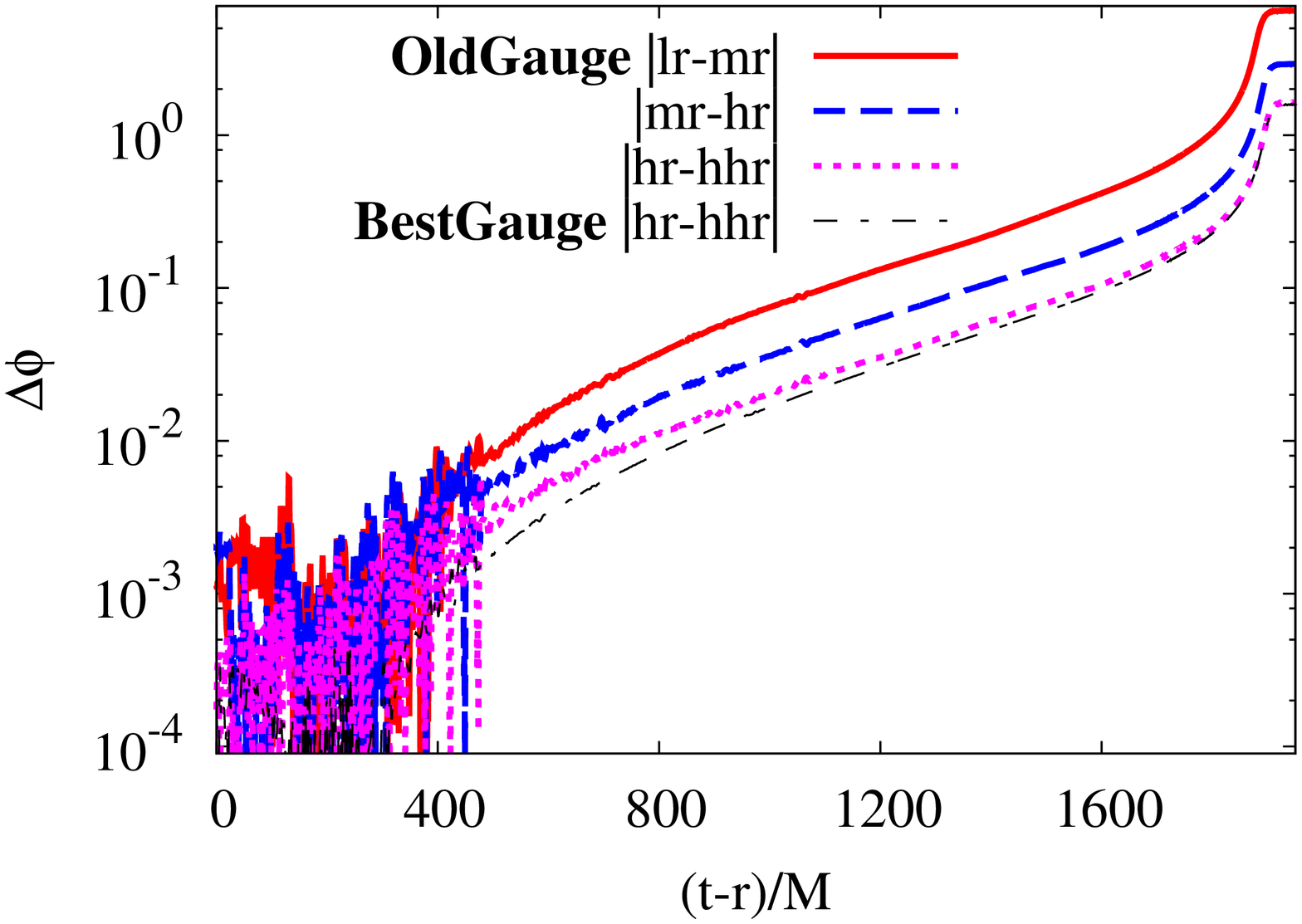}
\includegraphics[angle=270,width=0.45\textwidth]{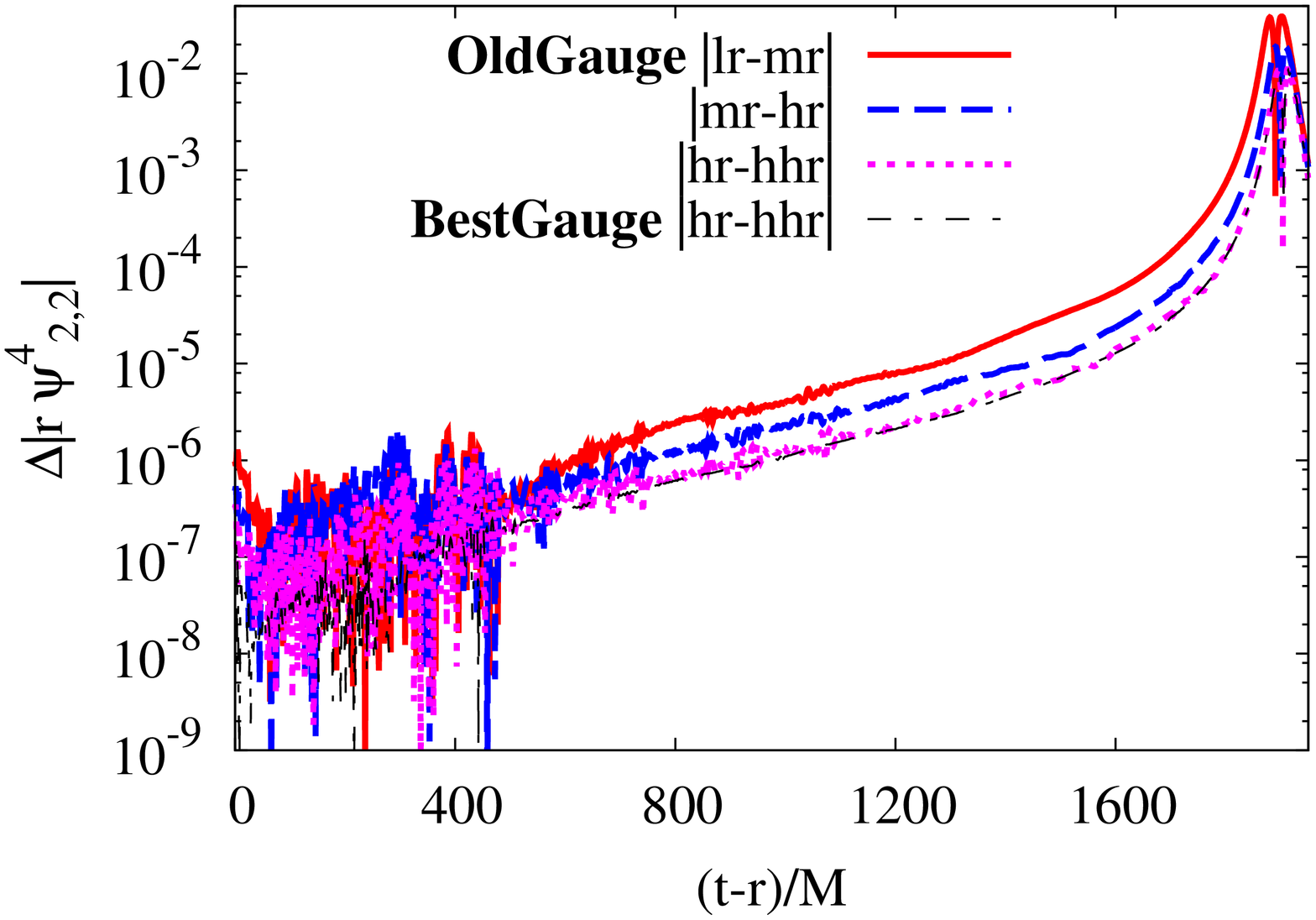} \\
\vspace{0.3cm}
\includegraphics[angle=270,width=0.45\textwidth]{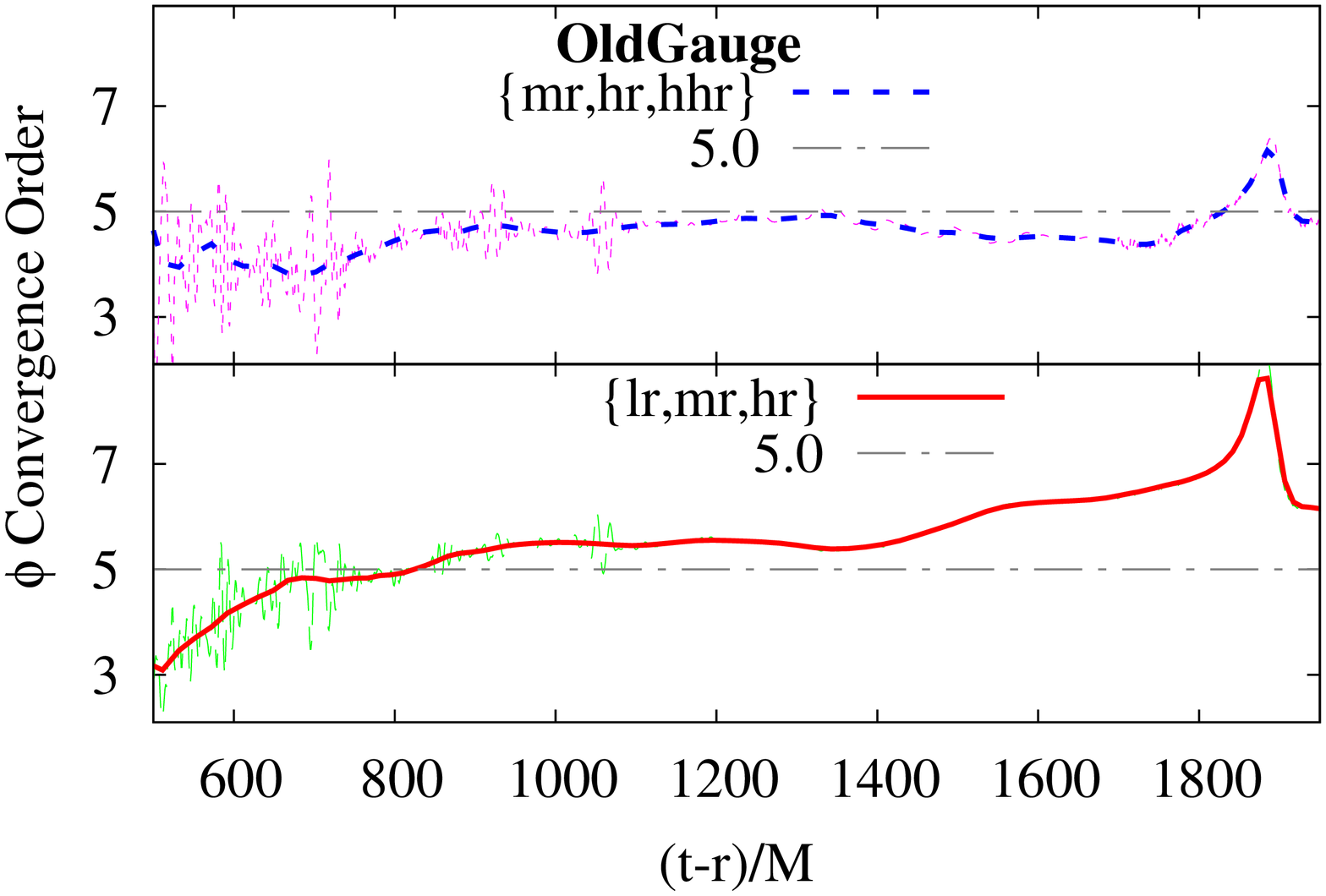}
\includegraphics[angle=270,width=0.45\textwidth]{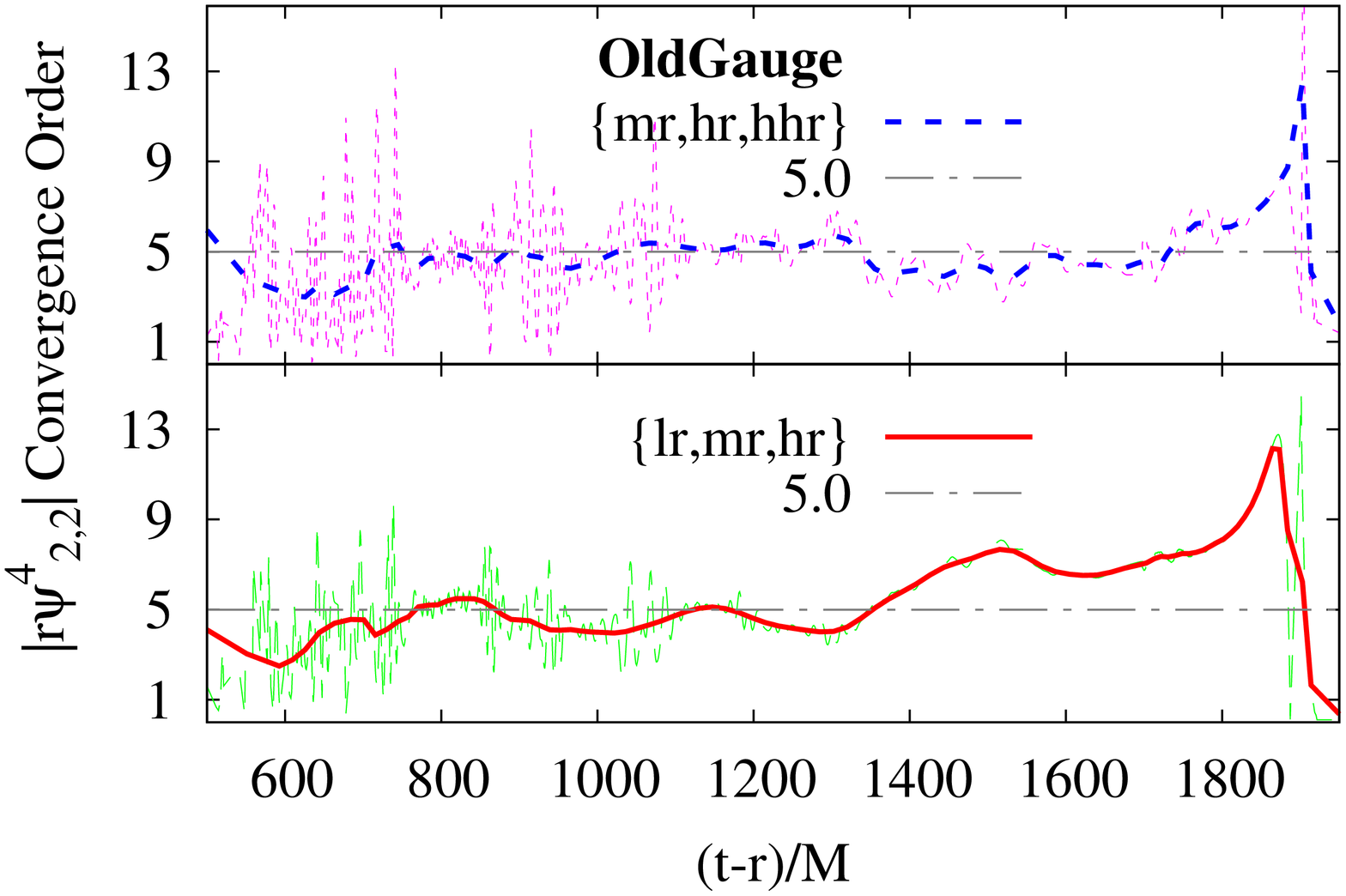} \\
\vspace{0.3cm}
\includegraphics[angle=270,width=0.45\textwidth]{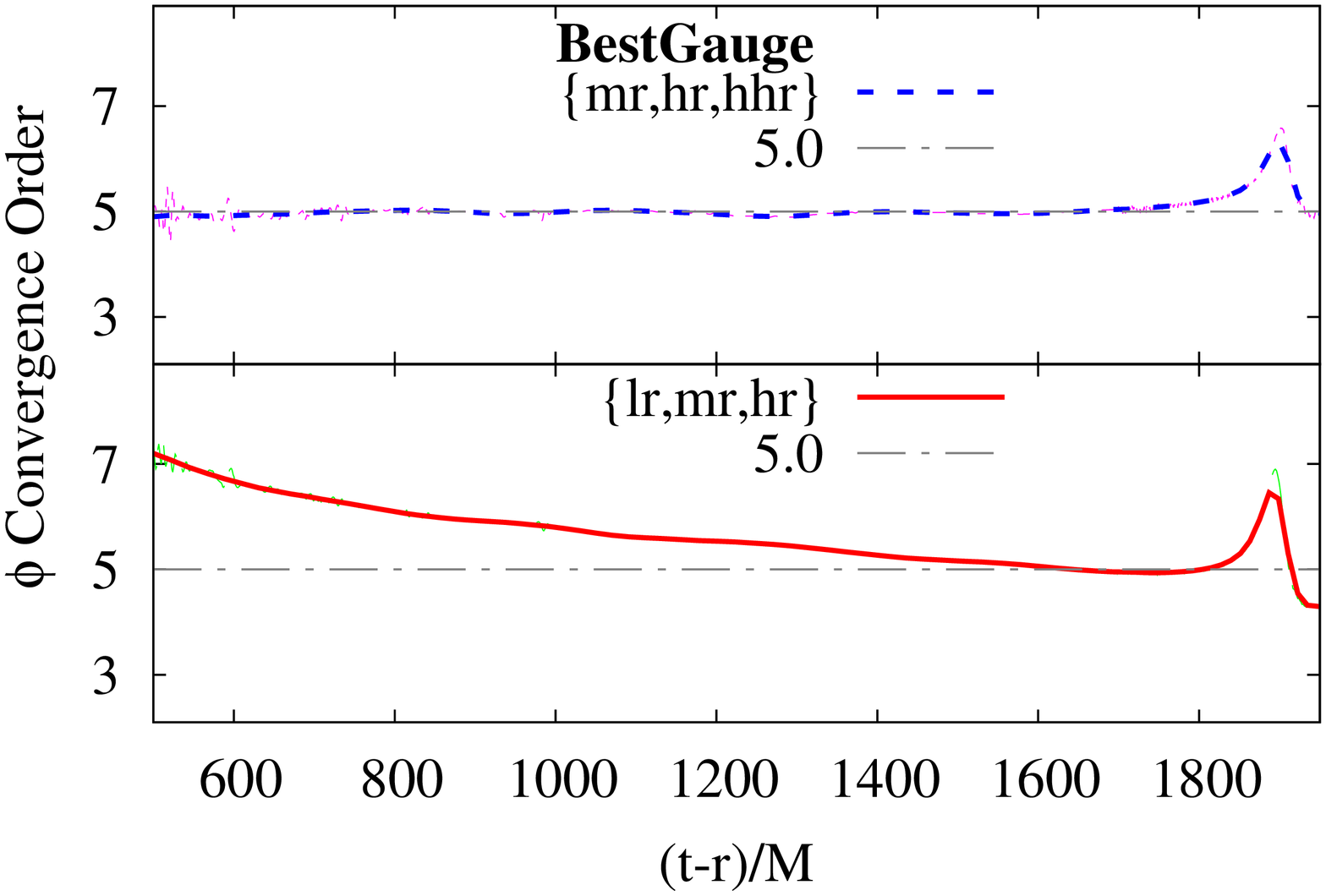}
\includegraphics[angle=270,width=0.45\textwidth]{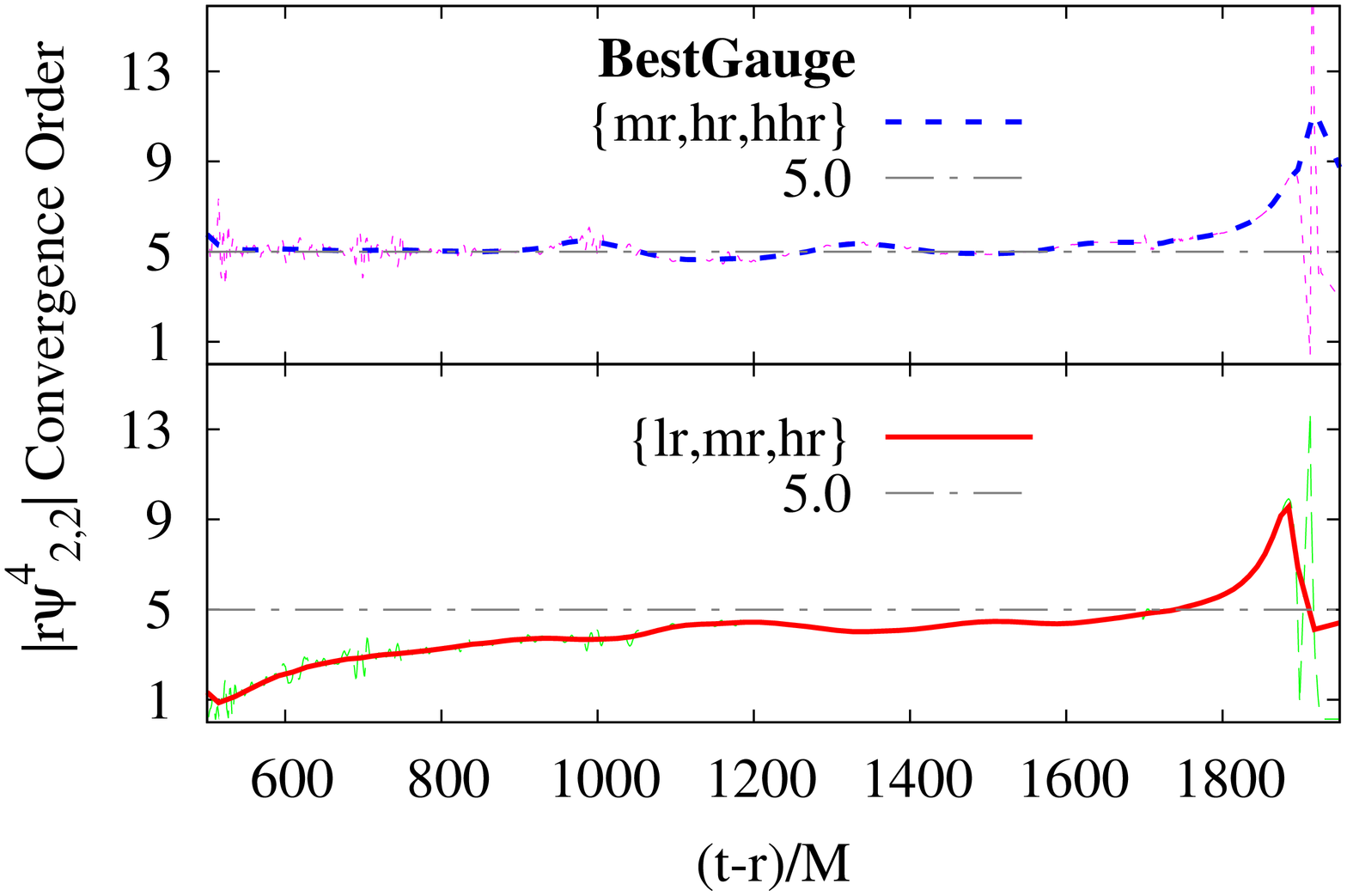}
\caption{$\psi^4_{2,2}$ convergence study, using {\it smoothed}
  amplitude and phase data.
  {\bf Upper-left (right) panel}:
  phase (amplitude) difference between pairs of resolutions, for OldGauge $|$lr-mr$|$
  (thick solid red), $|$mr-hr$|$ (thick dashed blue), and $|$hr-hhr$|$ (thick
  dotted magenta), as well as BestGauge  $|$hr-hhr$|$
  (g1+TR2+h+d5d7, thin dashed black). 
  {\bf Middle-left (right) panel}: OldGauge phase (amplitude)
  convergence order versus time, using results from \{lr,mr,hr\}
  resolutions (raw data: thin dashed green, B\'{e}zier-smoothed data:
  thick solid red) and \{mr,hr,hhr\}
  resolutions (raw data: thin dashed magenta, B\'{e}zier-smoothed data:
  thick dashed blue). 
  {\bf Lower-left (right) panel}: same as middle panels, except for
  BestGauge (g1+TR2+h+d5d7). 
  Please note that for
  ease of analysis, all raw amplitude and phase data in this figure
  were first smoothed with
  a moving-window quadratic least-squares fit method prior to computing
  convergence order, and where specified, B\'{e}zier smoothing was
  applied separately to the resulting convergence order data as well.
  Also, data $(t-r)/M<500$ have been removed from
  convergence order plots, as they are too noisy even after smoothing
  to show a clear convergence order (see top panels).
  $\psi^4_{2,2}$ data are measured at radius $r=68.6M$.}
\label{Fig:Psi4_conv_study}
\end{figure*}
\begin{figure*}[t]
\includegraphics[angle=270,width=0.45\textwidth]{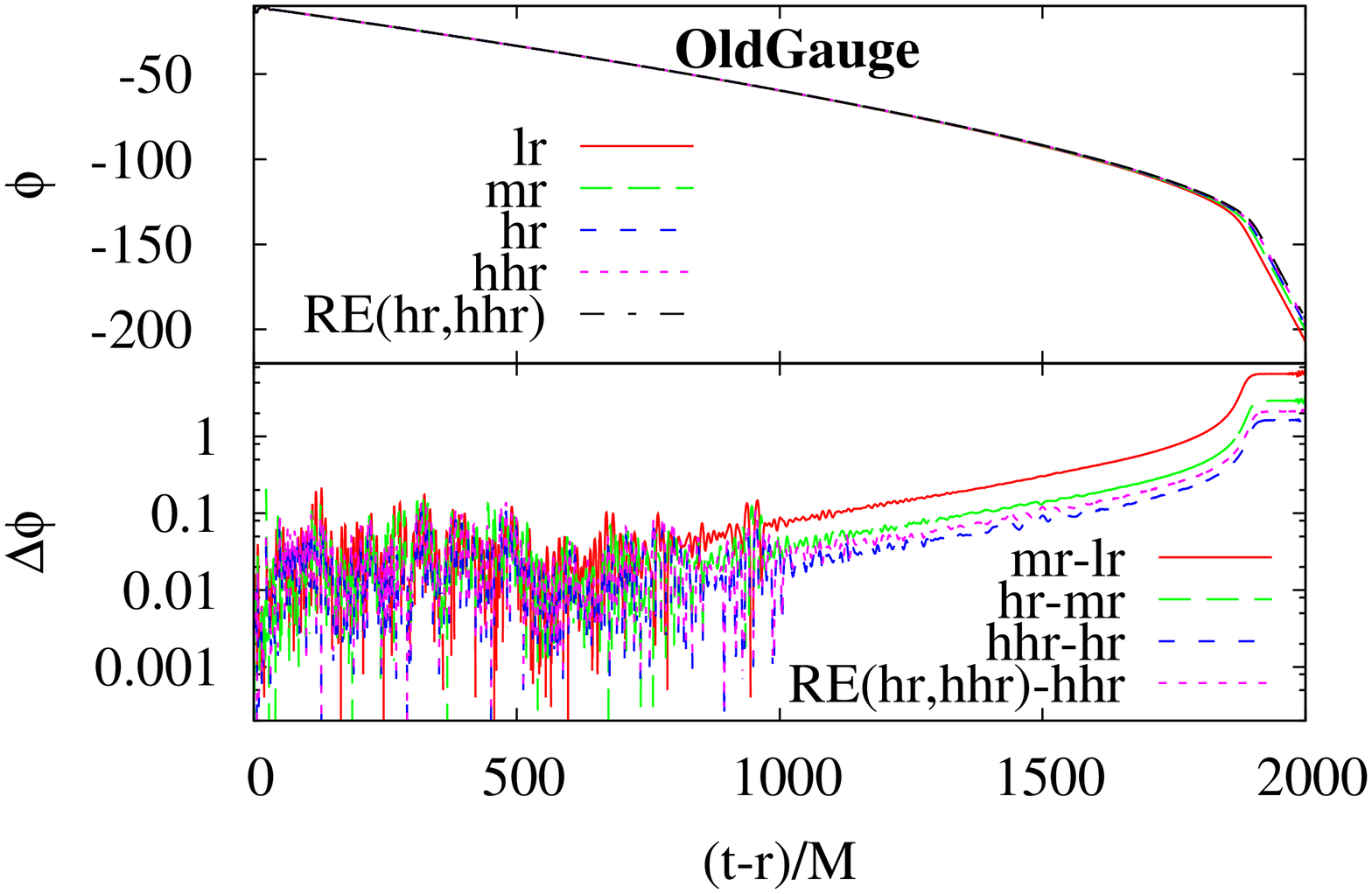}
\includegraphics[angle=270,width=0.45\textwidth]{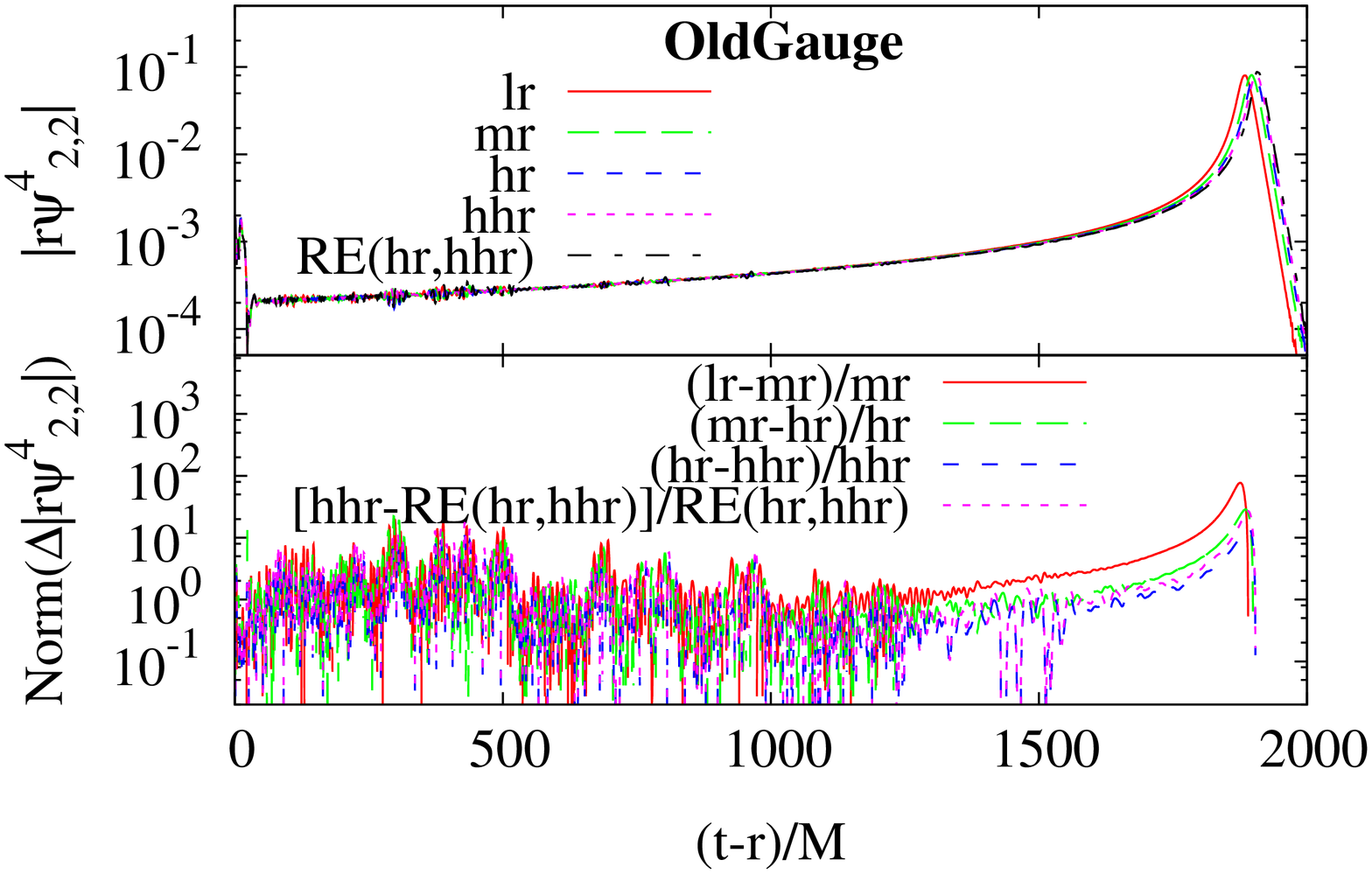} \\
\vspace{0.3cm}
\includegraphics[angle=270,width=0.45\textwidth]{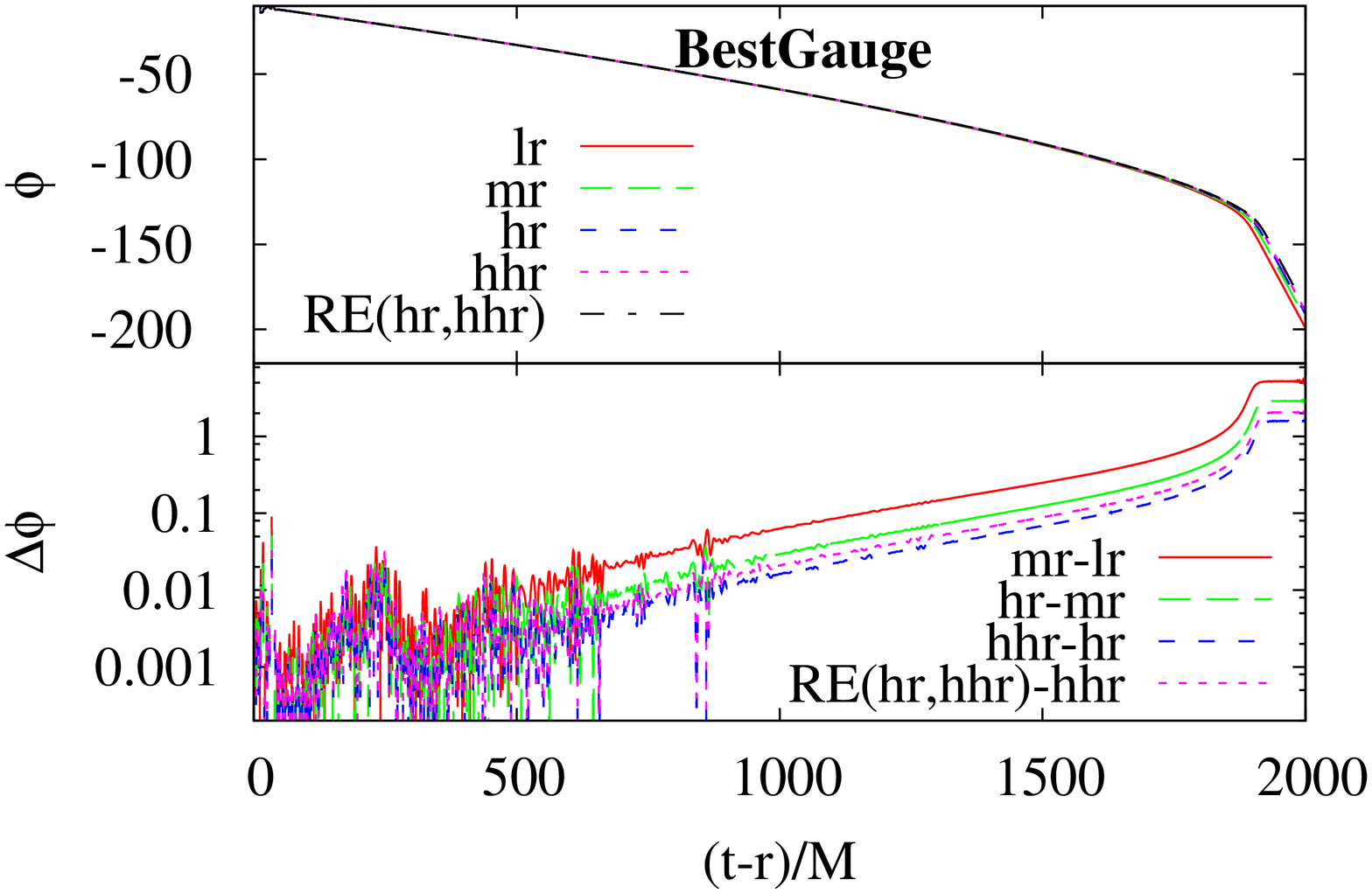}
\includegraphics[angle=270,width=0.45\textwidth]{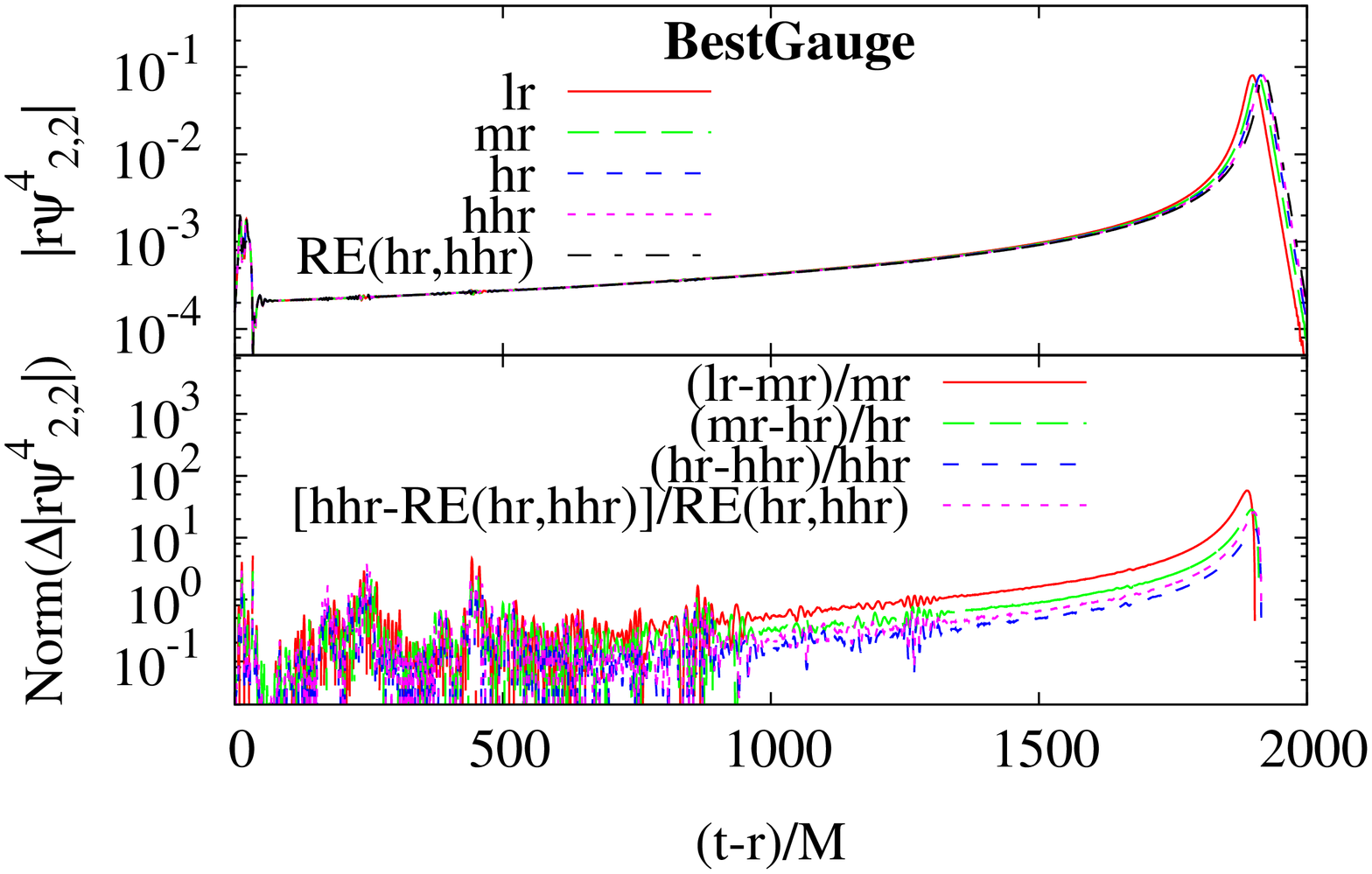} \\
\vspace{0.3cm}
\includegraphics[angle=270,width=0.45\textwidth]{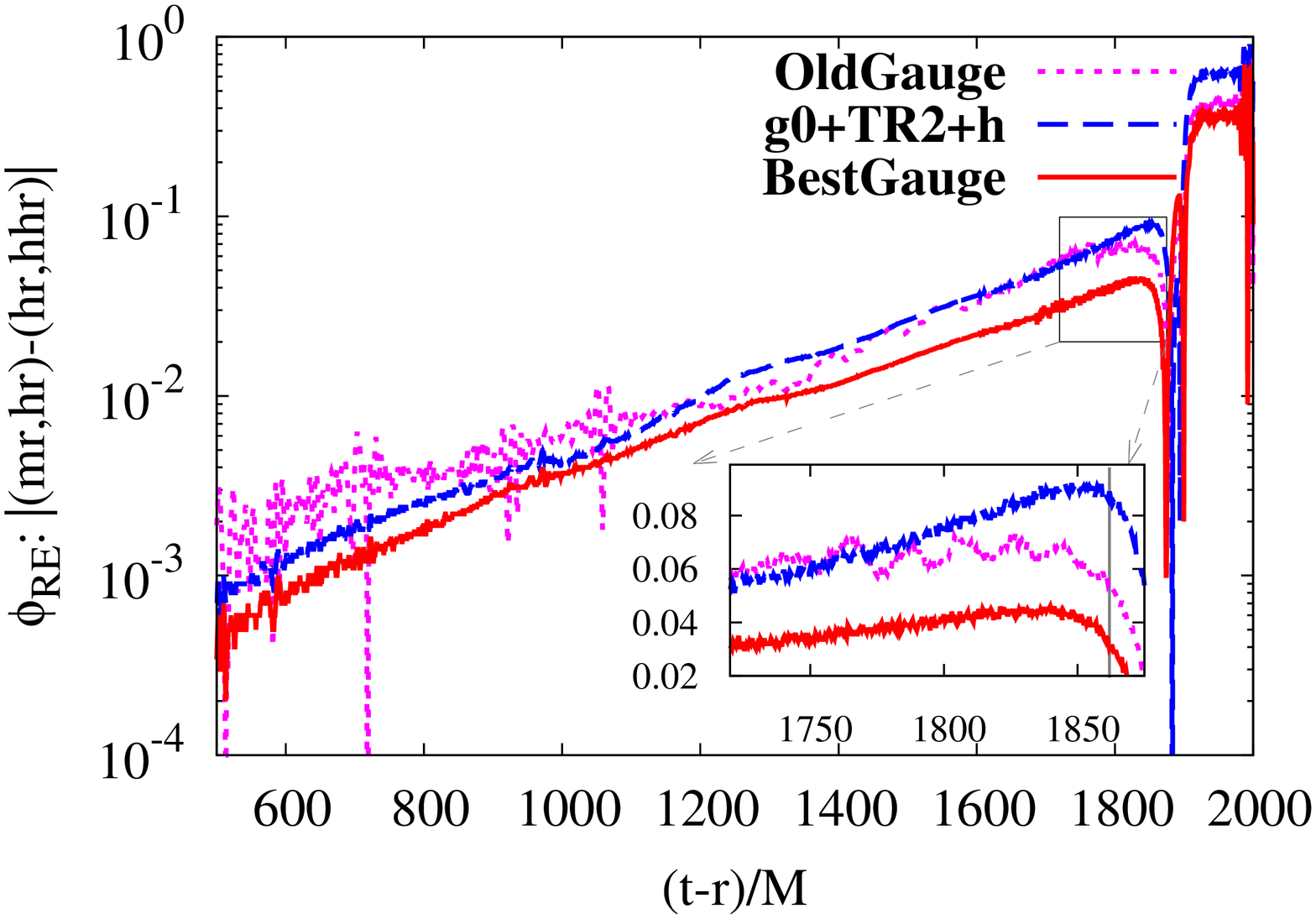}
\includegraphics[angle=270,width=0.45\textwidth]{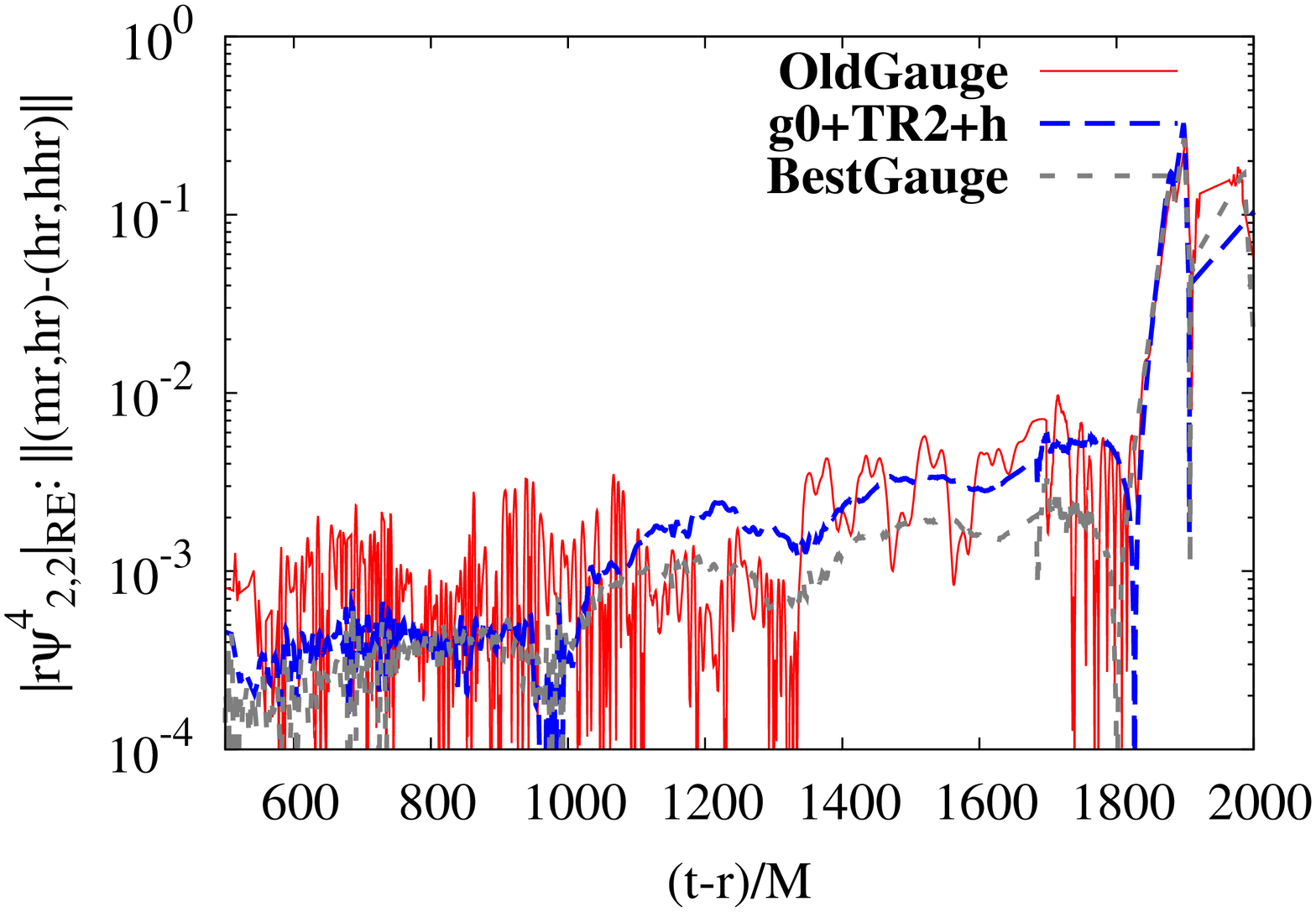}
\caption{$\psi^4_{2,2}$ convergence and error-analysis study, using
  {\it raw, unsmoothed} phase and amplitude data.
  {\bf Top panels}:
  Upper-left (right) plot: Phase (Amplitude) versus retarded time, for
  OldGauge case at resolutions ``lr'' (solid red), ``mr'' (long-dashed
  green), ``hr'' (medium-dashed blue), and ``hhr'' (dotted
  magenta). Richardson-extrapolated phase is overlaid, 
  using data from ``hr'' and ``hhr'' resolutions, and assuming
  constant fifth-order convergence (dash-dotted black). 
  Lower-left (right) plot: simple (not absolute value of) difference
  in phase (normalized amplitude) between resolutions.
  {\bf Middle panels}: same as top panels, but for case
  BestGauge (g1+TR2+h+d5d7).
  {\bf Bottom panels}: left (right) plot shows
  $|$RE(mr,hr)-RE(hr,hhr)$|$ for waveform phase (amplitude)
  versus retarded time, assuming fifth-order convergence, where
  RE(mr,hr) denotes the Richardson-extrapolated value combining phase
  (amplitude) numerical data at resolutions \{mr,hr\} with the
  assumption of fixed $n=5$ convergence order (Eq.~\ref{convergence}).
  Amplitude data are normalized by the higher-resolution value.
  Cases OldGauge (dotted magenta), g0+TR2+h
  (long-dashed blue), and BestGauge
  (g1+TR2+h+d5d7; solid red) are compared. Zoom inset on the left plot
  shows accumulated phase difference near the end of the inspiral,
  with vertical gray line denoting $M\omega_{2,2}=0.2$ (corresponding
  to a retarded time of $(t-r)/M\approx1866$). All $\psi^4_{2,2}$ data
  are measured at radius $r=68.6M$.
}
\label{Fig:Psi4_Rich_extrap_study}
\end{figure*}
\begin{figure*}[t]
\includegraphics[angle=270,width=0.45\textwidth]{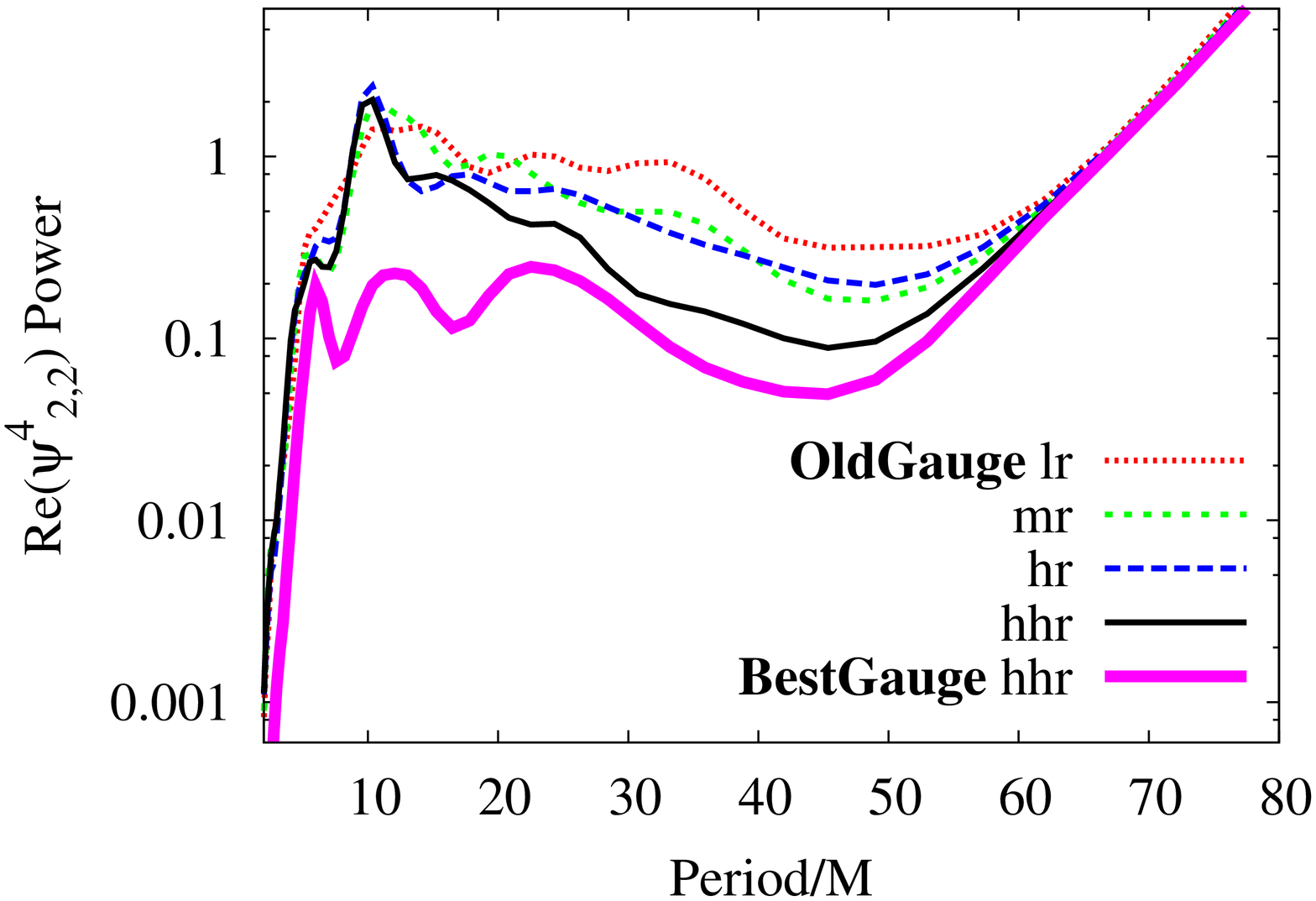}
\includegraphics[angle=270,width=0.45\textwidth]{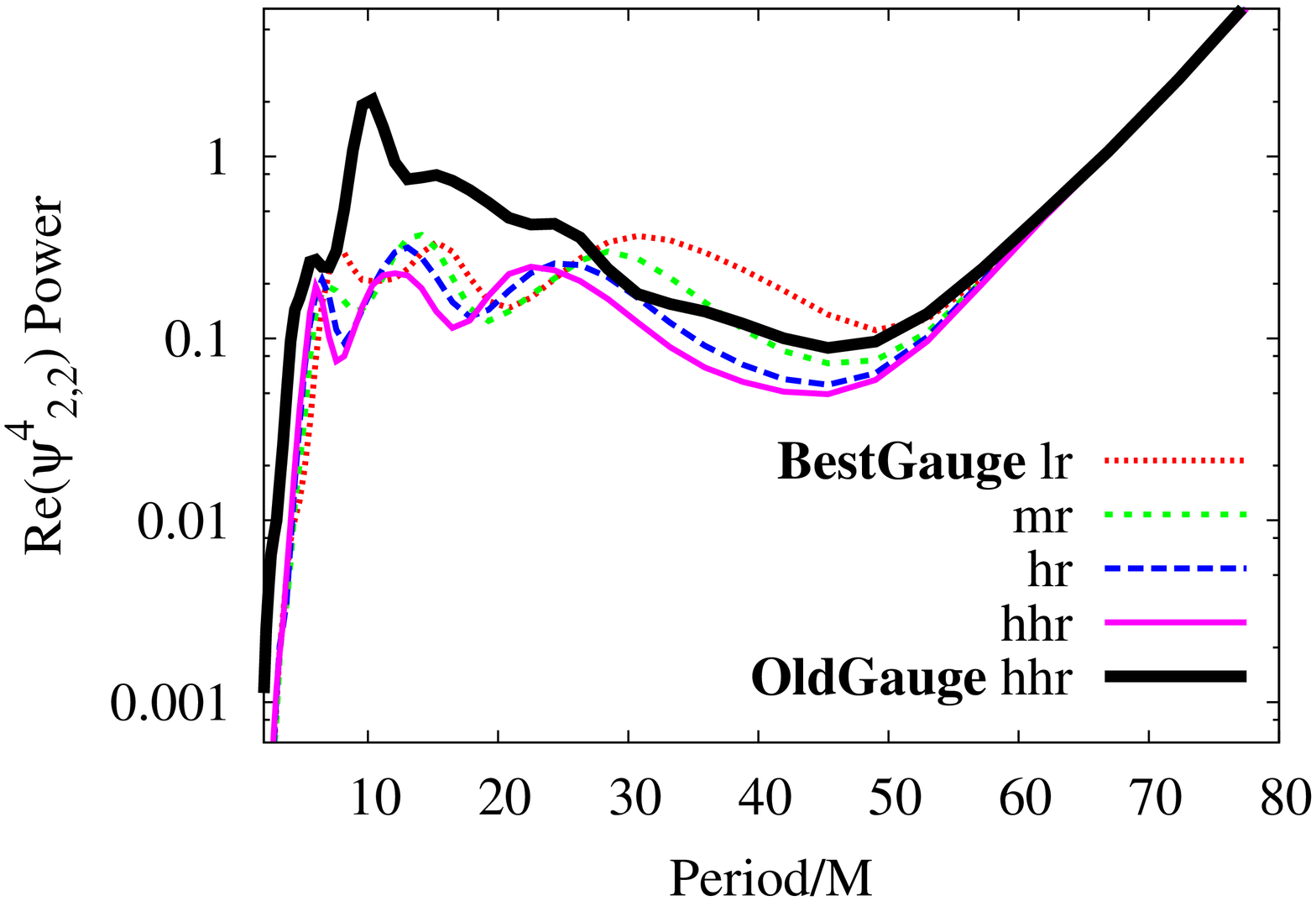}
\caption{Re($\psi^4_{2,2}$) noise convergence properties, during
  early inspiral [$58\lesssim(t-r)/M\lesssim1160$].
  {\bf Left panel}:
  B\'{e}zier-smoothed Power spectra of Re($\psi^4_{2,2}$) for case
  OldGauge, at resolutions ``lr'' (dotted red), ``mr''
  (short-dashed green), ``hr'' (dashed blue), and ``hhr'' (solid
  black). These are compared to the same quantity at ``hhr''
  resolution for case BestGauge
  (g1+TR2+h+d5d7; thick magenta).
  {\bf Right panel}: 
  B\'{e}zier-smoothed power spectra of Re($\psi^4_{2,2}$) for case
  BestGauge, at resolutions ``lr'' (dotted red), ``mr''
  (short-dashed green), ``hr'' (dashed blue), and ``hhr'' (solid
  magenta). These are compared to the same quantity at ``hhr''
  resolution for case OldGauge (thick black).
  Note that all Re($\psi^4_{2,2}$) data plotted are measured at radius
  $r=68.6M$, and for an example of unsmoothed data, see
  Fig.~\ref{Fig:Psi4_noise_reduction}. Re($\psi^4_{2,2}$) time series data were
  multiplied by the tapering function $\Erf(t;200,50) \times
  \Erf(-t;-1300,150)$ prior to the Fourier transform.}
\label{Fig:Psi4_waveform_noise_convergence}
\end{figure*}

Figure~\ref{Fig:Psi4_conv_study} presents a gravitational waveform
convergence study, separating $\psi^4_{2,2}(t)$ into amplitude and
phase separately, as these quantities could be influenced differently
by truncation error~\cite{Husa:2008}.

The top left (right) panel of Fig.~\ref{Fig:Psi4_conv_study} shows
$|Q_\text{numerical}(\Delta x_1)-Q_\text{numerical}(\Delta x_2)|$,
where $Q$ is the {\it smoothed} phase (amplitude) of $\psi^4_{2,2}(t)$,
and \{$\Delta x_1,\Delta x_2$\} denotes data at two adjacent resolutions in the set
\{lr,mr,hr,hhr\}. Notice that the \{hr,hhr\} phase difference is
slightly smaller during inspiral in the BestGauge
(g1+TR2+h+d5d7) than the OldGauge case. This is one measure that
demonstrates our gauge improvements reduce phase errors.

The remaining left (right) panels of Fig.~\ref{Fig:Psi4_conv_study}
plot convergence order $n$ in $\psi^4_{2,2}$ phase (amplitude) as a
function of time, comparing the OldGauge (middle panels) to the
BestGauge (g1+TR2+h+d5d7; bottom panels)
cases. At the three lowest resolutions \{lr,mr,hr\}, inconsistent
convergence between amplitude and phase is observed, regardless of
gauge choice. We attribute this to the under-resolution of the
spinning BH (Sec.~\ref{sec:Mirr_conv}). Cleaner convergence is
observed when computing $n$ from data at the three highest resolutions
\{mr,hr,hhr\}. However, convergence order in the OldGauge case
oscillates significantly between $n=4$ and $n=5$ until $t\approx
1300M$, and then drops to $n\approx4$ until merger. The same data from
our BestGauge case indicate extremely steady $n=5$ convergence order
with almost no oscillation, until merger.

At merger ($t\gtrsim1750M$), loss in convergence is observed
regardless of gauge choice, pointing to a non-convergent effect
not addressed by our gauge modifications. However, there is some
chance that this loss of convergence may be restored with another
careful time reparameterization near merger, which will allow the
gauge to more quickly settle as the gravitational fields undergo the
extremely rapid changes associated with merger. 

The consistent $n=5$ convergence observed in BestGauge is a highly
significant result, as it strongly indicates that our errors are
dominated by the fifth-order-accurate spatial interpolation errors at
grid refinement boundaries in BestGauge. We could have
reasonably expected this {\it a priori} given that the lower-order
temporal errors have been tamped down significantly by our choice of
timesteps well below the CFL limit. The convergence order
oscillations in OldGauge below $n=5$ are troubling, leading us to
question the reliability of Richardson-extrapolated estimates of
$\psi^4_{2,2}$ that assume a standard, fixed-in-time integer
convergence order when choosing OldGauge-like gauge conditions.

We believe that the OldGauge convergence order oscillations in
$\psi^4_{2,2}$ amplitude and phase may be related to the large amount of
power in $\psi^4_{2,2}$ noise, particularly at wave periods of
$P\sim 10M$ as shown in Fig.~\ref{Fig:Psi4_noise_reduction}, as this
noise has been largely removed in the BestGauge case. In
fact, in Sec.~\ref{sec:waveform_noise_conv} below, we show that
waveform noise at the dominant noise frequency ($P\approx 10M$) in
OldGauge {\it does not diminish with increasing resolution}, at the
resolutions chosen.

We end our discussion on Fig.~\ref{Fig:Psi4_conv_study} by cautioning
that our clean, fifth-order convergence in BestGauge has only been
demonstrated with the three highest resolutions, and that
higher-resolution studies will be useful in cementing our
conclusions. Although our gauge improvements have apparently
eliminated the non-convergent gravitational waveform noise (as shown
in Sec.~\ref{sec:waveform_noise_conv} below), some noise
does remain, and we anticipate that at high resolutions far beyond
that attempted here, further gauge improvements may be necessary to
maintain a uniform, integer convergence order.


We now turn our attention to the Richardson-extrapolated
waveforms. Given that our gauge improvements yield cleaner
gravitational-waveform convergence at higher resolutions, we may be able to
more precisely estimate the true value of $\psi^4_{2,2}$
through Richardson extrapolation. The Richardson-extrapolated value is
given by $Q_\text{RE}$ in Eq.~(\ref{convergence}), replacing $Q$ with
$\psi^4_{2,2}$ at all times $t$. Since we have established the
dominant waveform convergence order is cleanly $n=5$ for \{mr,hr,hhr\}
resolutions, we can compute $\psi^4_{2,2,\text{RE}}$ in
Eq.~(\ref{convergence}) by setting $n=5$ and combining numerical data
at two resolutions.

The top and middle panels of Fig.~\ref{Fig:Psi4_Rich_extrap_study} compare
{\it raw} $\psi^4_{2,2}$ amplitude and phase data at all four
resolutions to Richardson-extrapolated values at the highest two
resolutions \{hr,hhr\}. 
At $t\lesssim800M$ ($t\lesssim1200M$) in the OldGauge (BestGauge),
simple differences between data at adjacent resolutions are too noisy
to easily distinguish convergence properties (top panels). This can be
somewhat mitigated by careful smoothing, as was shown in
Fig.~\ref{Fig:Psi4_conv_study}, but both phase and amplitude data at
$t\lesssim500$ are too noisy to determine convergence order cleanly in
OldGauge, even with smoothing. Due to the large noise reductions
of BestGauge, the situation is greatly improved (middle panels), though determining
convergence order in phase (amplitude) remains difficult for times
$t\lesssim 600M$ ($t\lesssim 900M$) without smoothing. Notice that the
difference between Richardson-extrapolated phase and amplitude data
(as computed at \{hr,hhr\} resolutions)  and ``hhr'' resolution data
is larger than the difference between ``hr'' and ``hhr''
data.

The bottom left (right) panel of Fig.~\ref{Fig:Psi4_Rich_extrap_study}
shows differences in Richardson-extrapolated phase (amplitude) versus
time. One Richardson-extrapolated value combines data at \{mr,hr\}
resolutions---which we denote RE(mr,hr)---and the other combines
\{hr,hhr\}---denoted by RE(hr,hhr). Any nonzero value for
$|$RE(mr,hr)-RE(hr,hhr)$|$ may be caused by either deviations from
fifth-order convergence---which we have shown to be particularly
significant in OldGauge (Fig.~\ref{Fig:Psi4_conv_study})---or
truncation errors at non-dominant orders (cf. Eqs.~\ref{convergence}
and~\ref{convergence_exact}). Thus we consider
$|$RE(mr,hr)-RE(hr,hhr)$|$ as an error estimate for phase, and
$|$RE(mr,hr)-RE(hr,hhr)$|$/RE(hr,hhr) for amplitude.

Implied phase errors $|$RE(mr,hr)-RE(hr,hhr)$|$ increase
with time, but far more smoothly in the improved gauge cases, as
compared to the OldGauge case. Though phase errors are
roughly comparable in the OldGauge and g0+TR2+h cases, we find a
reduction in phase error of about $40\%$ in the BestGauge
(g1+TR2+h+d5d7) case throughout inspiral and
near merger. Normalized amplitude errors are much noisier, but appear
about 40--50\% smaller during inspiral in the BestGauge case as well,
as compared to the OldGauge and g0+TR2+h cases. However, during and
after merger, amplitude and phase errors for all three cases are large
and overlap. These large errors are expected, as the BBH merger
time changes as resolution is increased, and we do not account for
this effect by, e.g., shifting the waveforms so that their merger
times overlap.

We may use the two highest-resolution estimates for
$\psi^4_{2,2,\text{RE}}$ to compute accumulated amplitude and phase
errors at the end of inspiral, where we define the ``end of inspiral''
to be at the time in which gravitational wave frequency is
$M\omega=0.2$. An almost identical frequency was used in the NINJA
\cite{Ajith:2012az,Aasi:2014tra} and NRAR \cite{Hinder:2013oqa}
collaborations as the fiducial time at which to measure
waveform amplitude and phase errors, the only difference being that we
choose to normalize $\omega$ by total initial ADM mass instead of the
combined initial masses of the punctures. The difference in frequency
is less than one percent for runs presented here. We find accumulated
phase errors at $M\omega=0.2$ (vertical line in inset) of
approximately 0.046, 0.081, and 0.025 radians for OldGauge, g0+TR2+h,
and BestGauge cases, respectively. Thus the BestGauge gauge choice
reduces phase errors by about a factor of two. This is a rather
unfortunate time to measure phase error, as the OldGauge case phase
error does not increase as predictably as the BestGauge case, and just
happens to experience a local minimum in the rate of phase error
increase at $M\omega=0.2$. Had we measured phase errors slightly
earlier or perhaps at a higher resolution, we might have seen
further reduction of phase errors by BestGauge.

As pointed out earlier, normalized amplitude errors near merger are
comparable between the different cases, and we measure them at the
``end of inspiral'' to be 5.7\%, 6.0\%, and 5.8\% for OldGauge,
g0+TR2+h, and BestGauge cases, respectively. We conclude that
although amplitude errors are clearly smaller during inspiral in the
BestGauge case, they are comparable just prior to merger.

In Sec.~\ref{sec:noise_reduction}, we established that, when compared
to the OldGauge case, the BestGauge case significantly reduces noise
in gravitational waveforms, particularly at peak noise
frequencies. Further, we showed in the previous section that GWs
produced by BestGauge are very consistently fifth-order
convergent, while the OldGauge case experiences disturbing
oscillations in implied convergence order. We suspect that these
convergence order oscillations are related to the much noisier
waveforms in OldGauge, but we would na\"\i vely 
expect that the
noise should reduce with increasing resolution. The next section
demonstrates that at peak noise frequencies, the noise in OldGauge
{\it does not} diminish with increasing resolution.

\subsubsection{Waveform Noise Convergence Properties}
\label{sec:waveform_noise_conv}

Figure~\ref{Fig:Psi4_waveform_noise_convergence} presents a power
spectrum analysis of waveform noise convergence. The left panel shows
that the largest noise-related spike in OldGauge Re($\psi^4_{2,2}$) power
does not appear to diminish with increased
resolution, lending support to the notion that
poor waveform convergence in OldGauge is due to high-frequency
noise. In fact, we find that the power attributable to noise in
OldGauge at $P\approx 10M$ is about 1\% of the {\it maximum} power,
corresponding to the physical GW signal (at GW periods of half the
binary orbital period). Next notice that this large, non-converging
noise-related spike in the power spectrum in OldGauge runs has been
reduced by about an order of magnitude in our BestGauge gauge choice. This
is consistent with the time-domain data of
Fig.~\ref{Fig:Psi4_noise_reduction}, in which waveform
noise at ``lr'' resolution is shown to be hugely reduced in the
BestGauge case, as compared to OldGauge. In BestGauge, we observe
three distinct bumps in the power spectrum, likely associated with
noise. Unlike the dominant
noise-related spike at $P\approx 10M$ in OldGauge, these three bumps
in the BestGauge power spectra appear to drop more cleanly with
increasing resolution than OldGauge, and as resolution is
increased, the bumps appear to decrease in amplitude and shift toward
shorter wave periods. 


\section{Conclusions and Future Work}
\label{Conclusions}

It has been found that the standard BSSN/MP+AMR paradigm
yields inconsistent convergence in gravitational waveforms in BBH
evolutions. Without consistent convergence, it may be impossible to
produce reliable error estimates for these waveforms, particularly
when very high accuracies are needed. It has been hypothesized
\cite{Zlochower:2012fk} that this inconsistent convergence may be
related to short-wavelength waves being reflected from grid-refinement
boundaries, producing high-frequency noise and constraint violations.

We present a set of improvements to the moving-puncture gauge 
conditions that, with {\it negligible} increase in computational
expense, greatly reduce noise and improve convergence 
properties of gravitational waveforms from BBH inspiral and
mergers. These improvements are presented in the context of an NRAR
\cite{Hinder:2013oqa} BBH calculation that evolves a low-eccentricity BBH
system $\approx 11$ orbits to merger,
where one BH has an initial dimensionless spin of $0.3$ aligned with the
orbital angular momentum, and the other is nonspinning. Evolutions are performed at up to four resolutions with a
variety of gauge choices, starting with the ``standard'' moving
puncture gauge choice and adding improvements one by one until our
``best'' gauge choice is reached.

Our gauge improvements in part stem from the observation that
regularly-spaced spikes in Hamiltonian constraint 
violations are timed precisely to the grid-refinement boundary
crossings of an early outgoing wave traveling at coordinate speed
$\sqrt{2}$. Based on linear analyses \cite{Alcubierre:2002iq}, we
know of only one propagation mode
with that speed in the standard BSSN/MP formulation, which primarily
involves the lapse and is governed by the lapse evolution 
gauge condition. 

The initial outgoing lapse wave pulse possesses sharp features, which
are problematic because, as \cite{Zlochower:2012fk} 
points out, high-frequency waves crossing into a coarser AMR grid will
be partially reflected at the boundary, generating noise. Now the lapse
{\it and its derivatives} are strongly coupled to the gravitational
field evolution equations, so noise 
generated by this sharp outgoing lapse pulse crossing refinement
boundaries can be easily converted to noise in gravitational field
variables (such as GWs) and constraint-violating modes
(cf. \cite{Etienne:2011re}).

Guided by this, we focus our efforts primarily on modifications to the lapse
evolution equation, aimed at stretching and smoothing the initial
outgoing lapse wave pulse. We stretch the lapse wave by
monotonically increasing its speed as it propagates outward so that
the front of the lapse pulse propagates slightly faster than 
the back. We smooth the
lapse wave by adding both parabolic and stronger-than-usual
Kreiss-Oliger dissipation terms. The most significant improvements
spawn from the stretching of the initial lapse pulse. 

Despite the stretching of the initial lapse pulse, early-time
spikes in the constraint violations are not reduced. We are able to
significantly tamp down these spikes via a time reparameterization
that greatly accelerates the initial evolution of the lapse and shift
relative to the BSSN gravitational field evolution
variables. Effectively, this modification makes lapse waves propagate
at speeds up to $\sim 30$ times the coordinate speed of
light initially, enabling the lapse to respond much more quickly to
the rapidly-settling gravitational fields in the strong-field region
during the early evolution. 

Our ``best'' gauge condition (BestGauge) reduces Hamiltonian and
momentum constraint violations by factors of $\sim 20$ and $\sim 13$,
respectively. In addition, with this same gauge choice, noise in 
the dominant gravitational-wave mode---$\psi^4_{2,2}(t)$---is reduced
by nearly an order of magnitude, particularly at peak noise
frequencies. Such numerical noise can be far more problematic in
sub-dominant [i.e., $(l,m)\neq$(2,2)] modes, at times completely
obscuring the signal. As an example, with OldGauge, the $(l,m)$=(4,4)
mode is noise-dominated throughout much of the inspiral, but with the
new gauge improvements, this mode is much cleaner and more amenable to
analysis. We observe improvements, generally of a lesser degree, in
other sub-dominant modes as well.

Finally, we observe a significant reduction in ADM mass and angular
momentum noise at large radius,
particularly during inspiral and merger. Regardless of gauge choice,
we observe a large amount of noise in ADM angular momentum just after
merger, corresponding to a spike in momentum constraint at about this
time. We are uncertain of the cause for this noise/spike, but suspect they may
be due to either rapid coordinate and spacetime evolution associated with
BH merger, or possibly high-frequency {\it physical} waves propagating
outward into less-refined numerical grids. This warrants further
investigation, as the former hypothesis can be tested through a
late-time time reparameterization and the latter through grid
structure adjustments.

In addition to analyses of how constraint violations and noise are
affected by our new gauge choices, we perform a suite of constraint-violation,
irreducible-mass, and waveform ($\psi^4_{2,2}$) convergence
tests with a set of three gauge choices, starting from the ``standard''
moving-puncture gauge choice (OldGauge) to our ``best'' gauge
choice (BestGauge).

Using the ``standard'' moving-puncture gauge choice (OldGauge), the L2 norm of
Hamiltonian constraint violations converges to zero at between first
and second order with increasing resolution. If this L2 norm integral
excludes the wavezone region far from the binary, convergence order
increases to between fifth and sixth order, indicating that
subconvergence is related to effects far from the binary, on distant,
low-resolution grids. Meanwhile, with our improved gauge choices, we
observe between fifth- and sixth-order convergence in this diagnostic
{\it even when the integral extends to the outer boundary}. Though
momentum constraint violations at a given resolution are significantly
reduced with our new gauge choices, we do not observe an
improvement in the convergence to zero of momentum constraint
violations. This may be due to the fact that even in our ``best''
gauge choice, some high-frequency noise remains in our BBH
calculations.

Based on our waveform convergence analyses, we find that
$\psi^4_{2,2}(t)$ data at \{mr,hr,hhr\} resolutions yield approximate
fifth-order convergence, though large fluctuations around fifth order
are observed in the OldGauge case. These fluctuations are
significantly reduced when improved gauge conditions are adopted,
particularly in the BestGauge case. 

With such consistent fifth-order convergence observed in the highest
three resolutions in our best gauge choice (BestGauge), we then
analyze the difference between two Richardson-extrapolated
realizations of phase and normalized amplitude. One realization uses data at ``hr'' and ``hhr''
resolutions and the other ``mr'' and ``hr'' resolutions. Any nonzero
difference between these Richardson-extrapolated realizations of
phase and normalized amplitude can be attributed to fluctuations in
assumed convergence order $n$ from $n=5$ (which appears to be related
to noise) or to error from higher-order terms
(cf. Eqs.~\ref{convergence} and~\ref{convergence_exact}). Since these
differences in Richardson-extrapolated values are directly related to
errors, we use them as our error estimates for the amplitude and phase
(though it may be an unconventional choice;
cf. \cite{Hinder:2013oqa}). Comparing the ``standard'' moving-puncture gauge
choice to our ``best'' gauge, we find significant (factor of $\sim$ 2)
reductions in both amplitude and phase errors during inspiral. At the
end of the inspiral, when the frequency of the $\psi^4_{2,2}$ is
$M\omega=0.2$, we observe roughly a factor of two reduction in phase
errors in the BestGauge case but no significant improvement in
amplitude errors, as these are dominated near merger by small offsets
in the time of merger at different resolutions.

We then perform a GW noise analysis, comparing OldGauge and
BestGauge results at different resolutions
(see Fig.~\ref{Fig:Psi4_waveform_noise_convergence}). In the OldGauge case, the largest
spike in GW power directly related to numerical GW noise, at wave
periods $P\approx 10M$ (Fig.~\ref{Fig:Psi4_noise_reduction}), does not
drop with increasing resolution. Further, we find that
the noise-dominated power in this OldGauge spike is about 1\% of
the physical GW power maximum at $P=0.5P_\text{orb}$ throughout much
of the early inspiral. GW power at wave periods associated with noise
in OldGauge runs is reduced by nearly an order of
magnitude in the BestGauge runs, and unlike OldGauge, the power at
these wave periods appears to drop monotonically with
resolution, suggesting that noise in BestGauge converges away more
cleanly. Given the excellent amplitude- and phase-convergence
properties observed in the BestGauge case, we believe this GW power
spectrum analysis strongly indicates that poor convergence in the
``standard'' moving-puncture gauge conditions may indeed be related to
poorly-convergent GW noise generated by the initial sharp outgoing lapse
pulse. 

Future work will examine remaining uncertainties in these
evolutions, including the nearly simultaneous large spike in momentum
constraint violations and large noise in the ADM angular momentum
surface integral at the end of our evolutions. Though we have found
that our gauge improvements greatly reduce noise in
many quantities apparently generated by the sharp initial outgoing
lapse pulse, we have not completely eliminated the noise, and it is
unknown how much more phase and amplitude errors can be driven
downward with the current improvements. Without doubt, higher
resolutions and higher-order evolutions will be helpful in determining
this.

Though these new gauge conditions and techniques have been
presented in the context of moving-puncture BBH evolutions only, we
fully expect them to be beneficial in a wide variety of NR contexts,
including compact binary systems with matter or even BBH evolutions
using other $N$+1 NR formulations. With the era of gravitational-wave
astronomy now upon us, we hope this work will spur others to join the
search for gauge conditions and methods optimal for generating high-quality
gravitational waveforms within the MP+AMR context, as each improvement
will accelerate our community toward its goals in this exciting time.

\acknowledgments This paper was supported in part by NSF Grants
PHY-0963136 and PHY-1300903 as well as NASA Grants NNX11AE11G and
NNX13AH44G at the University of Illinois at Urbana-Champaign. BJK and
JGB were supported by NASA grants 09-ATP09-0136 and 11-ATP-046.  VP
gratefully acknowledges support from a Fortner Fellowship at
UIUC. This work used the Extreme Science and Engineering Discovery
Environment (XSEDE), which is supported by NSF grant number
OCI-1053575. A significant portion of the calculations presented here
were performed as part of the Blue Waters sustained-petascale
computing project, which is supported by the National Science
Foundation (award number OCI 07-25070) and the state of Illinois. Blue
Waters is a joint effort of the University of Illinois at
Urbana-Champaign and its National Center for Supercomputing
Applications.

\appendix
\section{Time Reparameterizations Are Gauge Choices}
\label{TR_Appendix}
In this paper, we have introduced two time reparameterization
techniques. The first, ``TR1'',
multiplies the BSSN field and MP gauge evolution equation RHSs by a
function $f(t)$ (Eq.~\ref{f_of_t}), and the second, ``TR2'',
multiplies only the BSSN field evolution equation RHSs by
$f(t)$. Here, ``RHS'' is defined as in Eq.~(\ref{RHS}).

Although TR1 and TR2 modify the RHSs of the BSSN equations, these
reparameterizations are equivalent to gauge choices, so that
the resulting set of equations are identical to Einstein's
equations. This is most easily seen from the 3+1 line element:
\beq
ds^2 = -\alpha^2 dt^2 + \gamma_{ij} (dx^i + \beta^i dt) (dx^j + \beta^j dt).
\eeq
Our time reparameterizations TR1 and TR2 are equivalent to the
transformation $dt\to f(t')dt'$, and this $f(t')$ factor may be
absorbed into the lapse $\alpha$ and shift $\beta^i$, such that
\beqn
\ta &=& f(t') \alpha  \\
\tb^i &=& f(t') \beta^i.
\eeqn
From here, the BSSN field evolution equations may be written {\it
  unmodified}, simply replacing the lapse, shift, and time with $\ta$,
$\tb$, and the new ``primed'' time coordinate, respectively, which is
equivalent to multiplying the RHS of the original BSSN formulation by
f(t). Removing the tildes and primes yields the original BSSN field
equations exactly.

Though the BSSN field equations remain unchanged, the standard MP
gauge evolution equations for $\ta$ and $\tb$ {\it are} modified for
$f(t)\neq 1$, though in the TR1 case the principal parts of all
equations will be identical.

In the TR2 case, however, the principal part analysis for $\ta$ yields
a wave equation with wave speeds identical to the standard moving
puncture lapse speed, but divided by $\sqrt{f(t')}$. I.e., TR2 greatly
increases the lapse characteristic wave speed at early times when
$f(t')$ is very small. Such large wave speeds would require very 
small timesteps in our numerical scheme, and as we have not
implemented a ``true'' adaptive timestepping numerical algorithm, we
opt not to evolve $\ta$ and $\tb$ as just described, and instead
choose to multiply the BSSN field evolution equation RHSs by
$f(t)$. As we have shown, TR2 is equivalent to a gauge
modification, but TR2 is far easier to implement, as it
requires no modification to our fixed-timestep time evolution
algorithms.


\bibliographystyle{apsrev4-1}
\bibliography{paper}

\end{document}